\documentclass[%
 reprint,
 amsmath,amssymb,
 aps,
]{revtex4-2}

\usepackage{graphicx}
\usepackage[dvipsnames]{xcolor}
\usepackage{dcolumn}
\usepackage{bm}
\usepackage{comment}
\usepackage{xspace}

\usepackage[version=3]{mhchem} 
\usepackage{subcaption}
\usepackage{amsmath}
\usepackage{mathtools}
\usepackage{soul}
\usepackage{url}

\newcommand{\pref}[1]{(\ref{#1})}
\newcommand{\epref}[1]{Eq.~(\ref{#1})}
\newcommand{\figref}[1]{Fig.~\ref{#1}}
\newcommand{\half}{\frac 12}
\newcommand{\ie}{\textit{i.e.}\xspace}
\newcommand{\etal}{\textit{et al.}\xspace}
\newcommand{\eg}{\textit{e.g.}}

\newcommand{\ai}{\textit{ab-initio}}

\newcommand{\si}[1]{Supplemental Material#1}
\newcommand{\HOMOo}{HOMO$^0$}
\newcommand{\LUMOo}{LUMO$^0$}

\newcommand{\revisiondiff}{0}
\newcommand{\revisionnew}{1}

\newcommand{\revisiontype}{\revisiondiff}
\newcommand{\revision}[2]{%
\if\revisiontype\revisiondiff
     {\color{Red}\st{#1}}{\color{Green}#2}
\else
     \if\revisiontype\revisionnew
         #2
     \else
         #1
     \fi
\fi}

\begin{document}


\title{A minimal model of inelastic tunneling of vibrating magnetic molecules on superconducting substrates}

\author{Athanasios Koliogiorgos}
 \email{athanasios.koliogiorgos@matfyz.cuni.cz}
\affiliation{%
Department of Condensed Matter Physics, Faculty of Mathematics and Physics, Charles University, Prague, Czech Republic
}%
\author{Richard Korytár}%
 \email{richard.korytar@ur.de}
\affiliation{%
Department of Condensed Matter Physics, Faculty of Mathematics and Physics, Charles University, Prague, Czech Republic
}%
\affiliation{%
Institute of Theoretical Physics, University of Regensburg, 93040 Regensburg, Germany
}

\date{\today}

\begin{abstract}
We present an efficient method of calculating the vibrational spectrum of
a magnetic molecule adsorbed on a superconductor, directly related to the first derivative of the tunneling $IV$ curve. The work is motivated by a recent scanning-tunneling spectroscopy of lead phthalocyanine on superconducting Pb(100), showing a wealth of vibrational excitations, the number of which highly exceeds molecular vibrations typically encountered on normal metals.
We design a minimal model which represents the inelastic transitions by the spectral function of a frontier orbital of the molecule in isolation. The model allows for an exact solution; otherwise the full correlated superconducting problem would be hard to treat. The model parameters are supplied from an \ai{} calculation, where the presence of the surface on the deformation of molecular geometry can be taken into account. The spectral function of the highest-occupied molecular orbital of the anionic PbPc$^{1-}$ shows the best agreement with the experimental reference among other molecular charge states and orbitals. The method allows to include multiple vibrational transitions straightforwardly.
\end{abstract}

\maketitle



\section{Introduction}
The experimental ability to contact single molecules (through scanning-tunneling, break-junction and other techniques) made it possible to study molecular many-body excitations by electron transport \cite{Scott2010,Scheer2017,Evers2020}. Unlike in optical spectroscopies, the excitation energy is provided by the tunneling electron, whose energy is controlled by the voltage bias $V$ across the junction. Magnetic molecules play a prominent role here because their low-bias electron transport spectra can be endowed with inelastic spin transitions or Kondo peaks, if the contacts are normal metals. If one or two contacts are superconducting, a narrow in-gap resonance, known as Yu-Shiba-Rusinov peak, may arise \cite{Heinrich2018}. In molecules, these spin-excitations often reflect interactions with other orbital and vibrational degrees of freedom \cite{Gauyacq2012,Rakhmilevitch2014, Minamitani2012, Ruby2016, Zalom2019}, thus providing insights into the many-body processes at molecular scale and the way the chemical environment
affects them \cite{Choi2017,Zonda2021,Li2020,Trivini2023}.

In a recent paper, Homberg \etal{} have reported resonances
in a PbPc molecule on a superconducting Pb(100) which they attributed to
vibrational excitations \cite{Homberg2022}.
These excitations appeared as peaks in the first derivative of the
current-voltage relation, $I(V)$, within the superconducting gap,
in a scanning-tunneling setup.  
The spectra of Ref.~\cite{Homberg2022} are exceptional as they 
deliver more than 40 sharp peaks in the $\mathrm dI/\mathrm dV$, compared to the commonly
encountered single peak in the normal-state 
$\mathrm d^2I/\mathrm dV^2$ of molecular junctions and adsorbates (see Ref.~\cite{Evers2020}
for a review).
Undoubtedly, the observation paves way to obtaining valuable information
about the adsorbate and its processes from the vibrational spectrum.
Nonetheless, the price to pay for such wealth of peaks is that the theoretical
description is very challenging. Apart
from the complexity of vibrational transitions in presence of an electronic
continuum, the system is superconducting and the molecule is paramagnetic due to Coulomb blockade.
At present, efficient solution of the problem is not available even for simplified model
Hamiltonians
that exclude the atomistic details. However, it is possible to gain valuable insights
into the vibrational transitions from pragmatic approaches that simplify the
problem to a form tractable by an \ai{} method. This is the attitude taken
in Ref.~\cite{Homberg2022}. In this manuscript we we offer an alternative \ai{} strategy.

The authors of Ref.~\cite{Homberg2022} employed density-functional theory (DFT)
calculations of the electronic and vibrational structure of the PbPc/Pb(100).
The results were fed into a non-equilibrium Green's function technique to
calculate the electric current as a function of voltage, $I(V)$,
which includes the electron-phonon coupling in a lowest-order expansion.
This technique operates in a normal-state metal, therefore, it includes
spurious renormalizations of the vibrations (\eg{} frequency shifts) from
 the (un-gapped) Fermi surface. Moreover, the first derivative, $\mathrm dI/\mathrm dV$,
produces step-like transitions instead of sharp peaks, because the density
of states around the Fermi level is continuous. Despite these conceptual
problems, therein reported quantitative agreement with the experiment is fairly 
good.

In this work we offer an alternative computational approach that avoids
the problem of the continuous density of states (spurious metallicity).
The central object of our approach is the electronic spectral
function of the molecule, which is directly related to the \textit{first}
derivative of the current in the tunneling limit of STM. We calculate
the vibrational transitions from given electronic orbitals of the molecule in
isolation. The resulting theoretical spectra of PbPc compare well with
the experiment and offer a new interpretation of correlated electronic transport through the adsorbate.

\section{Methodology}
\subsection{\label{sec:theory}Theoretical approach}
In scanning-tunneling spectroscopy, the STM tip acts as a probe of the local spectral 
function of the sample $A(\mathbf r, E)$, \ie{} the spectrum of electron addition and removal
processes at point $\mathbf r$ and energy $E$\cite{Tersoff1985,Meir1992,Coleman2015}.
The latter is directly proportional to
the differential conductance,
\begin{equation}
\label{eq:ldos}
\frac{\mathrm d I(V)}{\mathrm d V}\Biggl|_{\text{Tip over } \mathbf r} = b
A(\mathbf r, E_\mathrm F + V)
\end{equation}
where $E_\mathrm F$ is the Fermi energy of the sample and $V$ is the bias voltage.
The $A(\mathbf r, E)$ is independent on the chemical potential of the tip and the prefactor $b$
is weakly energy dependent.
The expression \pref{eq:ldos} is valid in the tunneling limit of a large tip-sample separation (tip height $z$);
the differential conductance provides information about the sample only, up to a $z$-dependent pre-factor.

If the STM tip is superconducting, the situation is more complicated because of Andreev processes involving
the transfer of Cooper pairs. It was shown by Ruby \etal{} \cite{Ruby2015} that in the large $z$ limit the
transport is dominated by single-electron tunneling. The latter is again represented by the
$A(\mathbf r, E)$, albeit it enters $I(V)$ convoluted by the singular 
tip density of states. To address electron-vibrational transitions of magnetic adsorbates, we
therefore focus on the local spectral function.

The physical basis that allows the observation of rich vibrational spectra in Ref.~\cite{Homberg2022} 
is the so called Andreev bound state  \cite{Balatsky2006},
resulting from interaction
of the molecular magnetic moment with the superconducting host, producing sharp peaks
in the spectral function inside the superconducting gap. These peaks are often
called Yu-Shiba-Rusinov (YSR) resonances. Furthermore, the existence of the
magnetic moment of the adsorbed molecule implies Coulomb blockade phenomenon\cite{Evers2020}.
On top of that, the molecular electronic degrees of freedom interact with
vibrations, allowing inelastic transitions in the electron addition (removal) processes.
The salient vibrational spectrum results from a combination of complex physical processes: interaction
of localized vibrations with delocalized electronic states, superconductivity
and magnetism. At present there are no controlled approximations to this
problem; useful insights into the spectra can only be gained from
\textit{toy-models} that simplify the problem considerably. We take this
pragmatic approach and construct a simplified model that allows part of the problem to be treated \ai{} to deliver insights taking into account the atomistic details of the electron-vibrational transitions of the molecular tunneling spectra.

The basis of our approach is the assumption that the vibrational spectrum results from
vibrational coupling with a discrete (in-gap) electronic resonance, labeled 
by $\alpha$. We do not adopt further
assumptions regarding the origin of the resonance $\alpha$. The latter could be viewed as an
electron-transporting frontier orbital, as it is common in molecular electronics \cite{Evers2020},
or the open-shell orbital hosting the Coulomb blockade and YSR resonance. 
We chose for the $\alpha$ one of the two
frontier orbitals and make an unbiased comparison between them.

\subsubsection{Definition of the minimal model}
Given the previously mentioned considerations, we truncate the physical problem
to the interaction of a single molecular orbital with molecular vibrations.
This familiar many-body problem offers direct access to the electronic
spectral function.
Given is a single electronic orbital (labeled by $\alpha$) and discrete independent vibrational excitations $v=1,\ldots N_{vib}$. The Hamiltonian reads
\begin{equation}
\label{eq:ham}
\hat H = E_\alpha \,\hat c^\dagger_\alpha \hat c^{\phantom\dagger}_\alpha +
+ \sum_v \hbar\omega_v \, \hat b^\dagger_v\hat b_v^{\phantom\dagger} +
\sum_{v} \lambda_{\alpha\alpha}^{v}\, \hat c^\dagger_\alpha \hat c^{\phantom\dagger}_\alpha
(\hat b_v + \hat b^\dagger_v),
\end{equation}
where $\hat c_\alpha,\hat b_\alpha$ are canonical annihilation operators of the electronic and vibrational degrees of freedom, $E_\alpha$ is the ``on-site'' energy of the orbital, $\omega_v$ are vibrational frequencies and  \(\lambda_{\alpha\alpha}^v\) are (diagonal) electron-vibrational (EV) coupling matrix elements.

In absence of EV coupling (\(\lambda_{\alpha\alpha}^\upsilon=0)\) the
electronic spectral function is $A^{(0)}_\alpha(E) = \delta(E - E_\alpha)$.  In
presence of finite EV coupling the zero-temperature spectral function is also known
exactly\cite{Mahan2000} \citep{Nitzan2006}, and reads
\begin{subequations}
\begin{align}
A_\alpha(E)\ &{=}\ A_\alpha^{(0)}(E) + A^{(2)}_\alpha(E) + A^{(4)}_\alpha(E) + \mathcal{O}\left(g_v^6\right)\\
A^{(2)}_\alpha(E)\ &{=}\ 
 \sum_{v} g_v^2\ \delta(E - \hbar\omega_v - E'_\alpha)\\
A^{(4)}_\alpha(E)\ &{=}\ 
 \half \sum_{v,v'} g_v^2g_{v'}^2\ \delta(E - \hbar\omega_{v'} - \hbar\omega_v - E'_\alpha).
\end{align}
\label{eq:sf}
\end{subequations}
The zero-order term denotes the unperturbed spectral function with
$E_\alpha$ substituted by a renormalized $E'_\alpha$. We shall not
reproduce the exact formula of the latter as it is of no importance in this work.
The remaining terms in \pref{eq:sf}
describe multiple EV transitions, controlled by the dimension-less coupling constant 
\begin{equation}
    g_v := {\lambda^{v}_{\alpha\alpha}}/{\hbar\omega_v}.
    \label{eq:g}
\end{equation} 
The second order term $A_\alpha^{(2)}(E)$, decorates the orbital resonance by
single vibrational excitations, displaced by the vibrational energy quanta
$\hbar\omega_v$ and scaled by the squared coupling constants. The fourth order
term $A^{(4)}_\alpha(E)$ describes electron addition (or removal) accompanied
by simultaneous excitations of two vibrations. Therefore it yields a resonance
displaced by the sum of two vibrational energies. Higher order terms are not
shown.
Finite temperature $T$ subjects the spectral function to corrections of the
order of the Boltzmann factor $e^{-\hbar\omega_v/k_\mathrm BT}$.
 The latter number results negligibly small for the 
experimentally relevant case, $T=4.2K$ (typical $\hbar\omega=40$ meV).

In usual inelastic electron tunneling spectroscopy (IETS) of molecules,
the EV coupling constants are usually of the
order of few percent (see \eg{} the discussion in \cite{Gauyacq2012}),
resulting spectral changes are small and two-vibrational excitations are
strongly suppressed. In \ai{} treatments of molecular
IETS\cite{Frederiksen2007}, two-vibrational excitations are discarded based on
this reasoning.
Here we include two-vibrational processes as shown in the \epref{eq:sf}, because some coupling
constants reach $\approx 0.15$ (for the case of gas phase neutral PbPc.)

The above defined simplified model allows us to treat a part of the problem by \ai{}: the molecular
vibrational and electronic spectra and their mutual interactions.
We employ the \epref{eq:sf} to calculate the spectral function of the PbPc on superconducting Pb(100). We shall use \ai{} DFT to feed the parameters $g_v, \omega_v$. We consider the molecule in isolation (our approach does not work with the electronic continuum of the supporting metal, because it would turn the molecular orbitals into wide resonances). The (renormalized) orbital energy $E_\alpha$ is not determined by the DFT as it results from the YSR physics; it will be treated as an unknown parameter which globally shifts the zero energy. We choose for $\alpha$ the frontier orbital, either the highest occupied molecular orbital (HOMO) or the lowest unoccupied molecular orbital (LUMO).
This approach does not describe the spatial dependence of the spectral
function [see \epref{eq:ldos}], because the $\alpha$ is an \textit{a-priori}
unknown parameter. In most cases, electronic transport through single molecules
at low bias is dominated by a single frontier orbital. We shall comment on this
later in Results.

\subsection{\label{sec:details}\textit{Ab-initio}
details}
\subsubsection{Geometry selection and relaxation}
The PbPc molecule was simulated in real space with DFT, in a finite structure
framework, using the program Turbomole, version 7.6\cite{turbomole, turbomole1989, turbomole1995, turbomole1998}.
First, a geometry optimization of the molecule was performed. Calculation of the electron-vibration (EV) coupling requires a high degree of accuracy in the DFT parameters.
Thus, for the geometry and subsequent calculations, a triple-zeta polarized basis set of the Ahlrichs group was employed, namely def2-TZVP\cite{karlsruheBasis}. When needed, the resolution-of-identity approximation was used, along with auxiliary basis sets \cite{rij1995, rij1997, rij2006}. The calculation was considered converged when the difference in energy between successive self-consistent field (SCF) cycles was smaller than \(10^{-8}\). Two functionals were used from two levels of theory: the Perdew-Burke-Ernzerhof (PBE) functional of Generalized Gradient Approximation (GGA), and the hybrid functional B3LYP\cite{pbeFunc,BeckeB3,LYP}. PBE is generally known to predict satisfactorily the structural properties while underestimating the band and HOMO-LUMO gap\cite{koliog1,koliog2}. B3LYP predicts more accurately the electronic structure and is being widely used in molecular simulations, and specifically in the simulation of PbPc molecule\cite{PbPc2,PbPc1}. Geometry optimization is followed by calculation of the electronic structure, vibrational modes and, finally, EV coupling. 

To approximately account for geometric distortions that PbPc suffers upon adsorption onto Pb(100), we introduce an additional type of PbPc geometry.
Namely, we relaxed PbPc in an equilibrium position on top of a 
non-periodic Pb(100) slab consisting of 114 atoms. The Pb(100) slab was converged without further relaxing the geometry, with an initial geometry based on bulk Pb(100), so that it approximates more accurately a 2D surface, as a fully relaxed slab comprising of 114 atoms has a much more distorted geometry compared to a 2D material. The coordinates of the slab remained fixed during the PbPc relaxation. A D3-DFT dispersion correction was used to account for the van der Waals forces between PbPc and Pb(100) \cite{Grimme2010}. In order to be able to calculate the binding energy of the compound, the D3 correction was also set for the separate components, so that all calculations were on the same level of theory. The slab atoms were discarded for the EV calculation. By allowing the molecule to fully relax on top of the Pb slab, we were able to determine its binding energy and its final exact position on the surface. The PbPc molecule was oriented with the Pb atom of the molecule pointing away from the substrate, according to the predominant configuration in the experimental study of \cite{Homberg2022}. In the experiment, the molecules adsorbed on the Pb(100) surface aggregate into 4 different orientations \cite{Homberg2022b}. In our calculations, the molecule after relaxation sits on the Pb surface in the way that is depicted in Fig.~S10 in the \si{} \cite{SM}. This orientation is one of the 4 in the above-mentioned experiment. 


In the experiment of Homberg \etal{} the PbPc presumably turns
anionic (and magnetic) when adsorbed as a 
monolayer\cite{Homberg2022,Homberg2022b}. The additional electron on the molecule
occupies the LUMO (e$_{\mathrm g}$) and the magnetic moment develops due to Coulomb
blockade. Therefore, we also calculate anionic PbPc$^{1-}$ with two kinds
of geometries. First, a geometry of the neutral PbPc relaxed on the surface.
We shall refer to this structure as the C$_{4v}$ PbPc/surf. The symmetry assignment
is justified in Results.
Second, the geometry of relaxed PbPc$^{1-}$ in vacuum. The latter undergoes a
Jahn-Teller distortion due to the twofold degeneracy of the $e_{\mathrm g}$ LUMO.
We refer to this structure as C$_{2v}$ PbPc$^{1-}$. We also calculated the doubly charged anion PbPc$^{2-}$.

\subsubsection{Electron-vibrational (EV) coupling}
We employ a method developed by B\"urkle \etal{} and implemented in the latest editions of Turbomole \cite{Burkle, aoforce2002, aoforce2002b, aoforce2004, aoforce2017}. The calculation proceeds in two steps:
First, from the standard calculation of vibrations we obtain the eigenvectors of the dynamical 
matrix \(\mathcal{C}_\chi^v\), where \(\chi\) is an index that refers to atom and spatial direction. This is converted to mass-normalized normal modes \(\mathcal{A}_\chi^v=\mathcal{C}_\chi^v/\sqrt{M_k}\) for each \(k\) atom, where \(M_k\) is the normalized mass.
We also obtain the vibrational frequencies \(\omega_v\) of each of the $v$ normal modes of vibration.
(For PbPc, the number of vibrational modes is 171.) Second, from the EV coupling calculation we obtain the first order derivative of the Kohn-Sham operator with respect to the atomic displacements: \(H^{e}_{\mu\nu,\chi}=\langle \mu|\frac{d\hat{H}^e}{d\chi}|\nu \rangle \), where $\mu,\nu$ 
are the basis functions. From these data, we calculate the EV coupling constants
\begin{equation}
\label{eq3}
    \lambda_{\mu\nu}^v = \left(\frac{\hbar}{2\omega_v} \right)^{1/2}\sum_{\chi}\langle \mu|\frac{d\hat{H}^e}{d\chi}|\nu \rangle \mathcal{A}_\chi^v.
\end{equation}
Following that, we calculate the dimensionless \(g_v\)  from the \epref{eq:sf}.

\section{\label{se:results}Results}

\subsection{Basic properties of PbPc and PbPc$^{1-}$ in isolation and in the potential of Pb(100)}
The molecule in isolation relaxes to a non planar geometry where the Pb atom is protruded above the Pc plane
(see inset of \figref{fig1}),
with a distance of Pb from the Pc plane of 2.15 \AA\ (PBE) and 2.35 \AA\ (B3LYP).
The orthogonal distance between the extremes of the Pc plane is 1.23 nm (PBE) and 1.22 nm (B3LYP).
The HOMO-LUMO gap is 1.363 eV (PBE) and 2.096 eV (B3LYP). 
These numbers are in agreement with previous reports \cite{PbPc2} and 
compare well with the experimental data 2.23 \AA, 1.23 nm and 2.01 eV, respectively\cite{scirep}.
We note a near-perfect agreement of the PBE structural parameters with the experiment.

The gas-phase PbPc enjoys a C$_{4v}$ symmetry that results in a doubly-degenerate lowest unoccupied molecular orbital (LUMO) (see \si{, Fig.~S5}) \cite{SM}. Upon charging with 1 electron, the molecule lowers its total energy by distorting to C$_{2v}$. As demonstrated in the \si{, Fig.~S9} \cite{SM}, this is mainly due to folding of one pair of opposite hexagonal lobes, \ie{} the B$_\text{1g}$ mode is the most prominent (cf. also Ref.~\cite{Tobik2007}). The energy of Jahn-Teller distortion is 66.7 meV.

The PbPc molecule in proximity to Pb(100) undergoes a significant structural change as can be seen in the inset of \figref{fig2}. 
Instead of the curved structure characteristic of gas phase PbPc, PbPc on Pb(100) turns to almost planar. The molecule after relaxation rests at a distance of approximately 3.5 \AA\
from the Pb(100) surface, in good comparison with the experimental distance of 3.7 \AA\ \cite{Homberg2022}. 
The top-view of the relaxed adsorption geometry is shown in \si{, Fig.~S10}, showing
no reduction of molecular symmetry \cite{SM}. We confirm the statement of Homberg \etal{} that
standalone adsorbate is subject to a negligible charge transfer. The total energy of the PbPc/Pb(100) pair was -23,856.905 Hartree. The energy of the standalone PbPc was -1,858.744 Hartree, while the energy of the Pb slab was -21,998.08 Hartree. Thus, the binding energy of the PbPc-Pb(100) compound was determined at 0.081 Hartree, or 2.20 eV.

\subsection{Analysis of spectra of PbPc}

\begin{figure}
\subcaptionbox{}{\includegraphics[width=0.45\textwidth]{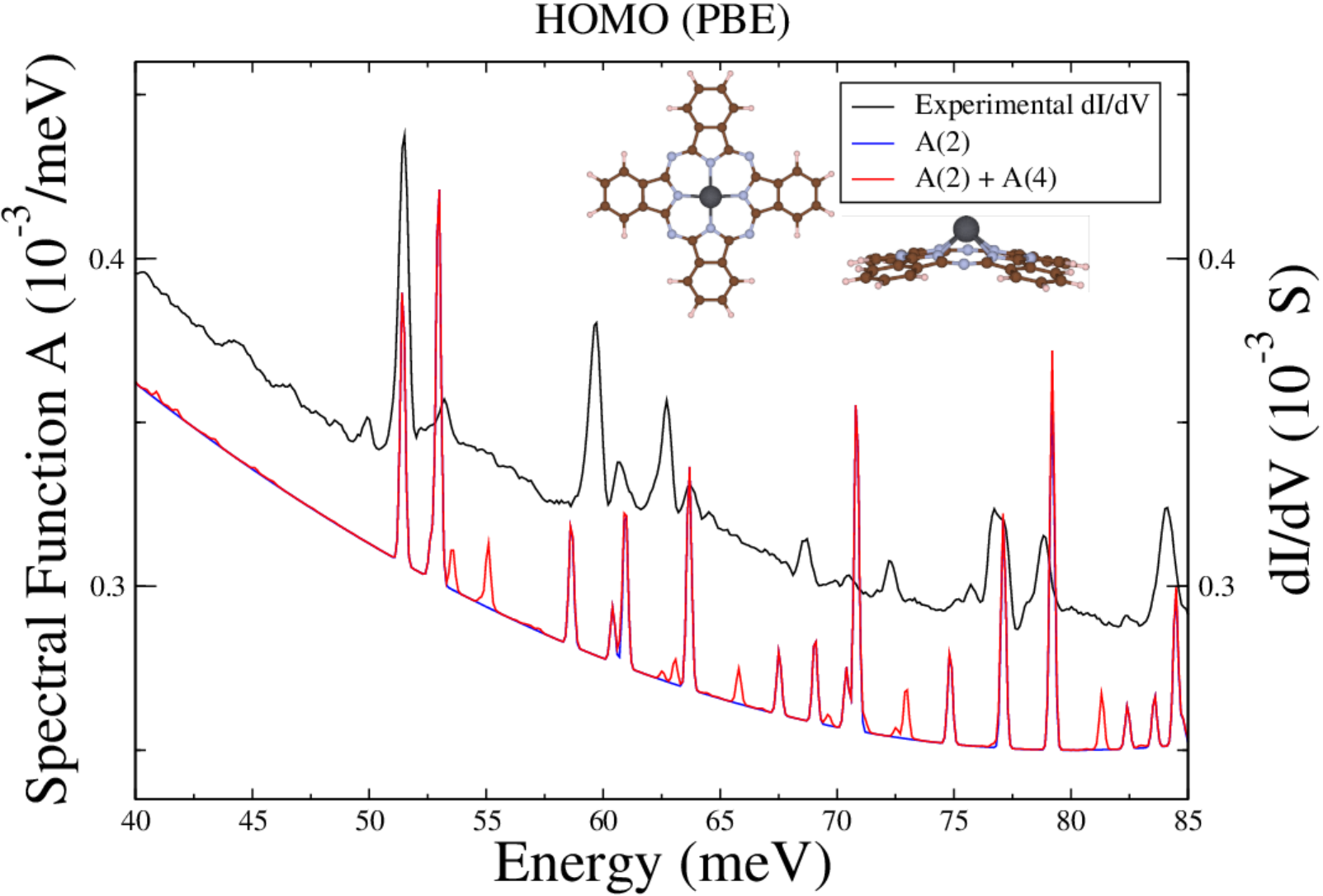}}
\subcaptionbox{}{\includegraphics[width=0.45\textwidth]{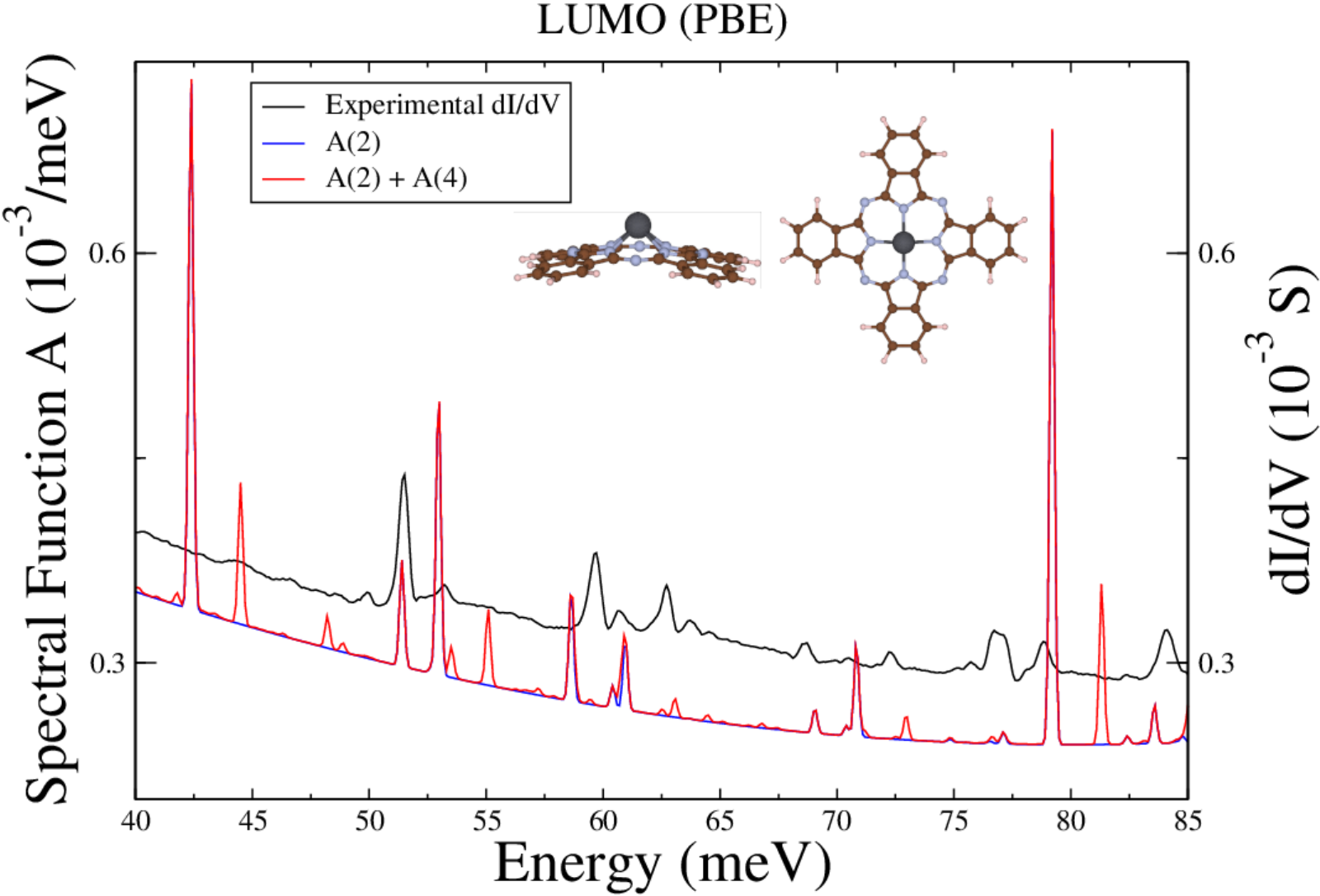}}
\centering
\caption{Spectral function describing single \([A^{(2)}_\alpha(\omega)]\) and double \([A^{(4)}_\alpha(\omega)]\) electron-vibrational transitions from the molecular orbitals $\alpha=$HOMO (a) and $\alpha=$LUMO (b) using the PBE functional, for the C$_{4v}$ PbPc relaxed in vacuum (see inset for the model). The experimental \(dI/dV\) spectrum from Ref.\cite{Homberg2022} is included. To facilitate comparison, a polynomial background has been added to the theoretical spectrum.}
\label{fig1}
\end{figure}

\figref{fig1} shows the HOMO and LUMO spectral functions of isolated PbPc alongside with the experimental d\(I\)/d\(V\) from Ref.~\cite{Homberg2022}.
To facilitate comparison, we have blueshifted the experimental spectrum
by 1 meV and we have added a $2^\text{rd}$ order polynomial that decreases slowly with energy.
The overall pre-factor of the theoretical spectrum was adjusted to facilitate the comparison with the experimental data. In the experiment it is largely controlled by the strength of the matrix element representing tunneling from the tip through the vacuum to the molecule. Therefore, only the relative peak heights are sample-specific. The theoretical spectrum
was convoluted by a Gaussian with broadening 0.1 meV. The dominant contributions to the
broadening of the peaks seen in the experiment
are the experimental resolution (of the van Hove singularity) 
and vibrational decay into the phonon continuum.
The latter effect should
be negligible for vibrational energies higher than the surface band limit. Here, we employ the Gaussian broadening
to facilitate the comparison. 
We also report the unconvoluted data, \textit{i.e.} the coupling constants $g_v$ 
(see \epref{eq:g}) in the \si{, Fig.~{S14}} \cite{SM}.

A quick glance through the graphs reveals two energy intervals with distinct performance of the theoretical model. 
Above $\approx 30$ meV it is possible to match certain theoretical peaks with
the experimental ones. Clustering of peaks seen in the experiment is
reproduced, although the model peak energies and relative peak heights do not
always correspond. We observe that the agreement with the experiment is better
for the HOMO than LUMO. For energies below $\approx 30$ meV the model exaggerates peak heights rendering peak identification impossible.
This threshold energy is above the phonon band width of lead $\approx 10$ meV \cite{Sklyadneva2012,Verstraete2008}.
The molecular vibrations in this low-energy end are either resonant with the vibrational continuum of the phonon band,
or strongly interacting with it, 
resulting in pronounced energy shifts and damping. A wider spectrum which includes the low energy range for the case of HOMO is presented in Fig.~S11 in the \si{} \cite{SM}.

Next, we turn to the anionic open-shell PbPc$^{1-}$ with the C$_{4v}$ PbPc/surf
structure relaxed on top of the Pb surface, as described in Methods.
We note that the
spin-unrestricted DFT calculation with odd electron count delivers a spin-polarized system.

The spectral functions of minority-spin HOMO and LUMO of 
PbPc$^{1-}$ with the   C$_{4v}$ PbPc/surf geometry are shown in \figref{fig2}. The agreement with the experiment is again better for the
HOMO than it is for LUMO and it is much better than for the molecule in isolation,
\figref{fig1}. The improvement is consistent with the assumption of Homberg \etal{}\cite{Homberg2022}
that PbPc develops a magnetic moment by becoming anionic.
The majority-spin spectral function, however, compares much worse, as can be seen in 
the \si{, Fig.~S2} \cite{SM}. We remark that this spin direction is
Coulomb-blockaded for electron transport in the open-shell Kohn-Sham DFT. The poor agreement in
Fig.~S2 compared to the minority-spin is therefore consistent with the open-shell state of the molecule. 

Finally, we turn to the anion's spectral functions with the C$_{2v}$ PbPc$^{1-}$ geometry,
Fig.~S12 and Fig.~S13 in the \si{} \cite{SM}. The comparison
with the experimental conductance is much worse in this case. This observation
is consistent with the fact that Jahn-Teller distortions tend to be quenched on
metallic surfaces, because the electron tunneling rate of the
e$_{\mathrm g}$ states
is much higher than the velocity of the distorting mode.

Summarizing, the spectral function of the minority-spin HOMO of 
PbPc$^{1-}$ with the C$_{4v}$ geometry shows the best agreement, consistently with
the assumptions of Ref.~\cite{Homberg2022}. A closer inspection of \figref{fig2}(a) shows
that out of the 22 most intense experimental peaks, 20 of them have a corresponding 
theoretical counterpart. The remaining 2 are not reproduced in the theory curve.
For each of the 20 mutually corresponding peaks we extract their energy centers,
$E^{exp}_i$ and $E^{th}_i$ and peak
heights with relative to the smooth background, $H^{exp}_i$ and $H^{th}_i$,
see \si{, Table I}. The average error of the peak energy, 0.286 meV, is very small.
We can rationalize the present success by recalling that the vibrational
energies are given by total energy differences, which are well represented
in DFT. In contrast, the electron-vibrational couplings involve excited-state
properties, that are approximated by Kohn-Sham states. Thus, it is perhaps 
not surprising that the peak heights have much larger error. 

In order to quantify the difference (error) between the experimental
and various theoretical curves, we introduce the mean deviation (MD) given by
\[
\mathrm{MD} =\sqrt{ \frac 1{20} \sum_{i=1}^{20} 
\left[ (E^{exp}_i-E^{th}_i)^{2}+(H^{exp}_i-H^{th}_i)^{2}W^{2}\right]}
\]
summing the deviations of the peak positions and heights. To account for differing
magnitudes, we introduced the weight factor $W=5000$.
The MD for the isolated PbPc is 0.44 and for
PbPc$^{1-}$/surf (minority spin) MD = 0.35 (data in \si{}, Table 1 \cite{SM}).

Electron-vibrational transitions in molecules obey selection rules imposed by 
symmetry. The symmetry of the adsorption site ensures that these
selection rules translate into the d$I/$d$V$ of the STM.
To see the effect of selection rules, we portray vibrations of the 5 most intense peaks of 
C$_{4v}$ PbPc$^{1-}$/surf in
Fig.~S15. These vibrations reduce the symmetry of the molecule but preserve 
the mirror plane that is perpendicular to the screen. We recall that due to an
unpaired electron, the C$_{4v}$ symmetry of the neutral PbPc is reduced (see Fig.~S6(b))
but the mirror plane is preserved.


\subsection{Two-vibrational excitations}
\begin{figure}
\subcaptionbox{}{\includegraphics[width=0.45\textwidth]{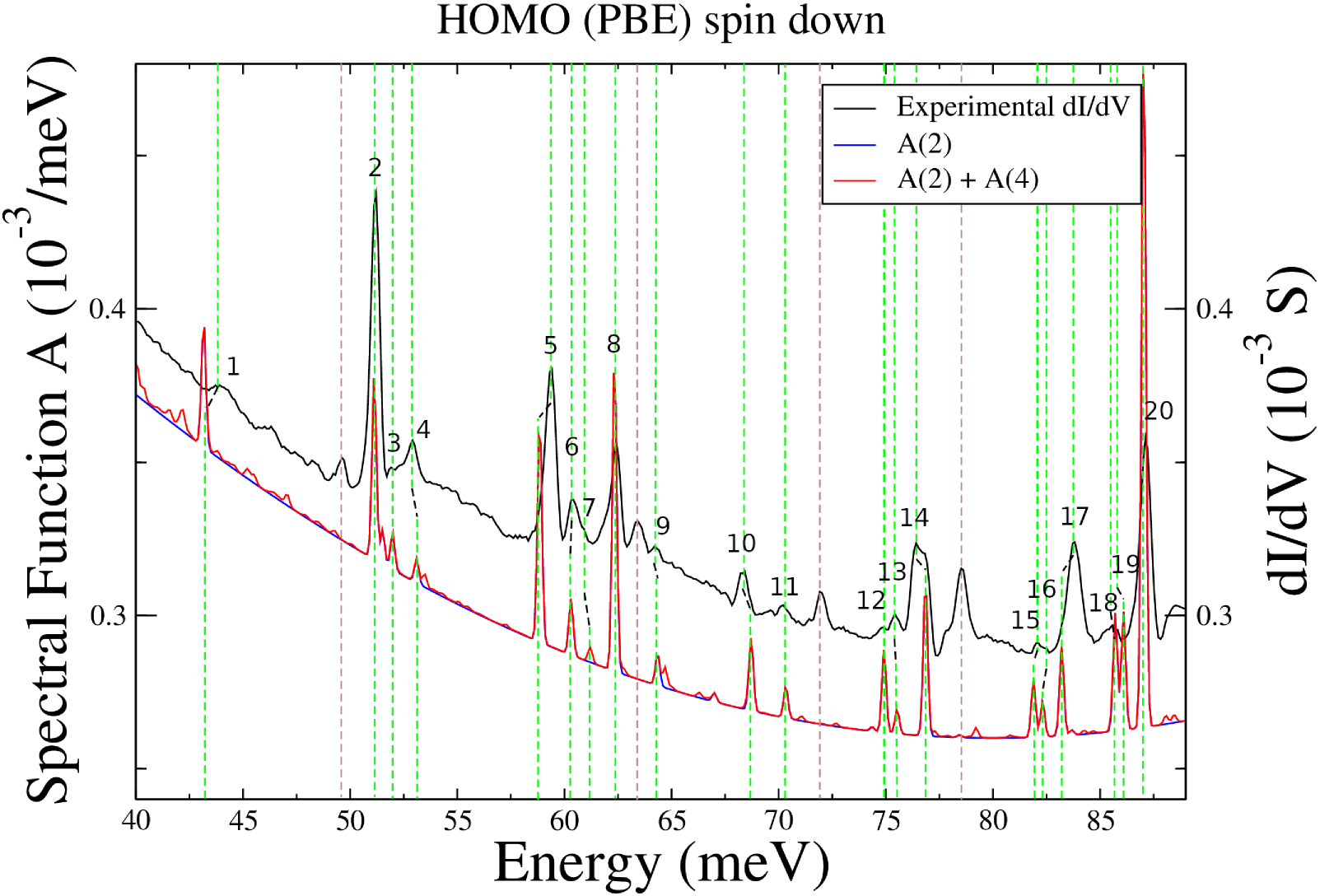}}
\subcaptionbox{}{\includegraphics[width=0.45\textwidth]{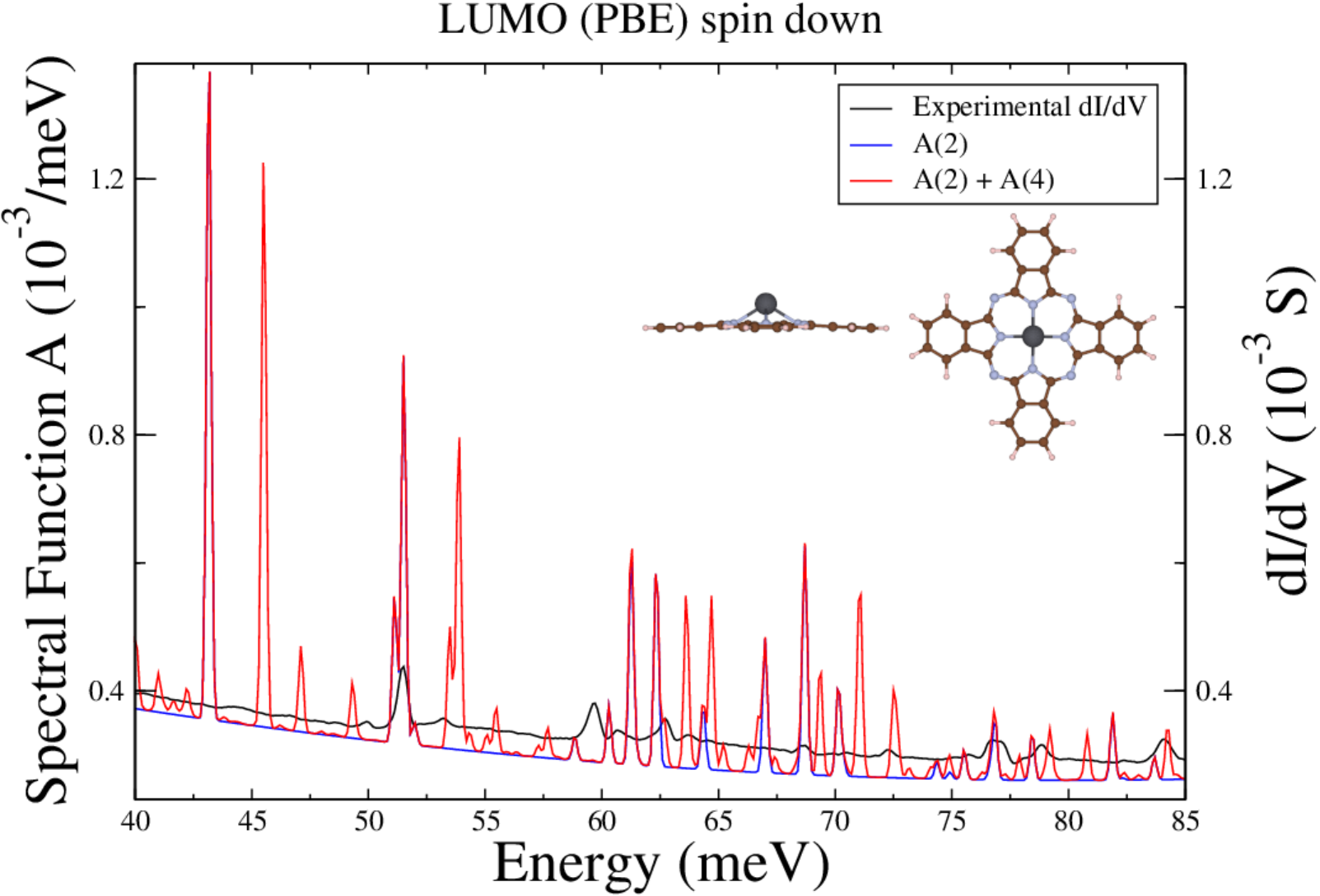}}
\centering
\caption{\label{fig2}
Spectral function describing single \([A^{(2)}_\alpha(\omega)]\) and double \([A^{(4)}_\alpha(\omega)]\) 
electron-vibrational transitions from the minority-spin molecular orbitals $\alpha=$HOMO (a) and LUMO (b) 
using the PBE functional, of PbPc$^{1-}$ with the C$_{4v}$ PbPc/surf structure.The 20 highest experimental peaks are numbered in (a). Green dotted lines indicate energies for which there is an evident correspondence between experimental and theoretical peaks. Brown dotted lines represent experimental or computational peaks for which there is no correspondence within 1 meV. The green dotted lines which are connected by a small black dotted line indicate cases where is a small shift in energy. Inset in (b) shows the C$_{4v}$ geometry of PbPc relaxed on top of Pb(100), employed in the calculation.}
\end{figure}

The plots of spectral functions not only show $A^{(2)}_\alpha(E)$, but also the inclusion
of two-vibration excitations (2VE) in $A^{(4)}_\alpha(E)$. The 2VE peaks are the red
peaks which are not reproduced by the blue trace.
Comparison shows that in all cases the
2VE contribute weakly and the weight of the 2VE peaks tends to decrease with $E$, consistently
with the presence of the frequency in the denominator of \pref{eq:g}.
We conclude that the 2VE are likely not very important in the measured data
of Homberg \etal{}.
However, the plots \figref{fig1}(a) with \figref{fig1}(b) show sizable 2VE
peaks, suggesting that their small weight in the \figref{fig2}(a) is rather
an exception.
Hence, the 2VE should be generally expected to pop up
in the spectrum in energy range given in similar systems. Note that the inelastic 2VE are routinely disregarded
in the standard IETS framework aimed at non-superconducting substrates.

\subsection{Further comparisons}
Calculations with B3LYP functional were also performed in the two systems as in Fig.~S7 and Fig.~S8 in the \si{},
but the results were consistently worse than results with PBE \cite{SM}.
The vibrational modes of the molecule strongly depend on the geometry and, since PBE produced slightly better results
with regards to the Pc planar geometry than B3LYP, this could explain the better estimation of EV coupling of PBE.

Next, the interpretation of the experiment involves certain ambiguity regarding the charge
of the molecule in question.
Namely, an adsorbed standalone PbPc on Pb(100) has negligible charge transfer, but the PbPc presumably turns
anionic (and magnetic) when adsorbed as a 
monolayer\cite{Homberg2022} (a deeper discussion is 
in the thesis \cite{Homberg2022b}). Therefore, the charge state of molecules
within a monolayer will likely be spatially dependent.
To complement the calculations of 0 and 1- charge states, we have also investigated the HOMO and 
HOMO-1 spectra of PbPc$^{2-}$. Note that
the latter is a closed shell and the two orbitals adiabaticaly connect with the LUMO and HOMO, respectively,
of PbPc$^{0}$. 
The comparison with measured curve is significantly worse; we direct the reader to
the \si{, Fig.~{S3}} \cite{SM}. Hence, the PbPc$^{1-}$ spectral function is the most consistent with the measurement, 
favoring the scenario of Homberg \etal{} given above.

\section{Discussion}
To make a final observation, we adopt a natural assumption that the HOMO orbital of the isolated molecule
(hereafter referred to as \HOMOo)
undergoes only small changes as we deform the geometry by mimicking the presence of the surface
or add electrons. Specifically,  \HOMOo{} is connected to the minority-spin HOMO of PbPc$^{1-}$
and HOMO-1 of PbPc$^{2-}$, as it follows from a Pauli principle.
In a similar fashion, the LUMO of the neutral specimen connects to the LUMO of the minority spin  PbPc$^{1-}$
and the HOMO of PbPc$^{2-}$. Moreover, it is connected to the HOMO of the majority spin PbPc$^{1-}$,
\ie{} the magnetic-moment carrying singly-occupied molecular orbital.
This statement is validated by comparing the spin density of PbPc$^{1-}$ with the molecular orbital isosurfaces
of neutral PbPc, shown in \si{, Figs.~S4-6}. The nodal structure of the two orbitals is the same.
We shall denote the orbitals connected to the LUMO of the neutral PbPc by \LUMOo.
From inspection of Figs.~\ref{fig1}, \ref{fig2} and S1 we conclude that
the spectral function of \HOMOo{}
is generally giving better agreement with the experiment than the \LUMOo.

An important feature of our method is the restriction to a single transporting molecular orbital.
It is known that if electronic transport is studied at low bias voltage $V$, 
it is valid to use a single frontier orbital as long as the other orbitals are energetically
well separated \cite{Evers2020}. Specifically, it must hold that $\Gamma_i/|E_i - E_\mathrm F| \ll 1$,
where $E_i - E_\mathrm F$ is the energy of the orbital $i$ from the Fermi level
and $\Gamma_i$ is its level broadening.
It appears that this condition is fulfilled for PbPc/Pb(100): in our calculations 
the level spacing between HOMO-1, HOMO, LUMO and LUMO+1 of PbPc$^0$ is of the order of 1 eV,
while $\Gamma \approx$ 30 meV \cite{Homberg2022} (see also the density of states in the Supplementary
File of the latter reference).

The off-diagonal part of the EV coupling $\lambda_{\mu\nu}^v$ allows,
in principle, transitions between orbitals. Such transitions are weighted additionally by 
a factor $\approx \Gamma_i/|E_i + \hbar\omega_v - E_\mathrm F|$ \cite{Lorente2020,Paulsson2005}.
Consequently, the off-diagonal EV terms can be neglected as long as one stays in the 
single transport-orbital regime. Our method allows to include more molecular orbitals,
as well as transitions involving off-diagonal EV terms. For the specific case of PbPc/Pb(100),
a two-orbital calculation would involve external parameters reflecting differences in
level alignment and coupling strengths of these orbitals to the substrate.

We observe that the good agreement of our \HOMOo{} results seems to be at odds
with the statement in Homberg \etal{} that \LUMOo{} is the main transport channel.
We speculate that the reason could be that the disordered character of the PbPc 
monolayer of Ref.~\cite{Homberg2022} can affect the level alignment in a way that
is not accounted for by the periodic calculation. Other reasons can be methodological
(d$I/$d$V$ vs d$^2I/$d$V^2$) or possibly multi-orbital.

It is known that the YSR resonance appears twice inside the gap:
for positive and negative voltages, and these resonances have different heights reflecting
differences in electron and hole tunneling. These differences
are highly tip-dependent and can
be seen as spectroscopically non-universal within the context of STM (see the overview in the Sec.~III.B of Ref.~\cite{Oppen2021}). The asymmetry of the YSR partners applies also
to the vibrational side-peaks. In principle, our techniques can
be extended to include such effects following the work of Ruby 
\etal{} \cite{Ruby2015}.

\section{Conclusions}
We have developed a minimal exactly-solvable model
of inelastic tunneling
spectrum of vibrating magnetic molecules adsorbed on superconductors. The purpose of the model
is to facilitate direct comparisons with experimental data by having
its Hamiltonian parameters supplied from
an \ai{} calculation. Although the interaction with Bogoliubov–de Gennes
quasi-particles is not included, the advantage of the model is that it
formally delivers vibrational peaks in the spectral function, which corresponds
to the first derivative of the $IV$-curve, the tunneling conductance.
This is in contrast to the output of common \ai{} inelastic
scanning tunneling spectra, which deliver second derivative of the $IV$-curve.
Moreover, multiple vibrational transitions are included at almost zero cost.

We apply the method to lead phthalocyanine on superconducting Pb(100) and
compare to the recently-published experimental data.
We obtain the best agreement for a negatively-charged PbPc$^{1-}$, consistent 
with the conjecture of adsorbed PbPc gaining magnetic moment by ionizing.
A detailed comparison of spectra from different molecular orbitals
convincingly points out the dominant role of the HOMO
in the electron-vibrational transitions. Since the magnetic moment resides
on the LUMO, our results reveal the distinct roles of the two orbitals
in the correlated low-energy behavior, guiding the design of more
elaborate many-body treatments.

\begin{acknowledgements}
We thank R. Berndt for drawing our attention to the experimental results and providing the measured data for comparison.
Discussions with M. B\"urkle, A. Donarini and C. S. Garoufalis are acknowledged.
We acknowledge support by the Ministry of Education, Youth and Sports of the Czech Republic through the project e-INFRA CZ (ID:90254) and through the project "e-Infrastruktura CZ" (e-INFRA CZ LM2018140) and the Czech Science foundation (project no. 22-22419S).
\end{acknowledgements}

\bibliography{references}

\begin{thebibliography}{52}%
\makeatletter
\providecommand \@ifxundefined [1]{%
 \@ifx{#1\undefined}
}%
\providecommand \@ifnum [1]{%
 \ifnum #1\expandafter \@firstoftwo
 \else \expandafter \@secondoftwo
 \fi
}%
\providecommand \@ifx [1]{%
 \ifx #1\expandafter \@firstoftwo
 \else \expandafter \@secondoftwo
 \fi
}%
\providecommand \natexlab [1]{#1}%
\providecommand \enquote  [1]{``#1''}%
\providecommand \bibnamefont  [1]{#1}%
\providecommand \bibfnamefont [1]{#1}%
\providecommand \citenamefont [1]{#1}%
\providecommand \href@noop [0]{\@secondoftwo}%
\providecommand \href [0]{\begingroup \@sanitize@url \@href}%
\providecommand \@href[1]{\@@startlink{#1}\@@href}%
\providecommand \@@href[1]{\endgroup#1\@@endlink}%
\providecommand \@sanitize@url [0]{\catcode `\\12\catcode `\$12\catcode `\&12\catcode `\#12\catcode `\^12\catcode `\_12\catcode `\%12\relax}%
\providecommand \@@startlink[1]{}%
\providecommand \@@endlink[0]{}%
\providecommand \url  [0]{\begingroup\@sanitize@url \@url }%
\providecommand \@url [1]{\endgroup\@href {#1}{\urlprefix }}%
\providecommand \urlprefix  [0]{URL }%
\providecommand \Eprint [0]{\href }%
\providecommand \doibase [0]{https://doi.org/}%
\providecommand \selectlanguage [0]{\@gobble}%
\providecommand \bibinfo  [0]{\@secondoftwo}%
\providecommand \bibfield  [0]{\@secondoftwo}%
\providecommand \translation [1]{[#1]}%
\providecommand \BibitemOpen [0]{}%
\providecommand \bibitemStop [0]{}%
\providecommand \bibitemNoStop [0]{.\EOS\space}%
\providecommand \EOS [0]{\spacefactor3000\relax}%
\providecommand \BibitemShut  [1]{\csname bibitem#1\endcsname}%
\let\auto@bib@innerbib\@empty
\bibitem [{\citenamefont {Scott}\ and\ \citenamefont {Natelson}(2010)}]{Scott2010}%
  \BibitemOpen
  \bibfield  {author} {\bibinfo {author} {\bibfnamefont {G.~D.}\ \bibnamefont {Scott}}\ and\ \bibinfo {author} {\bibfnamefont {D.}~\bibnamefont {Natelson}},\ }\bibfield  {title} {\bibinfo {title} {Kondo resonances in molecular devices},\ }\href@noop {} {\bibfield  {journal} {\bibinfo  {journal} {ACS Nano}\ }\textbf {\bibinfo {volume} {4}},\ \bibinfo {pages} {3560} (\bibinfo {year} {2010})}\BibitemShut {NoStop}%
\bibitem [{\citenamefont {Scheer}\ and\ \citenamefont {Cuevas}(2017)}]{Scheer2017}%
  \BibitemOpen
  \bibfield  {author} {\bibinfo {author} {\bibfnamefont {E.}~\bibnamefont {Scheer}}\ and\ \bibinfo {author} {\bibfnamefont {J.~C.}\ \bibnamefont {Cuevas}},\ }\href@noop {} {\emph {\bibinfo {title} {Molecular Electronics: An Introduction to Theory and Experiment}}}\ (\bibinfo  {publisher} {World Scientific},\ \bibinfo {year} {2017})\ \bibinfo {note} {2nd edition}\BibitemShut {NoStop}%
\bibitem [{\citenamefont {Evers}\ \emph {et~al.}(2020)\citenamefont {Evers}, \citenamefont {Koryt{\'a}r}, \citenamefont {Tewari},\ and\ \citenamefont {van Ruitenbeek}}]{Evers2020}%
  \BibitemOpen
  \bibfield  {author} {\bibinfo {author} {\bibfnamefont {F.}~\bibnamefont {Evers}}, \bibinfo {author} {\bibfnamefont {R.}~\bibnamefont {Koryt{\'a}r}}, \bibinfo {author} {\bibfnamefont {S.}~\bibnamefont {Tewari}},\ and\ \bibinfo {author} {\bibfnamefont {J.~M.}\ \bibnamefont {van Ruitenbeek}},\ }\bibfield  {title} {\bibinfo {title} {Advances and challenges in single-molecule electron transport},\ }\href {https://doi.org/10.1103/RevModPhys.92.035001} {\bibfield  {journal} {\bibinfo  {journal} {Rev. Mod. Phys.}\ }\textbf {\bibinfo {volume} {92}},\ \bibinfo {pages} {035001} (\bibinfo {year} {2020})}\BibitemShut {NoStop}%
\bibitem [{\citenamefont {Heinrich}\ \emph {et~al.}(2018)\citenamefont {Heinrich}, \citenamefont {Pascual},\ and\ \citenamefont {Franke}}]{Heinrich2018}%
  \BibitemOpen
  \bibfield  {author} {\bibinfo {author} {\bibfnamefont {B.~W.}\ \bibnamefont {Heinrich}}, \bibinfo {author} {\bibfnamefont {J.~I.}\ \bibnamefont {Pascual}},\ and\ \bibinfo {author} {\bibfnamefont {K.~J.}\ \bibnamefont {Franke}},\ }\bibfield  {title} {\bibinfo {title} {Single magnetic adsorbates on s-wave superconductors},\ }\href@noop {} {\bibfield  {journal} {\bibinfo  {journal} {Progress in Surface Science}\ }\textbf {\bibinfo {volume} {93}},\ \bibinfo {pages} {1} (\bibinfo {year} {2018})}\BibitemShut {NoStop}%
\bibitem [{\citenamefont {Gauyacq}\ \emph {et~al.}(2012)\citenamefont {Gauyacq}, \citenamefont {Lorente},\ and\ \citenamefont {Novaes}}]{Gauyacq2012}%
  \BibitemOpen
  \bibfield  {author} {\bibinfo {author} {\bibfnamefont {J.-P.}\ \bibnamefont {Gauyacq}}, \bibinfo {author} {\bibfnamefont {N.}~\bibnamefont {Lorente}},\ and\ \bibinfo {author} {\bibfnamefont {F.~D.}\ \bibnamefont {Novaes}},\ }\bibfield  {title} {\bibinfo {title} {Excitation of local magnetic moments by tunneling electrons},\ }\href {https://doi.org/10.1016/j.progsurf.2012.05.003} {\bibfield  {journal} {\bibinfo  {journal} {Prog. Surf. Sci.}\ }\textbf {\bibinfo {volume} {87}},\ \bibinfo {pages} {63} (\bibinfo {year} {2012})}\BibitemShut {NoStop}%
\bibitem [{\citenamefont {Rakhmilevitch}\ \emph {et~al.}(2014)\citenamefont {Rakhmilevitch}, \citenamefont {Koryt\'ar}, \citenamefont {Bagrets}, \citenamefont {Evers},\ and\ \citenamefont {Tal}}]{Rakhmilevitch2014}%
  \BibitemOpen
  \bibfield  {author} {\bibinfo {author} {\bibfnamefont {D.}~\bibnamefont {Rakhmilevitch}}, \bibinfo {author} {\bibfnamefont {R.}~\bibnamefont {Koryt\'ar}}, \bibinfo {author} {\bibfnamefont {A.}~\bibnamefont {Bagrets}}, \bibinfo {author} {\bibfnamefont {F.}~\bibnamefont {Evers}},\ and\ \bibinfo {author} {\bibfnamefont {O.}~\bibnamefont {Tal}},\ }\bibfield  {title} {\bibinfo {title} {Electron-{V}ibration {I}nteraction in the {P}resence of a {S}witchable {K}ondo {R}esonance {R}ealized in a {M}olecular {J}unction},\ }\href@noop {} {\bibfield  {journal} {\bibinfo  {journal} {Phys. Rev. Lett.}\ }\textbf {\bibinfo {volume} {113}},\ \bibinfo {pages} {236603} (\bibinfo {year} {2014})}\BibitemShut {NoStop}%
\bibitem [{\citenamefont {Minamitani}\ \emph {et~al.}(2012)\citenamefont {Minamitani}, \citenamefont {Tsukahara}, \citenamefont {Matsunaka}, \citenamefont {Kim}, \citenamefont {Takagi},\ and\ \citenamefont {Kawai}}]{Minamitani2012}%
  \BibitemOpen
  \bibfield  {author} {\bibinfo {author} {\bibfnamefont {E.}~\bibnamefont {Minamitani}}, \bibinfo {author} {\bibfnamefont {N.}~\bibnamefont {Tsukahara}}, \bibinfo {author} {\bibfnamefont {D.}~\bibnamefont {Matsunaka}}, \bibinfo {author} {\bibfnamefont {Y.}~\bibnamefont {Kim}}, \bibinfo {author} {\bibfnamefont {N.}~\bibnamefont {Takagi}},\ and\ \bibinfo {author} {\bibfnamefont {M.}~\bibnamefont {Kawai}},\ }\bibfield  {title} {\bibinfo {title} {Symmetry-driven novel kondo effect in a molecule},\ }\href@noop {} {\bibfield  {journal} {\bibinfo  {journal} {Phys. Rev. Lett.}\ }\textbf {\bibinfo {volume} {109}},\ \bibinfo {pages} {086602} (\bibinfo {year} {2012})}\BibitemShut {NoStop}%
\bibitem [{\citenamefont {Ruby}\ \emph {et~al.}(2016)\citenamefont {Ruby}, \citenamefont {Peng}, \citenamefont {von Oppen}, \citenamefont {Heinrich},\ and\ \citenamefont {Franke}}]{Ruby2016}%
  \BibitemOpen
  \bibfield  {author} {\bibinfo {author} {\bibfnamefont {M.}~\bibnamefont {Ruby}}, \bibinfo {author} {\bibfnamefont {Y.}~\bibnamefont {Peng}}, \bibinfo {author} {\bibfnamefont {F.}~\bibnamefont {von Oppen}}, \bibinfo {author} {\bibfnamefont {B.~W.}\ \bibnamefont {Heinrich}},\ and\ \bibinfo {author} {\bibfnamefont {K.~J.}\ \bibnamefont {Franke}},\ }\bibfield  {title} {\bibinfo {title} {Orbital picture of {Y}u-{S}hiba-{R}usinov multiplets},\ }\href@noop {} {\bibfield  {journal} {\bibinfo  {journal} {Phys. Rev. Lett.}\ }\textbf {\bibinfo {volume} {117}},\ \bibinfo {pages} {186801} (\bibinfo {year} {2016})}\BibitemShut {NoStop}%
\bibitem [{\citenamefont {Zalom}\ \emph {et~al.}(2019)\citenamefont {Zalom}, \citenamefont {De~Bruijckere}, \citenamefont {Gaudenzi}, \citenamefont {van~ver Zant}, \citenamefont {Novotný},\ and\ \citenamefont {Koryt{\'a}r}}]{Zalom2019}%
  \BibitemOpen
  \bibfield  {author} {\bibinfo {author} {\bibfnamefont {P.}~\bibnamefont {Zalom}}, \bibinfo {author} {\bibfnamefont {J.}~\bibnamefont {De~Bruijckere}}, \bibinfo {author} {\bibfnamefont {R.}~\bibnamefont {Gaudenzi}}, \bibinfo {author} {\bibfnamefont {H.~S.}\ \bibnamefont {van~ver Zant}}, \bibinfo {author} {\bibfnamefont {T.}~\bibnamefont {Novotný}},\ and\ \bibinfo {author} {\bibfnamefont {R.}~\bibnamefont {Koryt{\'a}r}},\ }\bibfield  {title} {\bibinfo {title} {Magnetically tuned {K}ondo effect in a molecular double quantum dot: Role of the anisotropic exchange},\ }\href@noop {} {\bibfield  {journal} {\bibinfo  {journal} {J. Phys. Chem. C}\ }\textbf {\bibinfo {volume} {123}},\ \bibinfo {pages} {11917} (\bibinfo {year} {2019})}\BibitemShut {NoStop}%
\bibitem [{\citenamefont {Choi}\ \emph {et~al.}(2017)\citenamefont {Choi}, \citenamefont {Abufager}, \citenamefont {Limot},\ and\ \citenamefont {Lorente}}]{Choi2017}%
  \BibitemOpen
  \bibfield  {author} {\bibinfo {author} {\bibfnamefont {D.-J.}\ \bibnamefont {Choi}}, \bibinfo {author} {\bibfnamefont {P.}~\bibnamefont {Abufager}}, \bibinfo {author} {\bibfnamefont {L.}~\bibnamefont {Limot}},\ and\ \bibinfo {author} {\bibfnamefont {N.}~\bibnamefont {Lorente}},\ }\bibfield  {title} {\bibinfo {title} {From tunneling to contact in a magnetic atom: The non-equilibrium {K}ondo effect},\ }\href@noop {} {\bibfield  {journal} {\bibinfo  {journal} {J. Chem. Phys.}\ }\textbf {\bibinfo {volume} {146}} (\bibinfo {year} {2017})}\BibitemShut {NoStop}%
\bibitem [{\citenamefont {\v{Z}onda}\ \emph {et~al.}(2021)\citenamefont {\v{Z}onda}, \citenamefont {Stetsovych}, \citenamefont {Koryt{\'a}r}, \citenamefont {Ternes}, \citenamefont {Temirov}, \citenamefont {Raccanelli}, \citenamefont {Tautz}, \citenamefont {Jel{\'\i}nek}, \citenamefont {Novotn\'{y}},\ and\ \citenamefont {\v{S}vec}}]{Zonda2021}%
  \BibitemOpen
  \bibfield  {author} {\bibinfo {author} {\bibfnamefont {M.}~\bibnamefont {\v{Z}onda}}, \bibinfo {author} {\bibfnamefont {O.}~\bibnamefont {Stetsovych}}, \bibinfo {author} {\bibfnamefont {R.}~\bibnamefont {Koryt{\'a}r}}, \bibinfo {author} {\bibfnamefont {M.}~\bibnamefont {Ternes}}, \bibinfo {author} {\bibfnamefont {R.}~\bibnamefont {Temirov}}, \bibinfo {author} {\bibfnamefont {A.}~\bibnamefont {Raccanelli}}, \bibinfo {author} {\bibfnamefont {F.~S.}\ \bibnamefont {Tautz}}, \bibinfo {author} {\bibfnamefont {P.}~\bibnamefont {Jel{\'\i}nek}}, \bibinfo {author} {\bibfnamefont {T.}~\bibnamefont {Novotn\'{y}}},\ and\ \bibinfo {author} {\bibfnamefont {M.}~\bibnamefont {\v{S}vec}},\ }\bibfield  {title} {\bibinfo {title} {Resolving ambiguity of the {K}ondo temperature determination in mechanically tunable single-molecule {K}ondo systems},\ }\href@noop {} {\bibfield  {journal} {\bibinfo  {journal} {J. Phys. Chem. C}\ }\textbf {\bibinfo {volume} {12}},\ \bibinfo {pages} {6320} (\bibinfo {year} {2021})}\BibitemShut
  {NoStop}%
\bibitem [{\citenamefont {Li}\ \emph {et~al.}(2020)\citenamefont {Li}, \citenamefont {Sanz}, \citenamefont {Castro-Esteban}, \citenamefont {Vilas-Varela}, \citenamefont {Friedrich}, \citenamefont {Frederiksen}, \citenamefont {Pe{\~n}a},\ and\ \citenamefont {Pascual}}]{Li2020}%
  \BibitemOpen
  \bibfield  {author} {\bibinfo {author} {\bibfnamefont {J.}~\bibnamefont {Li}}, \bibinfo {author} {\bibfnamefont {S.}~\bibnamefont {Sanz}}, \bibinfo {author} {\bibfnamefont {J.}~\bibnamefont {Castro-Esteban}}, \bibinfo {author} {\bibfnamefont {M.}~\bibnamefont {Vilas-Varela}}, \bibinfo {author} {\bibfnamefont {N.}~\bibnamefont {Friedrich}}, \bibinfo {author} {\bibfnamefont {T.}~\bibnamefont {Frederiksen}}, \bibinfo {author} {\bibfnamefont {D.}~\bibnamefont {Pe{\~n}a}},\ and\ \bibinfo {author} {\bibfnamefont {J.~I.}\ \bibnamefont {Pascual}},\ }\bibfield  {title} {\bibinfo {title} {Uncovering the triplet ground state of triangular graphene nanoflakes engineered with atomic precision on a metal surface},\ }\href@noop {} {\bibfield  {journal} {\bibinfo  {journal} {Phys. Rev. Lett.}\ }\textbf {\bibinfo {volume} {124}},\ \bibinfo {pages} {177201} (\bibinfo {year} {2020})}\BibitemShut {NoStop}%
\bibitem [{\citenamefont {Trivini}\ \emph {et~al.}(2023)\citenamefont {Trivini}, \citenamefont {Ortuzar}, \citenamefont {Vaxevani}, \citenamefont {Li}, \citenamefont {Bergeret}, \citenamefont {Cazalilla},\ and\ \citenamefont {Pascual}}]{Trivini2023}%
  \BibitemOpen
  \bibfield  {author} {\bibinfo {author} {\bibfnamefont {S.}~\bibnamefont {Trivini}}, \bibinfo {author} {\bibfnamefont {J.}~\bibnamefont {Ortuzar}}, \bibinfo {author} {\bibfnamefont {K.}~\bibnamefont {Vaxevani}}, \bibinfo {author} {\bibfnamefont {J.}~\bibnamefont {Li}}, \bibinfo {author} {\bibfnamefont {F.~S.}\ \bibnamefont {Bergeret}}, \bibinfo {author} {\bibfnamefont {M.~A.}\ \bibnamefont {Cazalilla}},\ and\ \bibinfo {author} {\bibfnamefont {J.~I.}\ \bibnamefont {Pascual}},\ }\bibfield  {title} {\bibinfo {title} {Cooper pair excitation mediated by a molecular quantum spin on a superconducting proximitized gold film},\ }\href@noop {} {\bibfield  {journal} {\bibinfo  {journal} {Phys. Rev. Lett.}\ }\textbf {\bibinfo {volume} {130}},\ \bibinfo {pages} {136004} (\bibinfo {year} {2023})}\BibitemShut {NoStop}%
\bibitem [{\citenamefont {Homberg}\ \emph {et~al.}(2022)\citenamefont {Homberg}, \citenamefont {Weismann}, \citenamefont {Markussen},\ and\ \citenamefont {Berndt}}]{Homberg2022}%
  \BibitemOpen
  \bibfield  {author} {\bibinfo {author} {\bibfnamefont {J.}~\bibnamefont {Homberg}}, \bibinfo {author} {\bibfnamefont {A.}~\bibnamefont {Weismann}}, \bibinfo {author} {\bibfnamefont {T.}~\bibnamefont {Markussen}},\ and\ \bibinfo {author} {\bibfnamefont {R.}~\bibnamefont {Berndt}},\ }\bibfield  {title} {\bibinfo {title} {Resonance-{E}nhanced {V}ibrational {S}pectroscopy of {M}olecules on a {S}uperconductor},\ }\href {https://doi.org/10.1103/PhysRevLett.129.116801} {\bibfield  {journal} {\bibinfo  {journal} {Phys. Rev. Lett.}\ }\textbf {\bibinfo {volume} {129}},\ \bibinfo {pages} {116801} (\bibinfo {year} {2022})}\BibitemShut {NoStop}%
\bibitem [{\citenamefont {Tersoff}\ and\ \citenamefont {Hamann}(1985)}]{Tersoff1985}%
  \BibitemOpen
  \bibfield  {author} {\bibinfo {author} {\bibfnamefont {J.}~\bibnamefont {Tersoff}}\ and\ \bibinfo {author} {\bibfnamefont {D.~R.}\ \bibnamefont {Hamann}},\ }\bibfield  {title} {\bibinfo {title} {Theory of the scanning tunneling microscope},\ }\href {https://doi.org/10.1103/PhysRevB.31.805} {\bibfield  {journal} {\bibinfo  {journal} {Phys. Rev. B}\ }\textbf {\bibinfo {volume} {31}},\ \bibinfo {pages} {805} (\bibinfo {year} {1985})}\BibitemShut {NoStop}%
\bibitem [{\citenamefont {Meir}\ and\ \citenamefont {Wingreen}(1992)}]{Meir1992}%
  \BibitemOpen
  \bibfield  {author} {\bibinfo {author} {\bibfnamefont {Y.}~\bibnamefont {Meir}}\ and\ \bibinfo {author} {\bibfnamefont {N.~S.}\ \bibnamefont {Wingreen}},\ }\bibfield  {title} {\bibinfo {title} {Landauer formula for the current through an interacting electron region},\ }\href {https://doi.org/10.1103/PhysRevLett.68.2512} {\bibfield  {journal} {\bibinfo  {journal} {Phys. Rev. Lett.}\ }\textbf {\bibinfo {volume} {68}},\ \bibinfo {pages} {2512} (\bibinfo {year} {1992})}\BibitemShut {NoStop}%
\bibitem [{\citenamefont {Coleman}(2015)}]{Coleman2015}%
  \BibitemOpen
  \bibfield  {author} {\bibinfo {author} {\bibfnamefont {P.}~\bibnamefont {Coleman}},\ }\bibinfo {title} {Introduction to {M}any-{B}ody {P}hysics}\ (\bibinfo  {publisher} {Cambridge University Press},\ \bibinfo {year} {2015})\ Chap.\ \bibinfo {chapter} {9.7.2}\BibitemShut {NoStop}%
\bibitem [{\citenamefont {Ruby}\ \emph {et~al.}(2015)\citenamefont {Ruby}, \citenamefont {Pientka}, \citenamefont {Peng}, \citenamefont {von Oppen}, \citenamefont {Heinrich},\ and\ \citenamefont {Franke}}]{Ruby2015}%
  \BibitemOpen
  \bibfield  {author} {\bibinfo {author} {\bibfnamefont {M.}~\bibnamefont {Ruby}}, \bibinfo {author} {\bibfnamefont {F.}~\bibnamefont {Pientka}}, \bibinfo {author} {\bibfnamefont {Y.}~\bibnamefont {Peng}}, \bibinfo {author} {\bibfnamefont {F.}~\bibnamefont {von Oppen}}, \bibinfo {author} {\bibfnamefont {B.~W.}\ \bibnamefont {Heinrich}},\ and\ \bibinfo {author} {\bibfnamefont {K.~J.}\ \bibnamefont {Franke}},\ }\bibfield  {title} {\bibinfo {title} {Tunneling processes into localized subgap states in superconductors},\ }\href {https://doi.org/10.1103/PhysRevLett.115.087001} {\bibfield  {journal} {\bibinfo  {journal} {Phys. Rev. Lett.}\ }\textbf {\bibinfo {volume} {115}},\ \bibinfo {pages} {087001} (\bibinfo {year} {2015})}\BibitemShut {NoStop}%
\bibitem [{\citenamefont {Balatsky}\ \emph {et~al.}(2006)\citenamefont {Balatsky}, \citenamefont {Vekhter},\ and\ \citenamefont {Zhu}}]{Balatsky2006}%
  \BibitemOpen
  \bibfield  {author} {\bibinfo {author} {\bibfnamefont {A.~V.}\ \bibnamefont {Balatsky}}, \bibinfo {author} {\bibfnamefont {I.}~\bibnamefont {Vekhter}},\ and\ \bibinfo {author} {\bibfnamefont {J.-X.}\ \bibnamefont {Zhu}},\ }\bibfield  {title} {\bibinfo {title} {Impurity-induced states in conventional and unconventional superconductors},\ }\href {https://doi.org/10.1103/RevModPhys.78.373} {\bibfield  {journal} {\bibinfo  {journal} {Rev. Mod. Phys.}\ }\textbf {\bibinfo {volume} {78}},\ \bibinfo {pages} {373} (\bibinfo {year} {2006})}\BibitemShut {NoStop}%
\bibitem [{\citenamefont {Mahan}(2000)}]{Mahan2000}%
  \BibitemOpen
  \bibfield  {author} {\bibinfo {author} {\bibfnamefont {G.~D.}\ \bibnamefont {Mahan}},\ }\href {https://doi.org/10.1007/978-1-4757-5714-9} {\emph {\bibinfo {title} {Many-{P}article {P}hysics}}}\ (\bibinfo  {publisher} {Springer Science \& Business Media},\ \bibinfo {year} {2000})\BibitemShut {NoStop}%
\bibitem [{\citenamefont {Nitzan}(2006)}]{Nitzan2006}%
  \BibitemOpen
  \bibfield  {author} {\bibinfo {author} {\bibfnamefont {A.}~\bibnamefont {Nitzan}},\ }\bibinfo {title} {Chemical {D}ynamics in {C}ondensed {P}hases}\ (\bibinfo  {publisher} {Oxford Graduate Texts},\ \bibinfo {year} {2006})\ Chap.\ \bibinfo {chapter} {12.5.3}, p.\ \bibinfo {pages} {444}\BibitemShut {NoStop}%
\bibitem [{\citenamefont {Frederiksen}\ \emph {et~al.}(2007)\citenamefont {Frederiksen}, \citenamefont {Paulsson}, \citenamefont {Brandbyge},\ and\ \citenamefont {Jauho}}]{Frederiksen2007}%
  \BibitemOpen
  \bibfield  {author} {\bibinfo {author} {\bibfnamefont {T.}~\bibnamefont {Frederiksen}}, \bibinfo {author} {\bibfnamefont {M.}~\bibnamefont {Paulsson}}, \bibinfo {author} {\bibfnamefont {M.}~\bibnamefont {Brandbyge}},\ and\ \bibinfo {author} {\bibfnamefont {A.-P.}\ \bibnamefont {Jauho}},\ }\bibfield  {title} {\bibinfo {title} {Inelastic transport theory from first principles: {M}ethodology and application to nanoscale devices},\ }\href {https://doi.org/10.1103/PhysRevB.75.205413} {\bibfield  {journal} {\bibinfo  {journal} {Phys. Rev. B}\ }\textbf {\bibinfo {volume} {75}},\ \bibinfo {pages} {205413} (\bibinfo {year} {2007})}\BibitemShut {NoStop}%
\bibitem [{\citenamefont {Balasubramani}\ \emph {et~al.}(2020)\citenamefont {Balasubramani}, \citenamefont {Chen}, \citenamefont {Coriani}, \citenamefont {Diedenhofen}, \citenamefont {Frank}, \citenamefont {Franzke}, \citenamefont {Furche}, \citenamefont {Grotjahn}, \citenamefont {Harding}, \citenamefont {H{\"a}ttig} \emph {et~al.}}]{turbomole}%
  \BibitemOpen
  \bibfield  {author} {\bibinfo {author} {\bibfnamefont {S.~G.}\ \bibnamefont {Balasubramani}}, \bibinfo {author} {\bibfnamefont {G.~P.}\ \bibnamefont {Chen}}, \bibinfo {author} {\bibfnamefont {S.}~\bibnamefont {Coriani}}, \bibinfo {author} {\bibfnamefont {M.}~\bibnamefont {Diedenhofen}}, \bibinfo {author} {\bibfnamefont {M.~S.}\ \bibnamefont {Frank}}, \bibinfo {author} {\bibfnamefont {Y.~J.}\ \bibnamefont {Franzke}}, \bibinfo {author} {\bibfnamefont {F.}~\bibnamefont {Furche}}, \bibinfo {author} {\bibfnamefont {R.}~\bibnamefont {Grotjahn}}, \bibinfo {author} {\bibfnamefont {M.~E.}\ \bibnamefont {Harding}}, \bibinfo {author} {\bibfnamefont {C.}~\bibnamefont {H{\"a}ttig}}, \emph {et~al.},\ }\bibfield  {title} {\bibinfo {title} {{TURBOMOLE}: Modular program suite for ab initio quantum-chemical and condensed-matter simulations},\ }\href {https://doi.org/10.1021/ja00040a007} {\bibfield  {journal} {\bibinfo  {journal} {J. Chem. Phys.}\ }\textbf {\bibinfo {volume} {152}},\ \bibinfo {pages} {184107} (\bibinfo {year}
  {2020})}\BibitemShut {NoStop}%
\bibitem [{\citenamefont {Ahlrichs}\ \emph {et~al.}(1989)\citenamefont {Ahlrichs}, \citenamefont {Bär}, \citenamefont {Häser}, \citenamefont {Horn},\ and\ \citenamefont {Kölmel}}]{turbomole1989}%
  \BibitemOpen
  \bibfield  {author} {\bibinfo {author} {\bibfnamefont {R.}~\bibnamefont {Ahlrichs}}, \bibinfo {author} {\bibfnamefont {M.}~\bibnamefont {Bär}}, \bibinfo {author} {\bibfnamefont {M.}~\bibnamefont {Häser}}, \bibinfo {author} {\bibfnamefont {H.}~\bibnamefont {Horn}},\ and\ \bibinfo {author} {\bibfnamefont {C.}~\bibnamefont {Kölmel}},\ }\bibfield  {title} {\bibinfo {title} {Electronic structure calculations on workstation computers: The program system turbomole},\ }\href {https://doi.org/10.1016/0009-2614(89)85118-8} {\bibfield  {journal} {\bibinfo  {journal} {Chem. Phys. Lett.}\ }\textbf {\bibinfo {volume} {162}},\ \bibinfo {pages} {165} (\bibinfo {year} {1989})}\BibitemShut {NoStop}%
\bibitem [{\citenamefont {Treutler}\ and\ \citenamefont {Ahlrichs}(1995)}]{turbomole1995}%
  \BibitemOpen
  \bibfield  {author} {\bibinfo {author} {\bibfnamefont {O.}~\bibnamefont {Treutler}}\ and\ \bibinfo {author} {\bibfnamefont {R.}~\bibnamefont {Ahlrichs}},\ }\bibfield  {title} {\bibinfo {title} {Efficient molecular numerical integration schemes},\ }\href {https://doi.org/10.1063/1.469408} {\bibfield  {journal} {\bibinfo  {journal} {J. Chem. Phys.}\ }\textbf {\bibinfo {volume} {102}},\ \bibinfo {pages} {346–354} (\bibinfo {year} {1995})}\BibitemShut {NoStop}%
\bibitem [{\citenamefont {Von~Arnim}\ and\ \citenamefont {Ahlrichs}(1998)}]{turbomole1998}%
  \BibitemOpen
  \bibfield  {author} {\bibinfo {author} {\bibfnamefont {M.}~\bibnamefont {Von~Arnim}}\ and\ \bibinfo {author} {\bibfnamefont {R.}~\bibnamefont {Ahlrichs}},\ }\bibfield  {title} {\bibinfo {title} {Performance of parallel {TURBOMOLE} for density functional calculations},\ }\href {https://doi.org/10.1002/(SICI)1096-987X(19981130)19:15<1746::AID-JCC7>3.0.CO;2-N} {\bibfield  {journal} {\bibinfo  {journal} {J. Comp. Chem.}\ }\textbf {\bibinfo {volume} {19}},\ \bibinfo {pages} {1746} (\bibinfo {year} {1998})}\BibitemShut {NoStop}%
\bibitem [{\citenamefont {Weigend}\ and\ \citenamefont {Ahlrichs}(2005)}]{karlsruheBasis}%
  \BibitemOpen
  \bibfield  {author} {\bibinfo {author} {\bibfnamefont {F.}~\bibnamefont {Weigend}}\ and\ \bibinfo {author} {\bibfnamefont {R.}~\bibnamefont {Ahlrichs}},\ }\bibfield  {title} {\bibinfo {title} {Balanced basis sets of split valence, triple zeta valence and quadruple zeta valence quality for {H} to {R}n: {D}esign and assessment of accuracy},\ }\href {https://doi.org/10.1039/B508541A} {\bibfield  {journal} {\bibinfo  {journal} {Phys. Chem. Chem. Phys.}\ }\textbf {\bibinfo {volume} {7}},\ \bibinfo {pages} {3297} (\bibinfo {year} {2005})}\BibitemShut {NoStop}%
\bibitem [{\citenamefont {Eichkorn}\ \emph {et~al.}(1995)\citenamefont {Eichkorn}, \citenamefont {Treutler}, \citenamefont {\"{O}hm}, \citenamefont {H\"{a}ser},\ and\ \citenamefont {Ahlrichs}}]{rij1995}%
  \BibitemOpen
  \bibfield  {author} {\bibinfo {author} {\bibfnamefont {K.}~\bibnamefont {Eichkorn}}, \bibinfo {author} {\bibfnamefont {O.}~\bibnamefont {Treutler}}, \bibinfo {author} {\bibfnamefont {H.}~\bibnamefont {\"{O}hm}}, \bibinfo {author} {\bibfnamefont {M.}~\bibnamefont {H\"{a}ser}},\ and\ \bibinfo {author} {\bibfnamefont {R.}~\bibnamefont {Ahlrichs}},\ }\bibfield  {title} {\bibinfo {title} {Auxiliary basis sets to approximate {C}oulomb potentials (chem. phys. letters 240 (1995) 283-290)},\ }\href {https://doi.org/10.1016/0009-2614(95)00838-U} {\bibfield  {journal} {\bibinfo  {journal} {Chem. Phys. Lett.}\ }\textbf {\bibinfo {volume} {242}},\ \bibinfo {pages} {652} (\bibinfo {year} {1995})}\BibitemShut {NoStop}%
\bibitem [{\citenamefont {Eichkorn}\ \emph {et~al.}(1997)\citenamefont {Eichkorn}, \citenamefont {Weigend}, \citenamefont {Treutler},\ and\ \citenamefont {Ahlrichs}}]{rij1997}%
  \BibitemOpen
  \bibfield  {author} {\bibinfo {author} {\bibfnamefont {K.}~\bibnamefont {Eichkorn}}, \bibinfo {author} {\bibfnamefont {F.}~\bibnamefont {Weigend}}, \bibinfo {author} {\bibfnamefont {O.}~\bibnamefont {Treutler}},\ and\ \bibinfo {author} {\bibfnamefont {R.}~\bibnamefont {Ahlrichs}},\ }\bibfield  {title} {\bibinfo {title} {Auxiliary basis sets for main row atoms and transition metals and their use to approximate {C}oulomb potentials},\ }\href {https://doi.org/10.1007/s002140050244} {\bibfield  {journal} {\bibinfo  {journal} {Theor. Chem. Acta}\ }\textbf {\bibinfo {volume} {97}},\ \bibinfo {pages} {119} (\bibinfo {year} {1997})}\BibitemShut {NoStop}%
\bibitem [{\citenamefont {Weigend}(2006)}]{rij2006}%
  \BibitemOpen
  \bibfield  {author} {\bibinfo {author} {\bibfnamefont {F.}~\bibnamefont {Weigend}},\ }\bibfield  {title} {\bibinfo {title} {Accurate {C}oulomb-fitting basis sets for {H} to {R}n},\ }\href {https://doi.org/10.1039/B515623H} {\bibfield  {journal} {\bibinfo  {journal} {Phys. Chem. Chem. Phys.}\ }\textbf {\bibinfo {volume} {8}},\ \bibinfo {pages} {1057} (\bibinfo {year} {2006})}\BibitemShut {NoStop}%
\bibitem [{\citenamefont {Perdew}\ \emph {et~al.}(1996)\citenamefont {Perdew}, \citenamefont {Burke},\ and\ \citenamefont {Ernzerhof}}]{pbeFunc}%
  \BibitemOpen
  \bibfield  {author} {\bibinfo {author} {\bibfnamefont {J.~P.}\ \bibnamefont {Perdew}}, \bibinfo {author} {\bibfnamefont {K.}~\bibnamefont {Burke}},\ and\ \bibinfo {author} {\bibfnamefont {M.}~\bibnamefont {Ernzerhof}},\ }\bibfield  {title} {\bibinfo {title} {Generalized {G}radient {A}pproximation {M}ade {S}imple},\ }\href {https://doi.org/10.1103/PhysRevLett.77.3865} {\bibfield  {journal} {\bibinfo  {journal} {Phys. Rev. Lett.}\ }\textbf {\bibinfo {volume} {77}},\ \bibinfo {pages} {3865} (\bibinfo {year} {1996})}\BibitemShut {NoStop}%
\bibitem [{\citenamefont {Becke}(1993)}]{BeckeB3}%
  \BibitemOpen
  \bibfield  {author} {\bibinfo {author} {\bibfnamefont {A.~D.}\ \bibnamefont {Becke}},\ }\bibfield  {title} {\bibinfo {title} {Density‐functional thermochemistry. {III}. {T}he role of exact exchange},\ }\href {https://doi.org/10.1063/1.464913} {\bibfield  {journal} {\bibinfo  {journal} {J. Chem. Phys.}\ }\textbf {\bibinfo {volume} {98}},\ \bibinfo {pages} {5648–5652} (\bibinfo {year} {1993})}\BibitemShut {NoStop}%
\bibitem [{\citenamefont {Lee}\ \emph {et~al.}(1988)\citenamefont {Lee}, \citenamefont {Yang},\ and\ \citenamefont {Parr}}]{LYP}%
  \BibitemOpen
  \bibfield  {author} {\bibinfo {author} {\bibfnamefont {C.}~\bibnamefont {Lee}}, \bibinfo {author} {\bibfnamefont {W.}~\bibnamefont {Yang}},\ and\ \bibinfo {author} {\bibfnamefont {R.~G.}\ \bibnamefont {Parr}},\ }\bibfield  {title} {\bibinfo {title} {Development of the {C}olle-{S}alvetti correlation-energy formula into a functional of the electron density},\ }\href {https://doi.org/10.1103/PhysRevB.37.785} {\bibfield  {journal} {\bibinfo  {journal} {Phys. Rev. B}\ }\textbf {\bibinfo {volume} {37}},\ \bibinfo {pages} {785} (\bibinfo {year} {1988})}\BibitemShut {NoStop}%
\bibitem [{\citenamefont {Koliogiorgos}\ \emph {et~al.}(2017)\citenamefont {Koliogiorgos}, \citenamefont {Baskoutas},\ and\ \citenamefont {Galanakis}}]{koliog1}%
  \BibitemOpen
  \bibfield  {author} {\bibinfo {author} {\bibfnamefont {A.}~\bibnamefont {Koliogiorgos}}, \bibinfo {author} {\bibfnamefont {S.}~\bibnamefont {Baskoutas}},\ and\ \bibinfo {author} {\bibfnamefont {I.}~\bibnamefont {Galanakis}},\ }\bibfield  {title} {\bibinfo {title} {Electronic and gap properties of lead-free perfect and mixed hybrid halide perovskites: An ab-initio study},\ }\href {https://doi.org/10.1016/j.commatsci.2017.06.026} {\bibfield  {journal} {\bibinfo  {journal} {Comp. Mater. Sci.}\ }\textbf {\bibinfo {volume} {138}},\ \bibinfo {pages} {92} (\bibinfo {year} {2017})}\BibitemShut {NoStop}%
\bibitem [{\citenamefont {Koliogiorgos}\ \emph {et~al.}(2018)\citenamefont {Koliogiorgos}, \citenamefont {Garoufalis}, \citenamefont {Baskoutas},\ and\ \citenamefont {Galanakis}}]{koliog2}%
  \BibitemOpen
  \bibfield  {author} {\bibinfo {author} {\bibfnamefont {A.}~\bibnamefont {Koliogiorgos}}, \bibinfo {author} {\bibfnamefont {C.~S.}\ \bibnamefont {Garoufalis}}, \bibinfo {author} {\bibfnamefont {S.}~\bibnamefont {Baskoutas}},\ and\ \bibinfo {author} {\bibfnamefont {I.}~\bibnamefont {Galanakis}},\ }\bibfield  {title} {\bibinfo {title} {Electronic and {O}ptical {P}roperties of {U}ltrasmall {ABX$_3$} ({A} = {C}s, {CH$_3$NH$_3$/B} = {G}e, {P}b, {S}n, {C}a, {S}r/{X} = {C}l, {B}r, {I}) {P}erovskite {Q}uantum {D}ots},\ }\href {https://doi.org/10.1021/acsomega.8b02525} {\bibfield  {journal} {\bibinfo  {journal} {ACS Omega}\ }\textbf {\bibinfo {volume} {3}},\ \bibinfo {pages} {92} (\bibinfo {year} {2018})}\BibitemShut {NoStop}%
\bibitem [{\citenamefont {Zhang}\ \emph {et~al.}(2005)\citenamefont {Zhang}, \citenamefont {Zhang}, \citenamefont {Liu}, \citenamefont {Bian},\ and\ \citenamefont {Jiang}}]{PbPc2}%
  \BibitemOpen
  \bibfield  {author} {\bibinfo {author} {\bibfnamefont {Y.}~\bibnamefont {Zhang}}, \bibinfo {author} {\bibfnamefont {X.}~\bibnamefont {Zhang}}, \bibinfo {author} {\bibfnamefont {Z.}~\bibnamefont {Liu}}, \bibinfo {author} {\bibfnamefont {Y.}~\bibnamefont {Bian}},\ and\ \bibinfo {author} {\bibfnamefont {J.}~\bibnamefont {Jiang}},\ }\bibfield  {title} {\bibinfo {title} {Structures and properties of 1,8,15,22-{T}etrasubstituted {P}hthalocyaninato-{L}ead {C}omplexes: {T}he {S}ubstitutional {E}ffect {S}tudy {B}ased on {D}ensity {F}unctional {T}heory {C}alculations},\ }\href {https://doi.org/10.1021/jp0511449} {\bibfield  {journal} {\bibinfo  {journal} {J. Phys. Chem. A}\ }\textbf {\bibinfo {volume} {109}},\ \bibinfo {pages} {6363–6370} (\bibinfo {year} {2005})}\BibitemShut {NoStop}%
\bibitem [{\citenamefont {Sumimoto}\ \emph {et~al.}(2012)\citenamefont {Sumimoto}, \citenamefont {Honda}, \citenamefont {Kawashima}, \citenamefont {Hori},\ and\ \citenamefont {Fujimoto}}]{PbPc1}%
  \BibitemOpen
  \bibfield  {author} {\bibinfo {author} {\bibfnamefont {M.}~\bibnamefont {Sumimoto}}, \bibinfo {author} {\bibfnamefont {T.}~\bibnamefont {Honda}}, \bibinfo {author} {\bibfnamefont {Y.}~\bibnamefont {Kawashima}}, \bibinfo {author} {\bibfnamefont {K.}~\bibnamefont {Hori}},\ and\ \bibinfo {author} {\bibfnamefont {H.}~\bibnamefont {Fujimoto}},\ }\bibfield  {title} {\bibinfo {title} {Theoretical and experimental investigation on the electronic properties of the shuttlecock shaped and the double-decker structured metal phthalocyanines, {MP}c and {M}({P}c)2 ({M} = {S}n and {P}b)},\ }\href {https://doi.org/10.1039/c2dt30187c} {\bibfield  {journal} {\bibinfo  {journal} {Dalton Trans.}\ }\textbf {\bibinfo {volume} {41}},\ \bibinfo {pages} {7141} (\bibinfo {year} {2012})}\BibitemShut {NoStop}%
\bibitem [{\citenamefont {Grimme}\ \emph {et~al.}(2010)\citenamefont {Grimme}, \citenamefont {Antoni}, \citenamefont {Ehrlich},\ and\ \citenamefont {Krieg}}]{Grimme2010}%
  \BibitemOpen
  \bibfield  {author} {\bibinfo {author} {\bibfnamefont {S.}~\bibnamefont {Grimme}}, \bibinfo {author} {\bibfnamefont {J.}~\bibnamefont {Antoni}}, \bibinfo {author} {\bibfnamefont {S.}~\bibnamefont {Ehrlich}},\ and\ \bibinfo {author} {\bibfnamefont {H.}~\bibnamefont {Krieg}},\ }\bibfield  {title} {\bibinfo {title} {A consistent and accurate ab initio parametrization of density functional dispersion correction ({DFT-D}) for the 94 elements {H}-{P}u},\ }\href {https://doi.org/10.1063/1.3382344} {\bibfield  {journal} {\bibinfo  {journal} {J. Chem. Phys.}\ }\textbf {\bibinfo {volume} {132}},\ \bibinfo {pages} {154104} (\bibinfo {year} {2010})}\BibitemShut {NoStop}%
\bibitem [{\citenamefont {Homberg}(2022)}]{Homberg2022b}%
  \BibitemOpen
  \bibfield  {author} {\bibinfo {author} {\bibfnamefont {J.~B.}\ \bibnamefont {Homberg}},\ }\emph {\bibinfo {title} {Yu-Shiba-Rusinov States of Molecules on Pb(100)}},\ \href {https://macau.uni-kiel.de/servlets/MCRFileNodeServlet/macau_derivate_00004173/Jan_Homberg.pdf} {Ph.D. thesis},\ \bibinfo  {school} {Christian-Albrechts-Universität zu Kiel}, \bibinfo {address} {Mathematisch-Naturwissenschaftliche Fakultät} (\bibinfo {year} {2022})\BibitemShut {NoStop}%
\bibitem [{SM()}]{SM}%
  \BibitemOpen
  \href@noop {} {\bibinfo {title} {See {S}upplemental {M}aterial at [{URL} will be inserted by publisher] for a display of the {HOMO} and {LUMO} isosurface plots of the {P}b{P}c and {P}b{P}c$^{1-}$/surf; spin density plots of {P}b{P}c$^{1-}$; {J}ahn-{T}eller distortion of {P}b{P}c$^{1-}$; top view of {P}b{P}c relaxed on a {P}b(100) surface; the spectral function of {P}b{P}c with the {P}b{P}c/surf geometry; of {P}b{P}c$^{1-}$ with {P}b{P}c/surf geometry (majority spin); of {P}b{P}c$^{2-}$; of {P}b{P}c with {B3LYP} functional; of {P}b{P}c with the {P}b{P}c/surf geometry with {B3LYP} functional; of {P}b{P}c for a larger energy range; plot of {EV} coupling constants; spectral function of {HOMO} of {P}b{P}c$^{1-}$/surf together with respective vibrational modes; a table with the data for 20 peaks for {HOMO} of {P}b{P}c$^{1-}$/surf.}}\BibitemShut {Stop}%
\bibitem [{\citenamefont {B\"{u}rkle}\ \emph {et~al.}(2013)\citenamefont {B\"{u}rkle}, \citenamefont {Viljas}, \citenamefont {Hellmuth}, \citenamefont {Scheer}, \citenamefont {Weigend}, \citenamefont {Schön},\ and\ \citenamefont {Pauly}}]{Burkle}%
  \BibitemOpen
  \bibfield  {author} {\bibinfo {author} {\bibfnamefont {M.}~\bibnamefont {B\"{u}rkle}}, \bibinfo {author} {\bibfnamefont {J.~K.}\ \bibnamefont {Viljas}}, \bibinfo {author} {\bibfnamefont {T.~J.}\ \bibnamefont {Hellmuth}}, \bibinfo {author} {\bibfnamefont {E.}~\bibnamefont {Scheer}}, \bibinfo {author} {\bibfnamefont {F.}~\bibnamefont {Weigend}}, \bibinfo {author} {\bibfnamefont {G.}~\bibnamefont {Schön}},\ and\ \bibinfo {author} {\bibfnamefont {F.}~\bibnamefont {Pauly}},\ }\bibfield  {title} {\bibinfo {title} {Influence of vibrations on electron transport through nanoscale contacts},\ }\href {https://doi.org/10.1002/pssb.201350212} {\bibfield  {journal} {\bibinfo  {journal} {Phys. Stat. Sol. b}\ }\textbf {\bibinfo {volume} {250}},\ \bibinfo {pages} {2468} (\bibinfo {year} {2013})}\BibitemShut {NoStop}%
\bibitem [{\citenamefont {Deglmann}\ \emph {et~al.}(2002)\citenamefont {Deglmann}, \citenamefont {Furche},\ and\ \citenamefont {Ahlrichs}}]{aoforce2002}%
  \BibitemOpen
  \bibfield  {author} {\bibinfo {author} {\bibfnamefont {P.}~\bibnamefont {Deglmann}}, \bibinfo {author} {\bibfnamefont {F.}~\bibnamefont {Furche}},\ and\ \bibinfo {author} {\bibfnamefont {R.}~\bibnamefont {Ahlrichs}},\ }\bibfield  {title} {\bibinfo {title} {An efficient implementation of second analytical derivatives for density functional methods},\ }\href {https://doi.org/10.1016/S0009-2614(02)01084-9} {\bibfield  {journal} {\bibinfo  {journal} {Chem. Phys. Lett.}\ }\textbf {\bibinfo {volume} {362}},\ \bibinfo {pages} {511} (\bibinfo {year} {2002})}\BibitemShut {NoStop}%
\bibitem [{\citenamefont {Deglmann}\ and\ \citenamefont {Furche}(2002)}]{aoforce2002b}%
  \BibitemOpen
  \bibfield  {author} {\bibinfo {author} {\bibfnamefont {P.}~\bibnamefont {Deglmann}}\ and\ \bibinfo {author} {\bibfnamefont {F.}~\bibnamefont {Furche}},\ }\bibfield  {title} {\bibinfo {title} {Efficient characterization of stationary points on potential energy surfaces},\ }\href {https://doi.org/10.1063/1.1523393} {\bibfield  {journal} {\bibinfo  {journal} {J. Chem. Phys.}\ }\textbf {\bibinfo {volume} {117}},\ \bibinfo {pages} {9535–9538} (\bibinfo {year} {2002})}\BibitemShut {NoStop}%
\bibitem [{\citenamefont {Deglmann}\ \emph {et~al.}(2004)\citenamefont {Deglmann}, \citenamefont {May}, \citenamefont {Furche},\ and\ \citenamefont {Ahlrichs}}]{aoforce2004}%
  \BibitemOpen
  \bibfield  {author} {\bibinfo {author} {\bibfnamefont {P.}~\bibnamefont {Deglmann}}, \bibinfo {author} {\bibfnamefont {K.}~\bibnamefont {May}}, \bibinfo {author} {\bibfnamefont {F.}~\bibnamefont {Furche}},\ and\ \bibinfo {author} {\bibfnamefont {R.}~\bibnamefont {Ahlrichs}},\ }\bibfield  {title} {\bibinfo {title} {Nuclear second analytical derivative calculations using auxiliary basis set expansions},\ }\href {https://doi.org/10.1016/j.cplett.2003.11.080} {\bibfield  {journal} {\bibinfo  {journal} {Chem. Phys. Lett.}\ }\textbf {\bibinfo {volume} {384}},\ \bibinfo {pages} {103} (\bibinfo {year} {2004})}\BibitemShut {NoStop}%
\bibitem [{\citenamefont {Reiter}\ \emph {et~al.}(2017)\citenamefont {Reiter}, \citenamefont {K\"{u}hn},\ and\ \citenamefont {Weigend}}]{aoforce2017}%
  \BibitemOpen
  \bibfield  {author} {\bibinfo {author} {\bibfnamefont {K.}~\bibnamefont {Reiter}}, \bibinfo {author} {\bibfnamefont {M.}~\bibnamefont {K\"{u}hn}},\ and\ \bibinfo {author} {\bibfnamefont {F.}~\bibnamefont {Weigend}},\ }\bibfield  {title} {\bibinfo {title} {Vibrational circular dichroism spectra for large molecules and molecules with heavy elements},\ }\href {https://doi.org/10.1063/1.4974897} {\bibfield  {journal} {\bibinfo  {journal} {J. Chem. Phys.}\ }\textbf {\bibinfo {volume} {146}},\ \bibinfo {pages} {054102} (\bibinfo {year} {2017})}\BibitemShut {NoStop}%
\bibitem [{\citenamefont {Kato}\ \emph {et~al.}(2022)\citenamefont {Kato}, \citenamefont {Yoshizawa}, \citenamefont {Nakaya}, \citenamefont {Kitagawa}, \citenamefont {Okamoto}, \citenamefont {Yamada}, \citenamefont {Yoshino},\ and\ \citenamefont {Jun}}]{scirep}%
  \BibitemOpen
  \bibfield  {author} {\bibinfo {author} {\bibfnamefont {M.}~\bibnamefont {Kato}}, \bibinfo {author} {\bibfnamefont {H.}~\bibnamefont {Yoshizawa}}, \bibinfo {author} {\bibfnamefont {M.}~\bibnamefont {Nakaya}}, \bibinfo {author} {\bibfnamefont {Y.}~\bibnamefont {Kitagawa}}, \bibinfo {author} {\bibfnamefont {K.}~\bibnamefont {Okamoto}}, \bibinfo {author} {\bibfnamefont {T.}~\bibnamefont {Yamada}}, \bibinfo {author} {\bibfnamefont {M.}~\bibnamefont {Yoshino}},\ and\ \bibinfo {author} {\bibfnamefont {O.}~\bibnamefont {Jun}},\ }\bibfield  {title} {\bibinfo {title} {Unraveling the reasons behind lead phthalocyanine acting as a good absorber for near-infrared sensitive devices},\ }\href {https://doi.org/10.1038/s41598-022-12990-z} {\bibfield  {journal} {\bibinfo  {journal} {Sci. Rep.}\ }\textbf {\bibinfo {volume} {12}},\ \bibinfo {pages} {8810} (\bibinfo {year} {2022})}\BibitemShut {NoStop}%
\bibitem [{\citenamefont {T{\'o}bik}\ and\ \citenamefont {Tosatti}(2007)}]{Tobik2007}%
  \BibitemOpen
  \bibfield  {author} {\bibinfo {author} {\bibfnamefont {J.}~\bibnamefont {T{\'o}bik}}\ and\ \bibinfo {author} {\bibfnamefont {E.}~\bibnamefont {Tosatti}},\ }\bibfield  {title} {\bibinfo {title} {Jahn--{T}eller effect in the magnesium phthalocyanine anion},\ }\href {https://doi.org/10.1016/j.molstruc.2006.12.051} {\bibfield  {journal} {\bibinfo  {journal} {Journal of molecular structure}\ }\textbf {\bibinfo {volume} {838}},\ \bibinfo {pages} {112} (\bibinfo {year} {2007})}\BibitemShut {NoStop}%
\bibitem [{\citenamefont {Sklyadneva}\ \emph {et~al.}(2012)\citenamefont {Sklyadneva}, \citenamefont {Heid}, \citenamefont {Bohnen}, \citenamefont {Echenique},\ and\ \citenamefont {Chulkov}}]{Sklyadneva2012}%
  \BibitemOpen
  \bibfield  {author} {\bibinfo {author} {\bibfnamefont {I.~Y.}\ \bibnamefont {Sklyadneva}}, \bibinfo {author} {\bibfnamefont {R.}~\bibnamefont {Heid}}, \bibinfo {author} {\bibfnamefont {K.-P.}\ \bibnamefont {Bohnen}}, \bibinfo {author} {\bibfnamefont {P.~M.}\ \bibnamefont {Echenique}},\ and\ \bibinfo {author} {\bibfnamefont {E.~V.}\ \bibnamefont {Chulkov}},\ }\bibfield  {title} {\bibinfo {title} {Surface phonons on pb(111)},\ }\href {https://doi.org/10.1088/0953-8984/24/10/104004} {\bibfield  {journal} {\bibinfo  {journal} {J. Phys.: Condens. Matter}\ }\textbf {\bibinfo {volume} {24}},\ \bibinfo {pages} {104004} (\bibinfo {year} {2012})}\BibitemShut {NoStop}%
\bibitem [{\citenamefont {Verstraete}\ \emph {et~al.}(2008)\citenamefont {Verstraete}, \citenamefont {Torrent}, \citenamefont {Jollet}, \citenamefont {Z\'erah},\ and\ \citenamefont {Gonze}}]{Verstraete2008}%
  \BibitemOpen
  \bibfield  {author} {\bibinfo {author} {\bibfnamefont {M.~J.}\ \bibnamefont {Verstraete}}, \bibinfo {author} {\bibfnamefont {M.}~\bibnamefont {Torrent}}, \bibinfo {author} {\bibfnamefont {F.}~\bibnamefont {Jollet}}, \bibinfo {author} {\bibfnamefont {G.}~\bibnamefont {Z\'erah}},\ and\ \bibinfo {author} {\bibfnamefont {X.}~\bibnamefont {Gonze}},\ }\bibfield  {title} {\bibinfo {title} {Density functional perturbation theory with spin-orbit coupling: {P}honon band structure of lead},\ }\href {https://doi.org/10.1103/PhysRevB.78.045119} {\bibfield  {journal} {\bibinfo  {journal} {Phys. Rev. B}\ }\textbf {\bibinfo {volume} {78}},\ \bibinfo {pages} {045119} (\bibinfo {year} {2008})}\BibitemShut {NoStop}%
\bibitem [{\citenamefont {Lorente}\ and\ \citenamefont {Persson}(2000)}]{Lorente2020}%
  \BibitemOpen
  \bibfield  {author} {\bibinfo {author} {\bibfnamefont {N.}~\bibnamefont {Lorente}}\ and\ \bibinfo {author} {\bibfnamefont {M.}~\bibnamefont {Persson}},\ }\bibfield  {title} {\bibinfo {title} {Theory of single molecule vibrational spectroscopy and microscopy},\ }\href {https://doi.org/10.1103/PhysRevLett.85.2997} {\bibfield  {journal} {\bibinfo  {journal} {Phys. Rev. Lett.}\ }\textbf {\bibinfo {volume} {85}},\ \bibinfo {pages} {2997} (\bibinfo {year} {2000})}\BibitemShut {NoStop}%
\bibitem [{\citenamefont {Paulsson}\ \emph {et~al.}(2005)\citenamefont {Paulsson}, \citenamefont {Frederiksen},\ and\ \citenamefont {Brandbyge}}]{Paulsson2005}%
  \BibitemOpen
  \bibfield  {author} {\bibinfo {author} {\bibfnamefont {M.}~\bibnamefont {Paulsson}}, \bibinfo {author} {\bibfnamefont {T.}~\bibnamefont {Frederiksen}},\ and\ \bibinfo {author} {\bibfnamefont {M.}~\bibnamefont {Brandbyge}},\ }\bibfield  {title} {\bibinfo {title} {Modeling inelastic phonon scattering in atomic- and molecular-wire junctions},\ }\href {https://doi.org/10.1103/PhysRevB.72.201101} {\bibfield  {journal} {\bibinfo  {journal} {Phys. Rev. B}\ }\textbf {\bibinfo {volume} {72}},\ \bibinfo {pages} {201101} (\bibinfo {year} {2005})}\BibitemShut {NoStop}%
\bibitem [{\citenamefont {von Oppen}\ and\ \citenamefont {Franke}(2021)}]{Oppen2021}%
  \BibitemOpen
  \bibfield  {author} {\bibinfo {author} {\bibfnamefont {F.}~\bibnamefont {von Oppen}}\ and\ \bibinfo {author} {\bibfnamefont {K.~J.}\ \bibnamefont {Franke}},\ }\bibfield  {title} {\bibinfo {title} {Yu-shiba-rusinov states in real metals},\ }\href@noop {} {\bibfield  {journal} {\bibinfo  {journal} {Phys. Rev. B}\ }\textbf {\bibinfo {volume} {103}},\ \bibinfo {pages} {205424} (\bibinfo {year} {2021})}\BibitemShut {NoStop}%
\end{thebibliography}%

\end{document}


\title{A minimal model of inelastic tunneling of vibrating magnetic molecules on superconducting substrates}

\author{Athanasios Koliogiorgos}
\affiliation{Department of Condensed Matter Physics, Faculty of Mathematics and Physics, Charles University, Prague, Czech Republic}
\email{athanasios.koliogiorgos@matfyz.cuni.cz}

\author{Richard Korytár}
\affiliation{Department of Condensed Matter Physics, Faculty of Mathematics and Physics, Charles University, Prague, Czech Republic}
\affiliation{Institute of Theoretical Physics, University of Regensburg, 93040 Regensburg, Germany}
\email{richard.korytar@ur.de}

\maketitle

\section{Supplemental Material}

\begin{figure}[ht]
\centering
\begin{subfigure}[b]{0.45\textwidth}
    \centering
    \includegraphics[width=\textwidth]{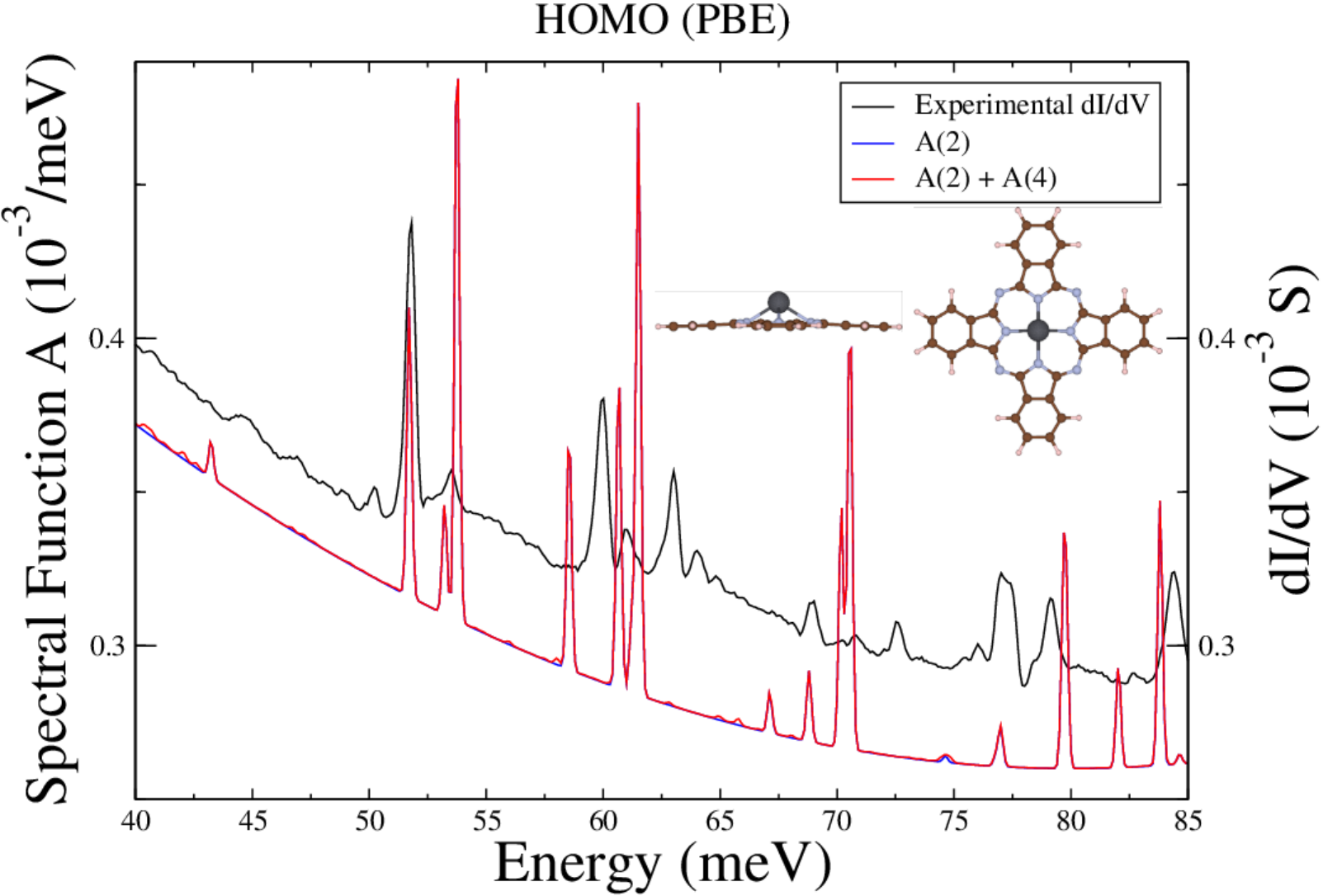}
    \caption{}
    \label{figS1a}
\end{subfigure}
\begin{subfigure}[b]{0.45\textwidth}
    \centering
    \includegraphics[width=\textwidth]{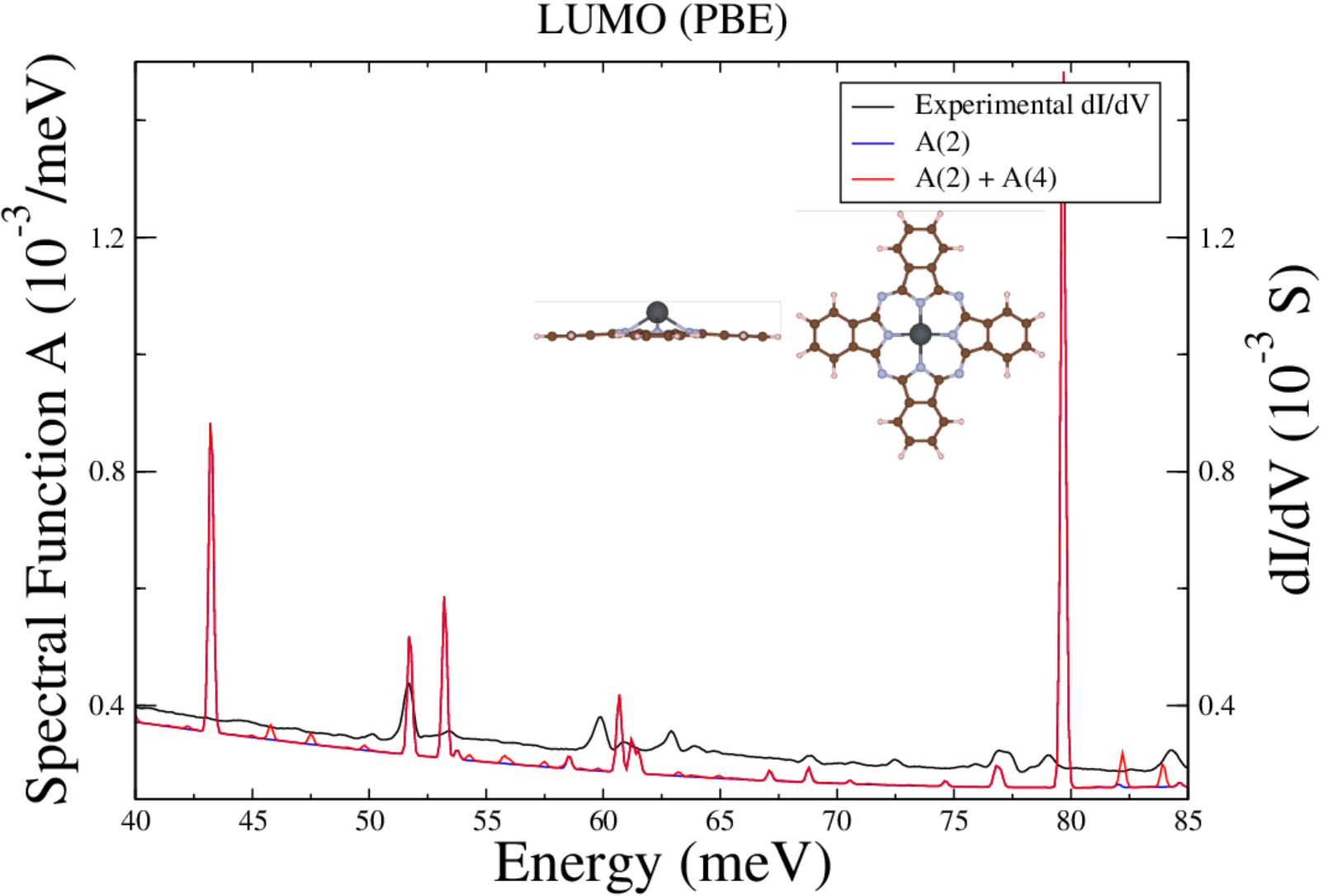}
    \caption{}
    \label{figS1b}
\end{subfigure}
\caption{Spectral function of PbPc describing single \([A^{(2)}_\alpha(\omega)]\) and double \([A^{(4)}_\alpha(\omega)]\) electron-vibrational transitions from the molecular orbitals $\alpha=$HOMO (a) and LUMO (b) using the PBE functional, compared with the experimental conductance spectrum. The geometry of PbPc was relaxed on top of Pb(100) and it is shown in the inset.}
\label{figS1}
\end{figure}

\begin{figure}[ht]
\centering
\begin{subfigure}[b]{0.45\textwidth}
    \centering
    \includegraphics[width=\textwidth]{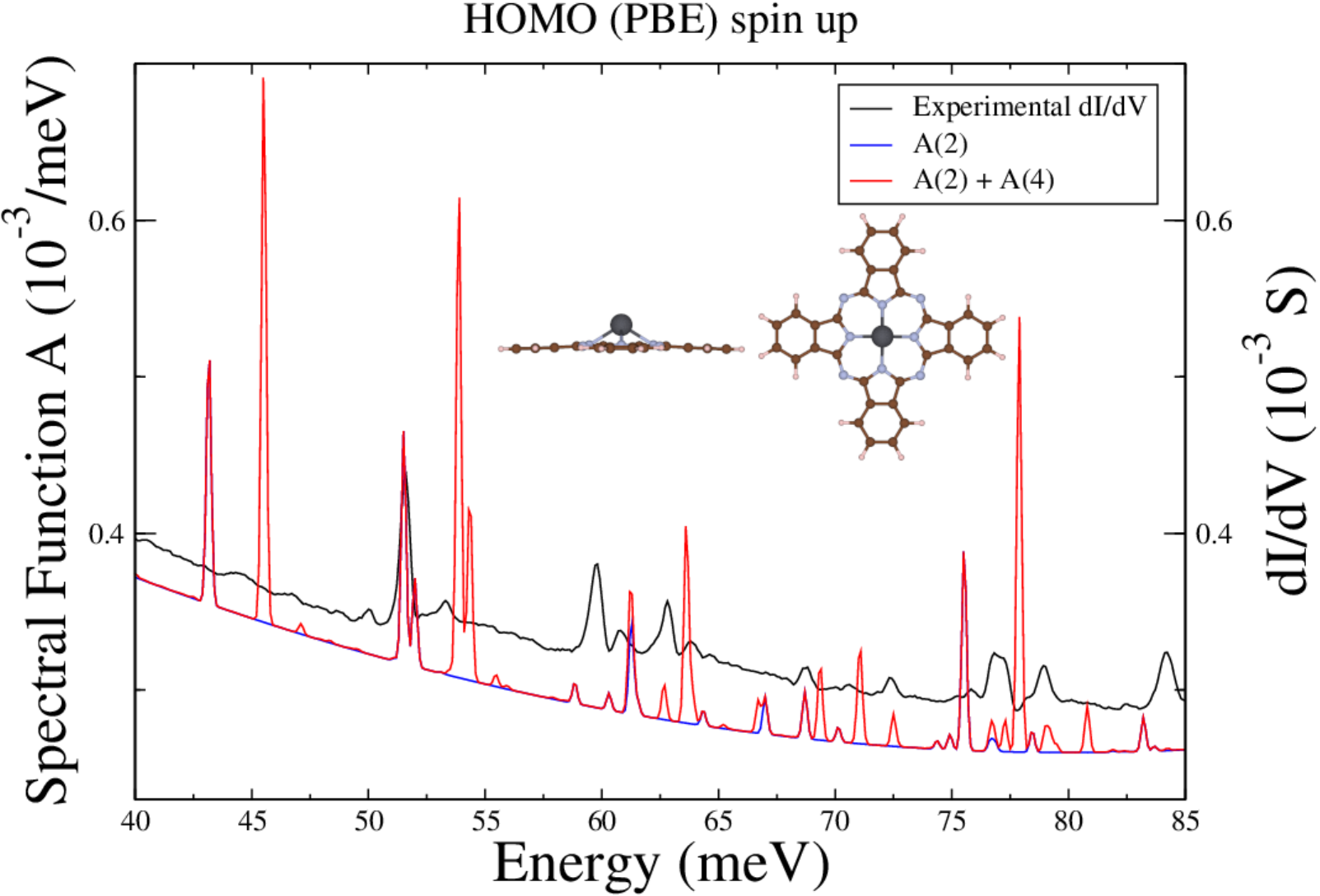}
    \caption{}
    \label{figS2a}
\end{subfigure}
\begin{subfigure}[b]{0.45\textwidth}
    \centering
    \includegraphics[width=\textwidth]{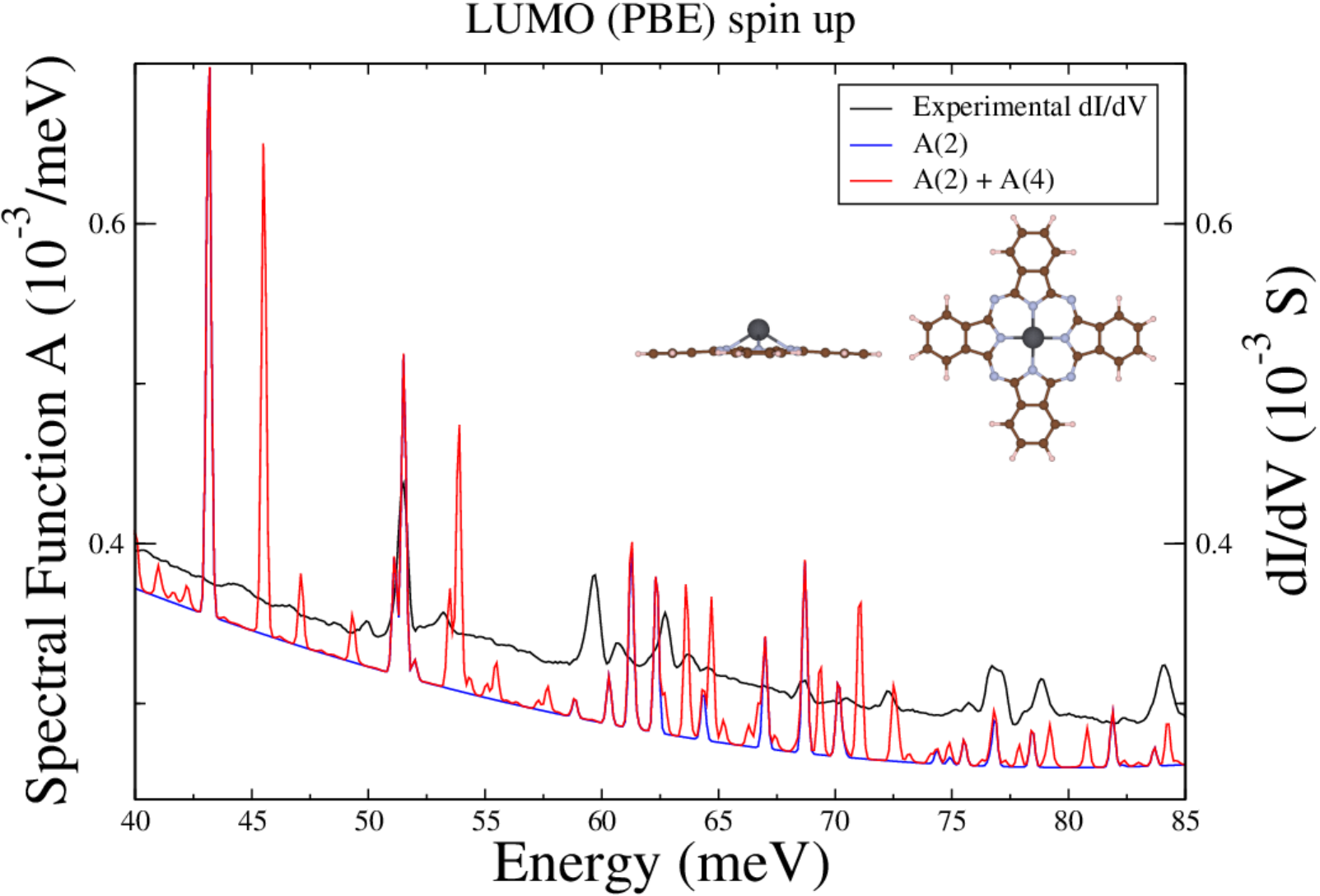}
    \caption{}
    \label{figS2b}
\end{subfigure}
\caption{Spectral function of PbPc$^{1-}$ describing single \([A^{(2)}_\alpha(\omega)]\) and double \([A^{(4)}_\alpha(\omega)]\) electron-vibrational transitions from the majority spin molecular orbitals $\alpha=$HOMO (a) and LUMO (b) using the PBE functional, compared with the experimental conductance spectrum. The C$_{4v}$ PbPc/surf geometry of the molecule appears in the inset.}
\label{figS2}
\end{figure}

\begin{figure}[ht]
\centering
\begin{subfigure}[b]{0.45\textwidth}
    \centering
    \includegraphics[width=\textwidth]{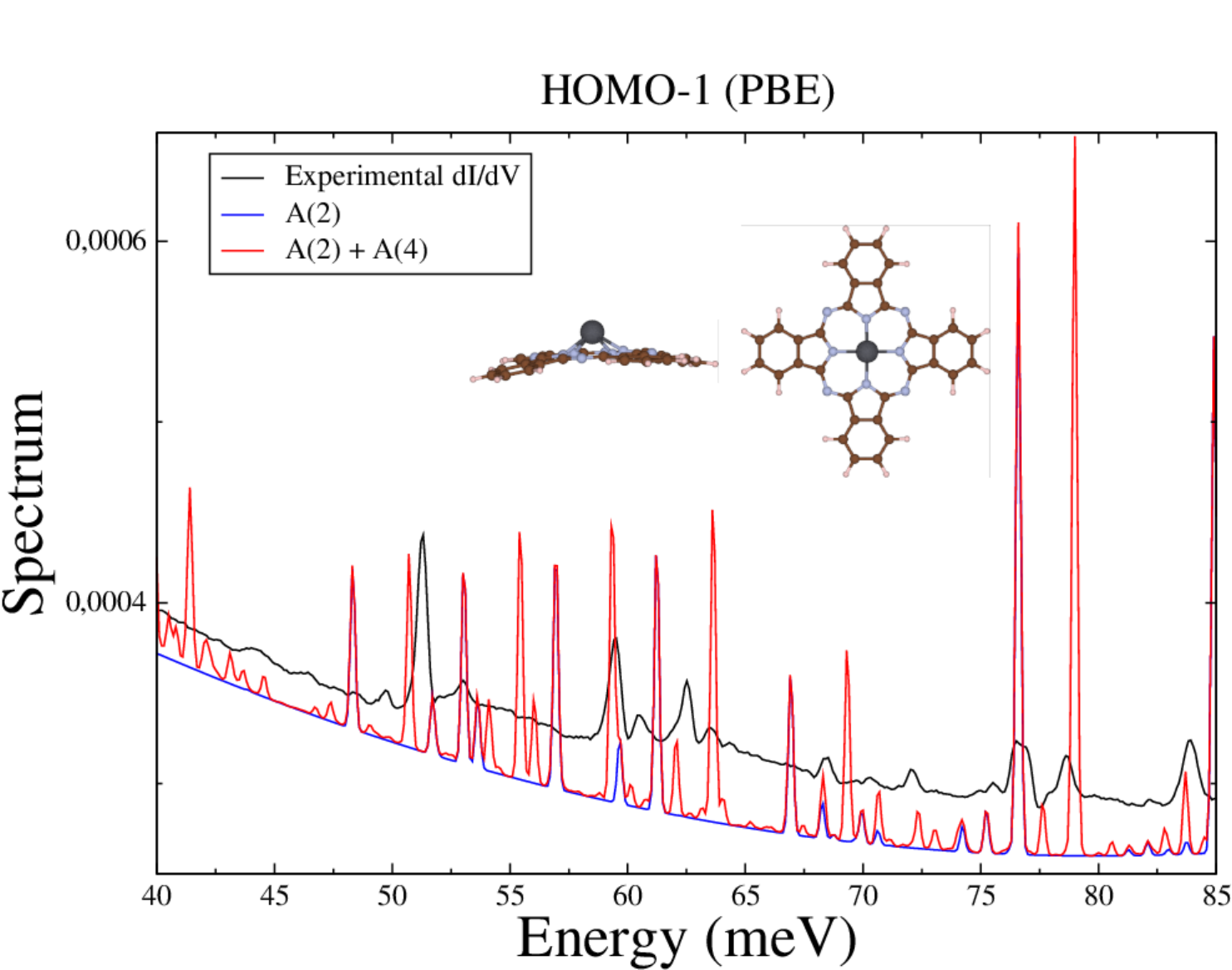}
    \caption{}
    \label{figS3a}
\end{subfigure}
\begin{subfigure}[b]{0.45\textwidth}
    \centering
    \includegraphics[width=\textwidth]{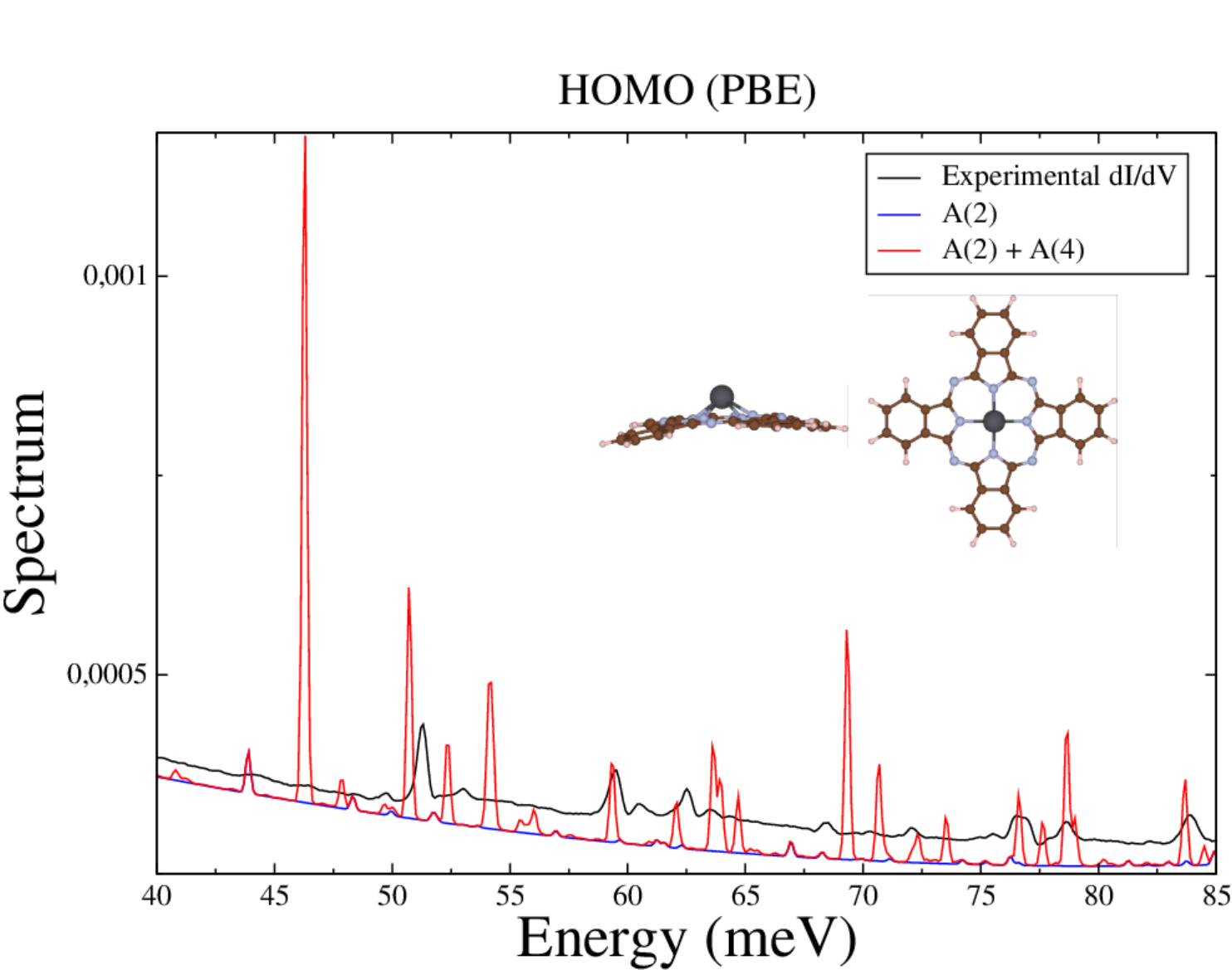}
    \caption{}
    \label{figS3b}
\end{subfigure}
\caption{Spectral function of gas phase PbPc$^{2-}$ with PBE functional, for HOMO-1 (a) and HOMO (b). The HOMO-1 level is equivalent to the HOMO in the neutral PbPc. The geometry of the molecule is curved similar to PbPc, but more asymmetrical. The much lower agreement with the experimental d$I$/d$V$ compared to PbPc$^{1-}$ indicates that the charge transfer taking place in proximity to Pb(100) is closer to 1${-}$ charge rather than 2${-}$.}
\label{figS3}
\end{figure}

\begin{figure}[ht]
\centering
\begin{subfigure}[b]{0.45\textwidth}
    \centering
    \includegraphics[width=\textwidth]{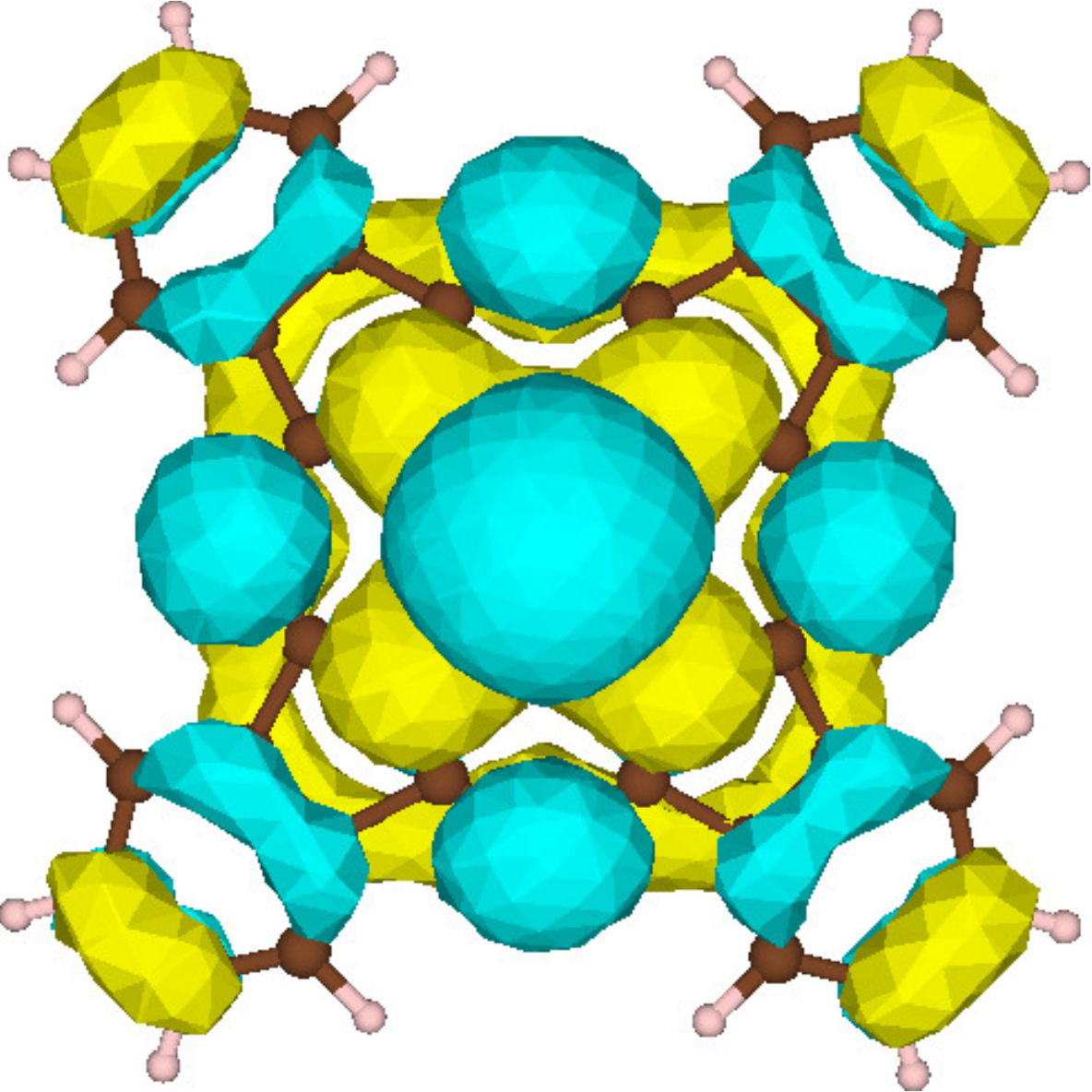}
    \caption{}
    \label{figS4a}
\end{subfigure}
\begin{subfigure}[b]{0.45\textwidth}
    \centering
    \includegraphics[width=\textwidth]{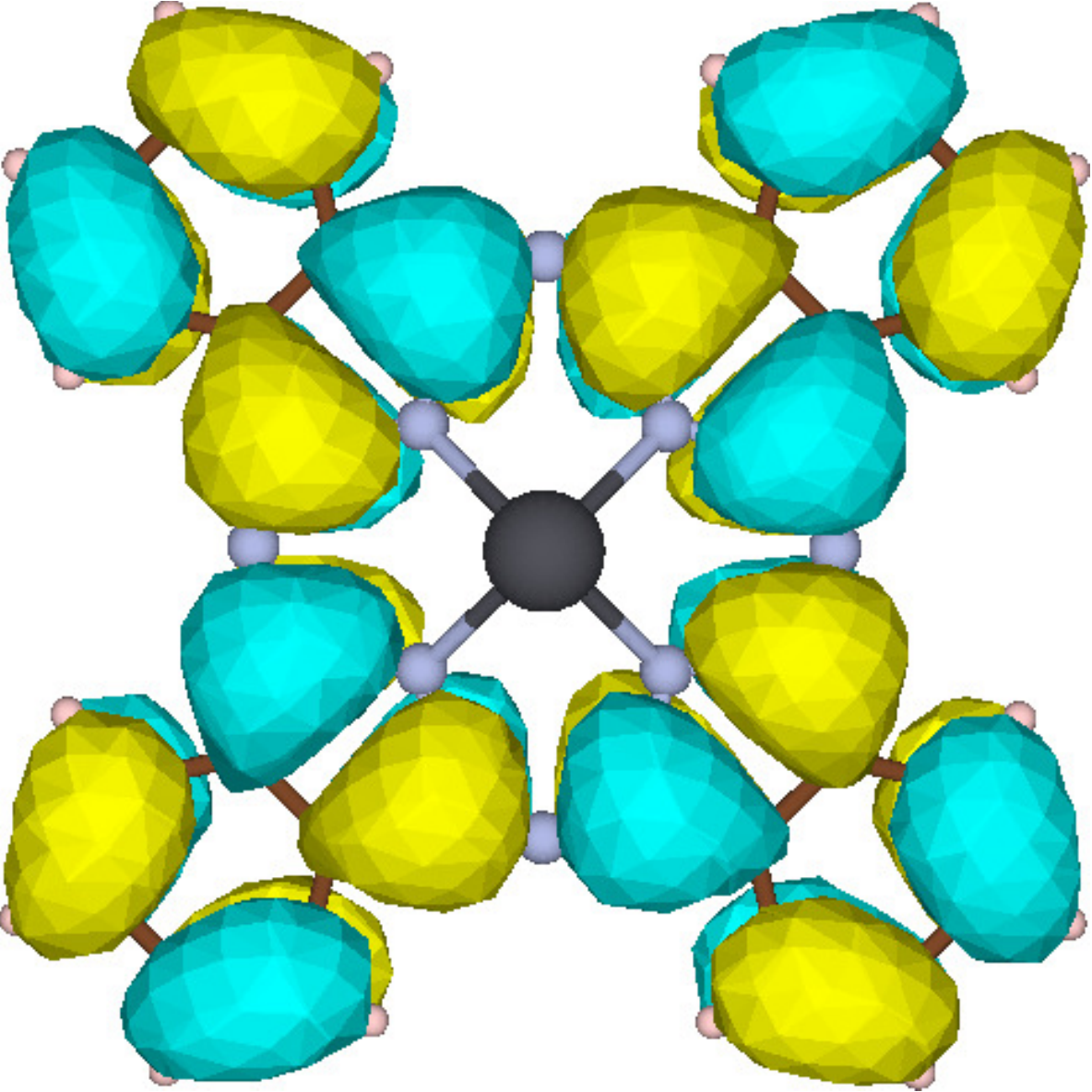}
    \caption{}
    \label{figS4b}
\end{subfigure}
\caption{Isosurface plots of the two highest occupied molecular orbitals for neutral PbPc: HOMO-1 (a) and HOMO (b).}
\label{figS4}
\end{figure}

\begin{figure}[ht]
\centering
\begin{subfigure}[b]{0.45\textwidth}
    \centering
    \includegraphics[width=\textwidth]{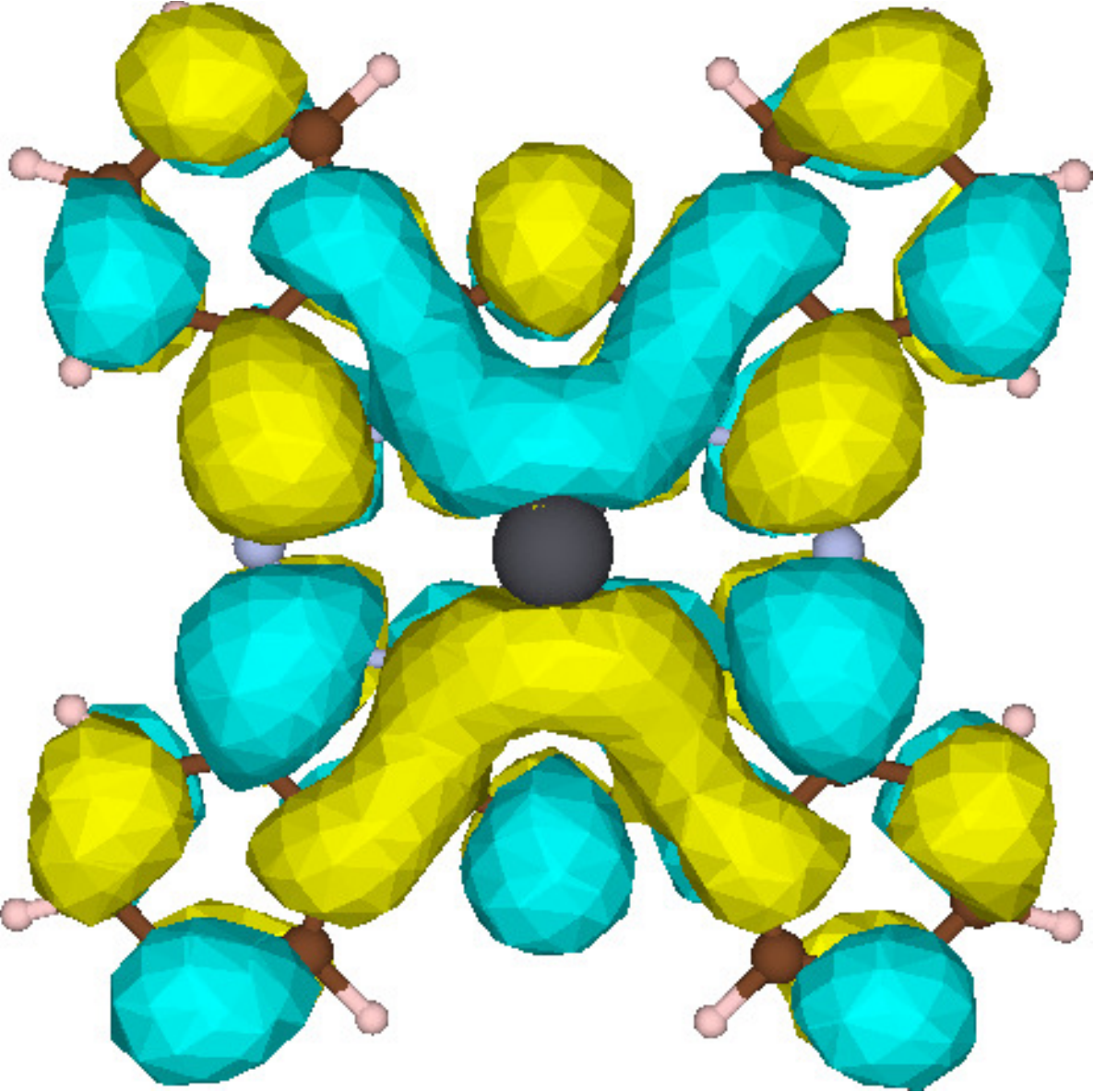}
    \caption{}
    \label{figS5a}
\end{subfigure}
\begin{subfigure}[b]{0.45\textwidth}
    \centering
    \includegraphics[width=\textwidth]{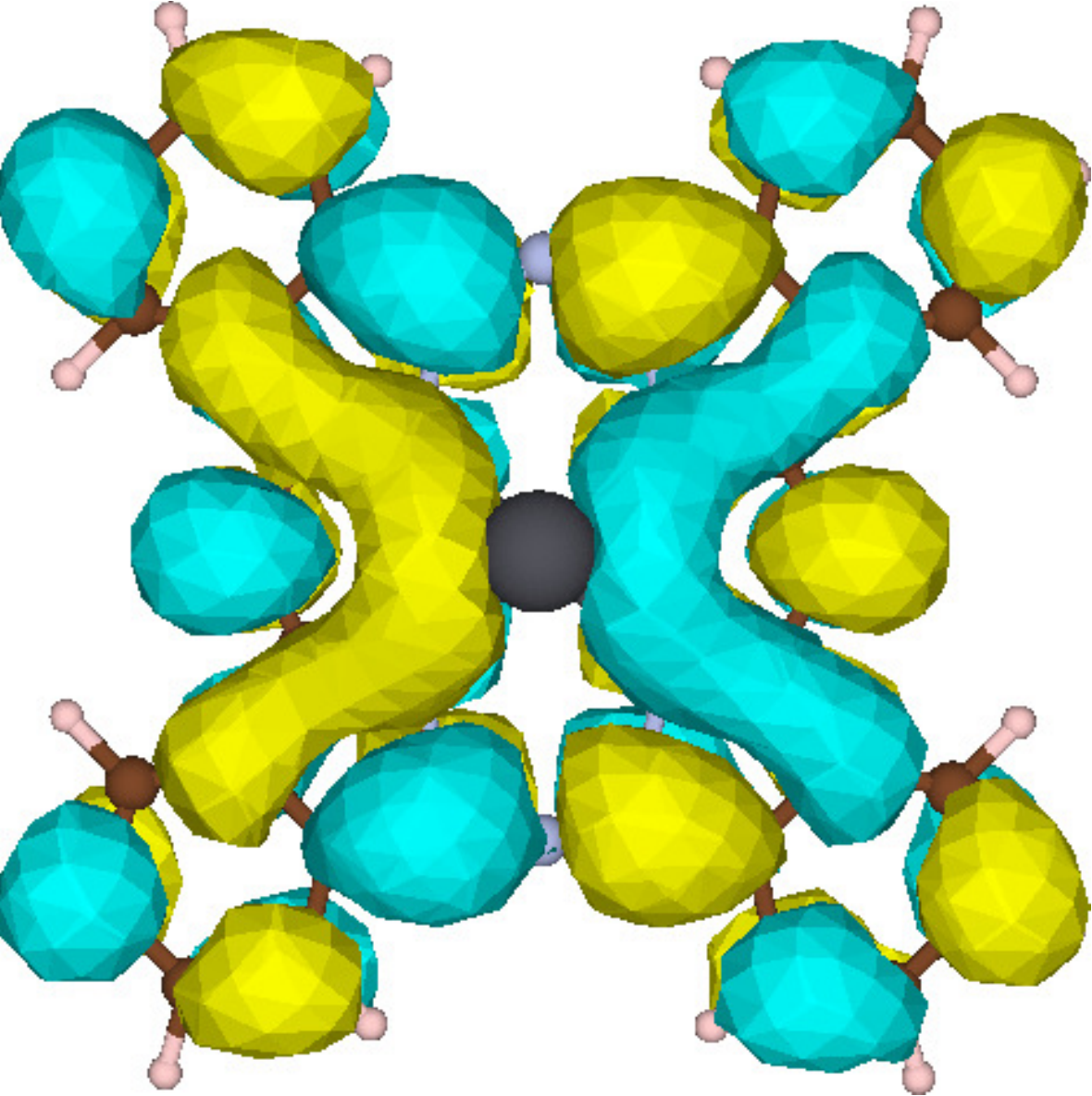}
    \caption{}
    \label{figS5b}
\end{subfigure}
\caption{Isosurface plots of the two e$_{\mathrm g}$ LUMOs for the PbPc in isolation. In contrast with HOMO, in LUMO there is more charge concentration near the Pb atom.}
\label{figS5}
\end{figure}

\begin{figure}[ht]
\centering
\begin{subfigure}[b]{0.45\textwidth}
    \centering
    \includegraphics[width=\textwidth]{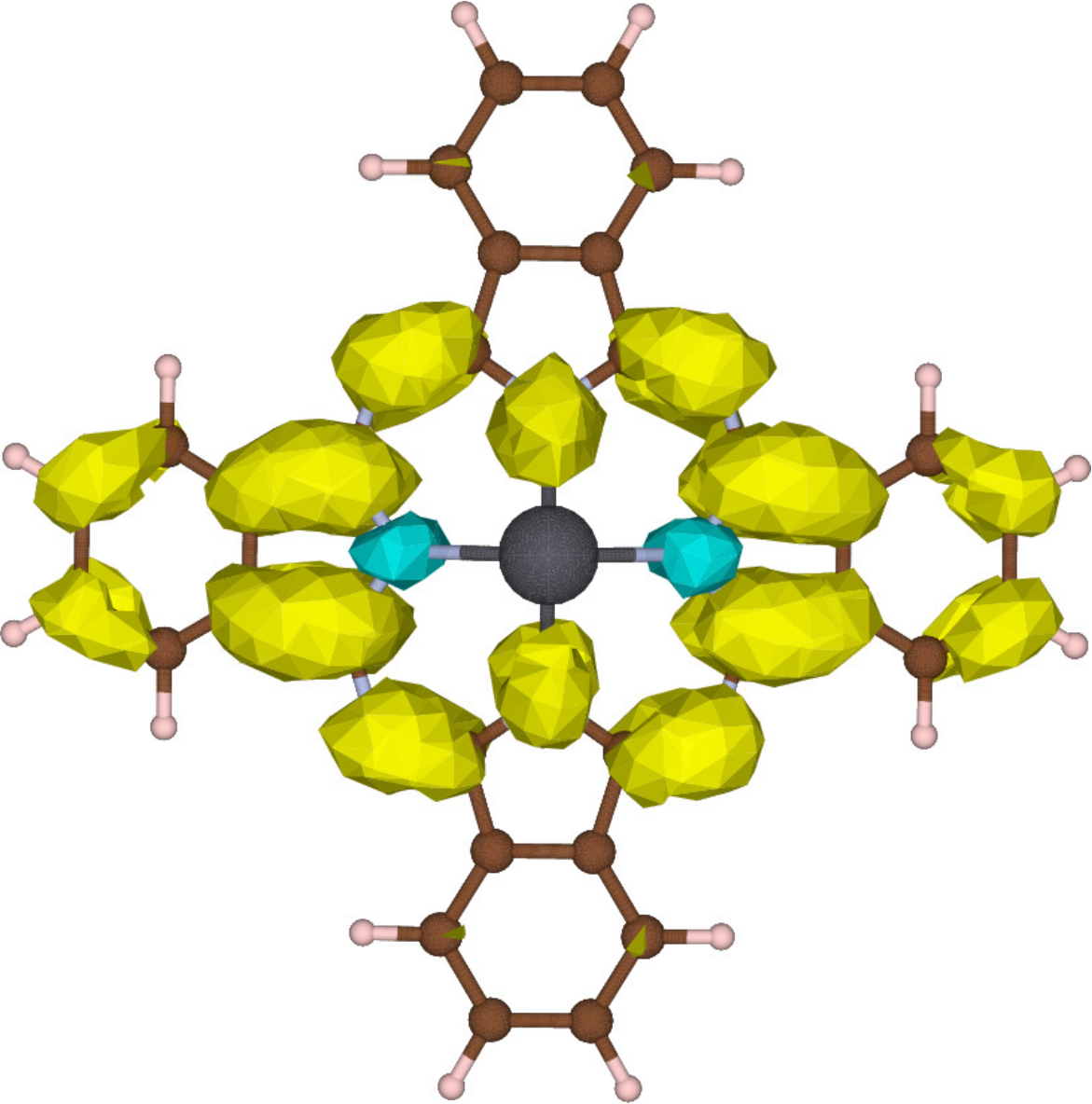}
    \caption{}
    \label{figS6a}
\end{subfigure}
\begin{subfigure}[b]{0.45\textwidth}
    \centering
    \includegraphics[width=\textwidth]{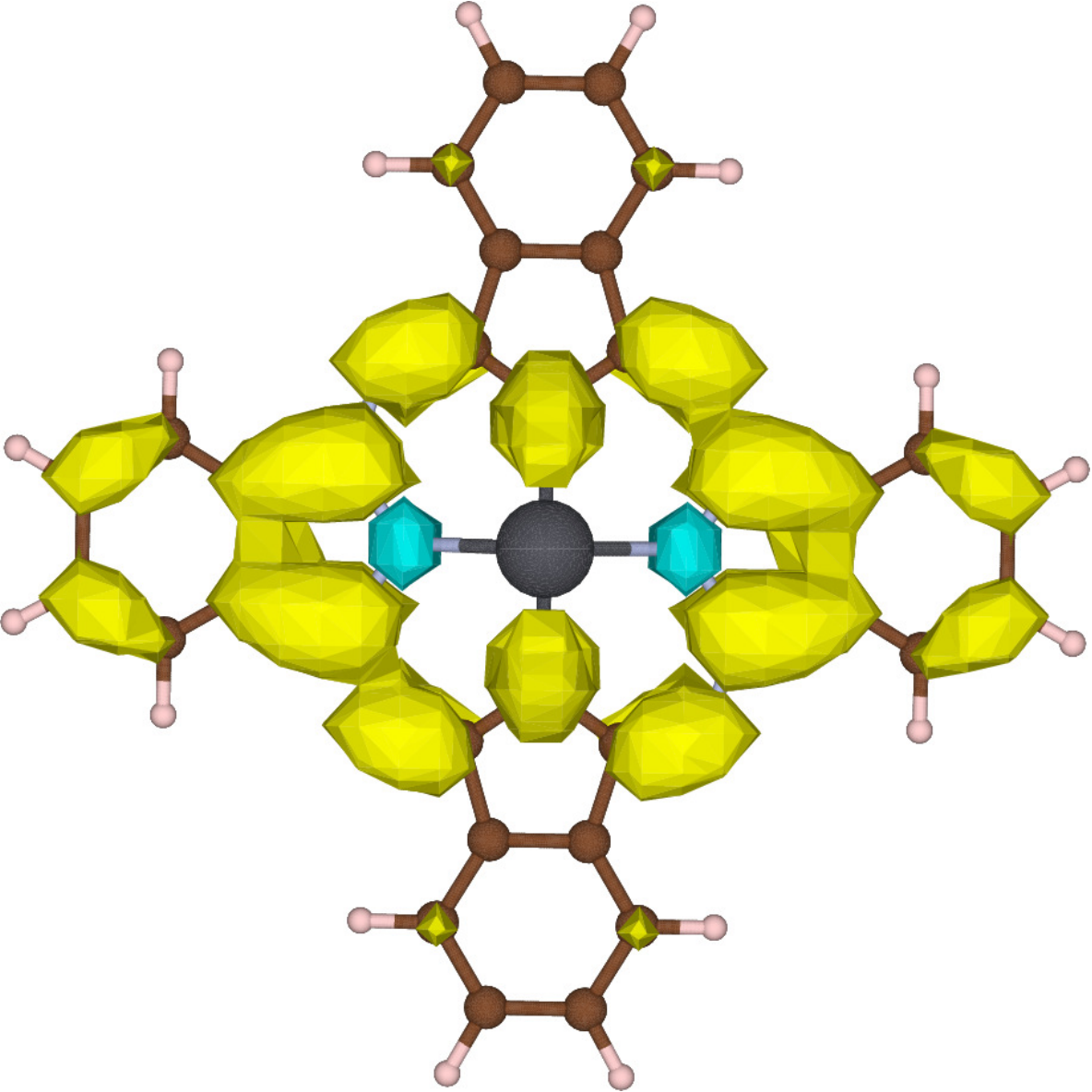}
    \caption{}
    \label{figS6b}
\end{subfigure}
\caption{Isosurface plot of the spin density of the gas phase C$_{2v}$ PbPc$^{1-}$ (a) and the PbPc$^{1-}$ with the C$_{4v}$ PbPc/surf geometry (b). Notice that in both cases the spin density can be seen as originating from a linear combination of the two LUMO e$_{\mathrm g}$ orbitals but with vanishing HOMO contribution.}
\label{figS6}
\end{figure}

\begin{figure}[ht]
\centering
\begin{subfigure}[b]{0.45\textwidth}
    \centering
    \includegraphics[width=\textwidth]{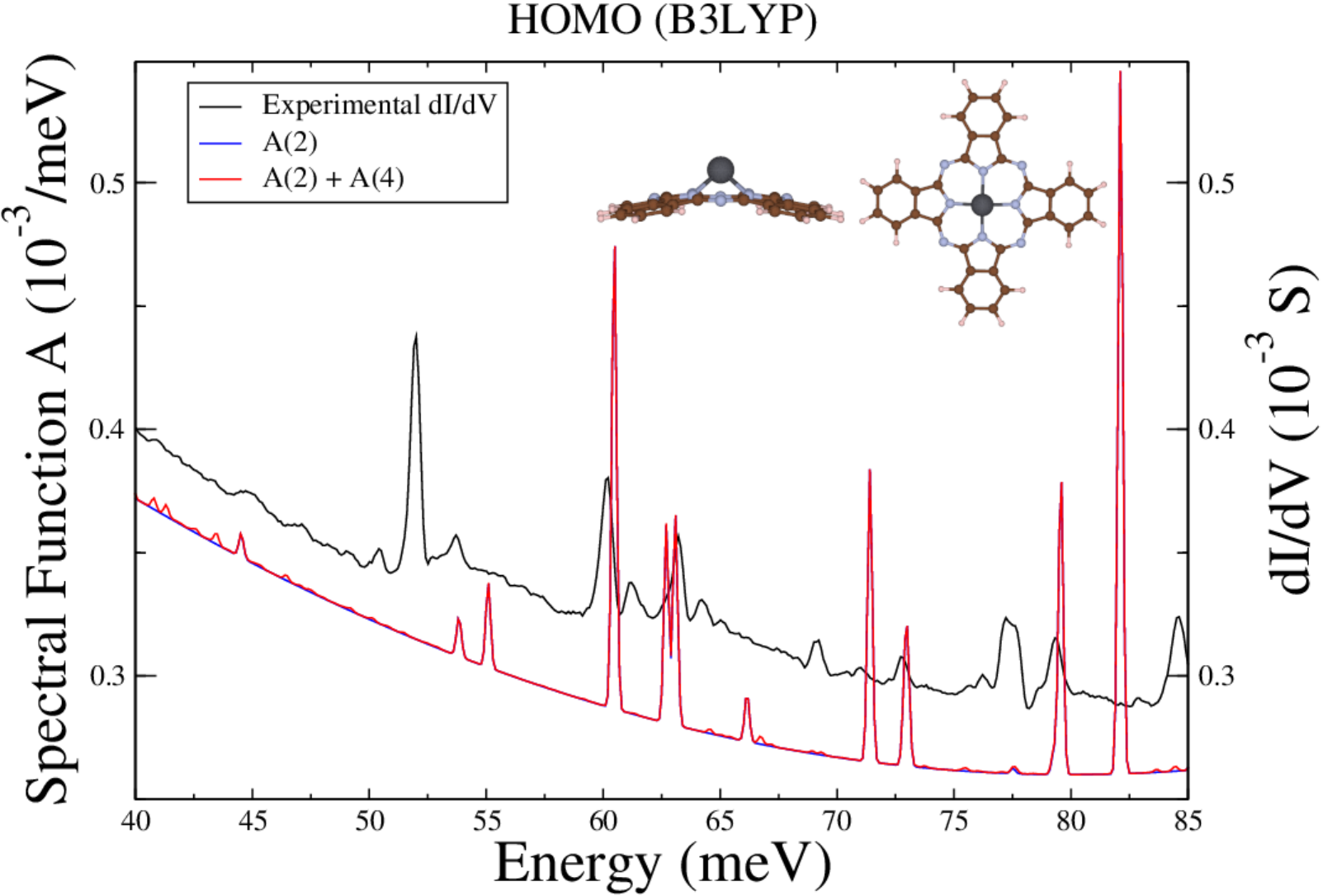}
    \caption{}
    \label{figS7a}
\end{subfigure}
\begin{subfigure}[b]{0.45\textwidth}
    \centering
    \includegraphics[width=\textwidth]{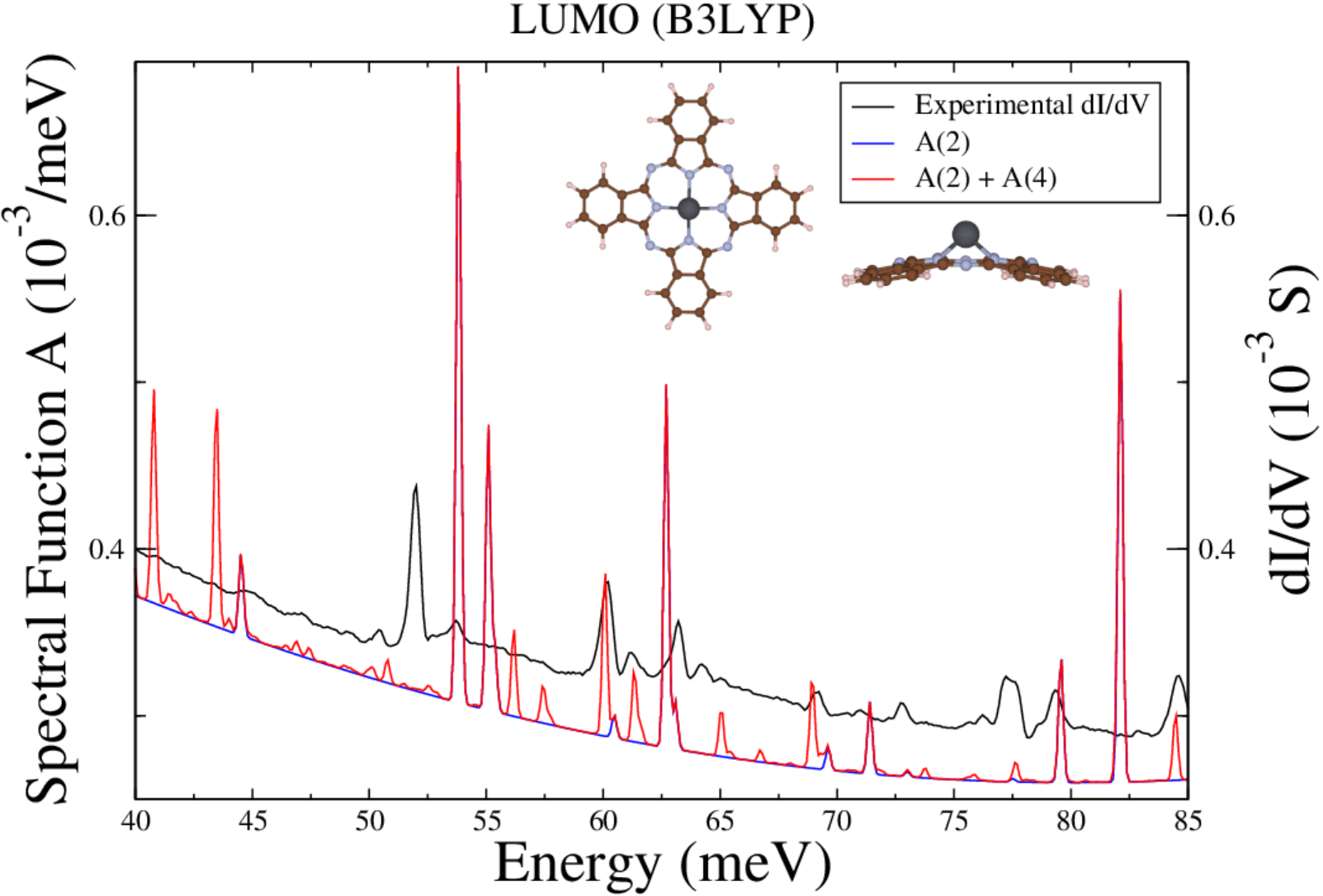}
    \caption{}
    \label{figS7b}
\end{subfigure}
\caption{Spectral function of the gas phase PbPc describing single \([A^{(2)}_\alpha(\omega)]\) and double \([A^{(4)}_\alpha(\omega)]\) electron-vibrational transitions from the molecular orbitals $\alpha=$HOMO (a) and LUMO (b) using the B3LYP functional, compared with the experimental conductance spectrum. The optimized geometry of the molecule appears in the inset.}
\label{figS7}
\end{figure}

\begin{figure}[ht]
\centering
\begin{subfigure}[b]{0.45\textwidth}
    \centering
    \includegraphics[width=\textwidth]{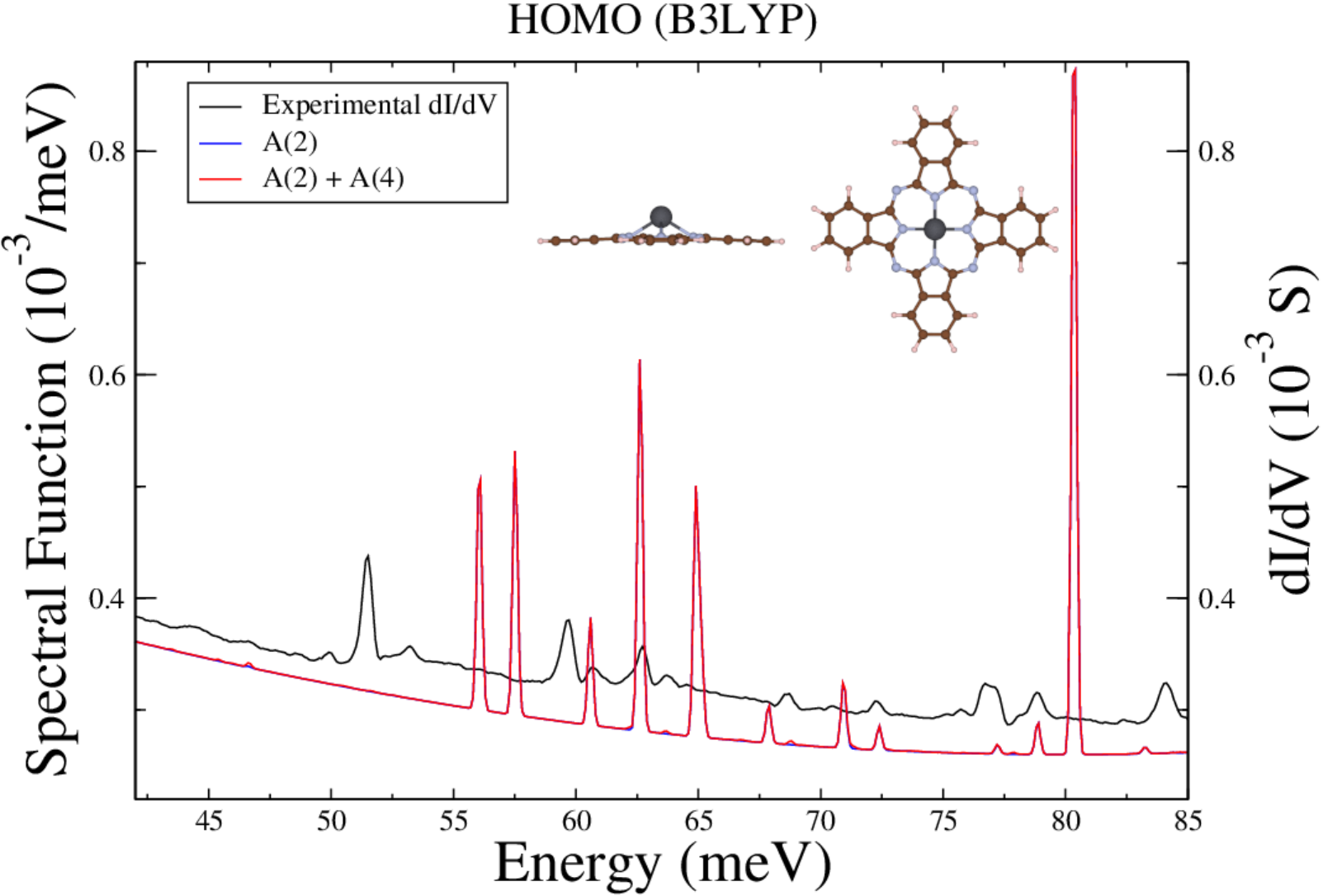}
    \caption{}
    \label{figS8a}
\end{subfigure}
\begin{subfigure}[b]{0.45\textwidth}
    \centering
    \includegraphics[width=\textwidth]{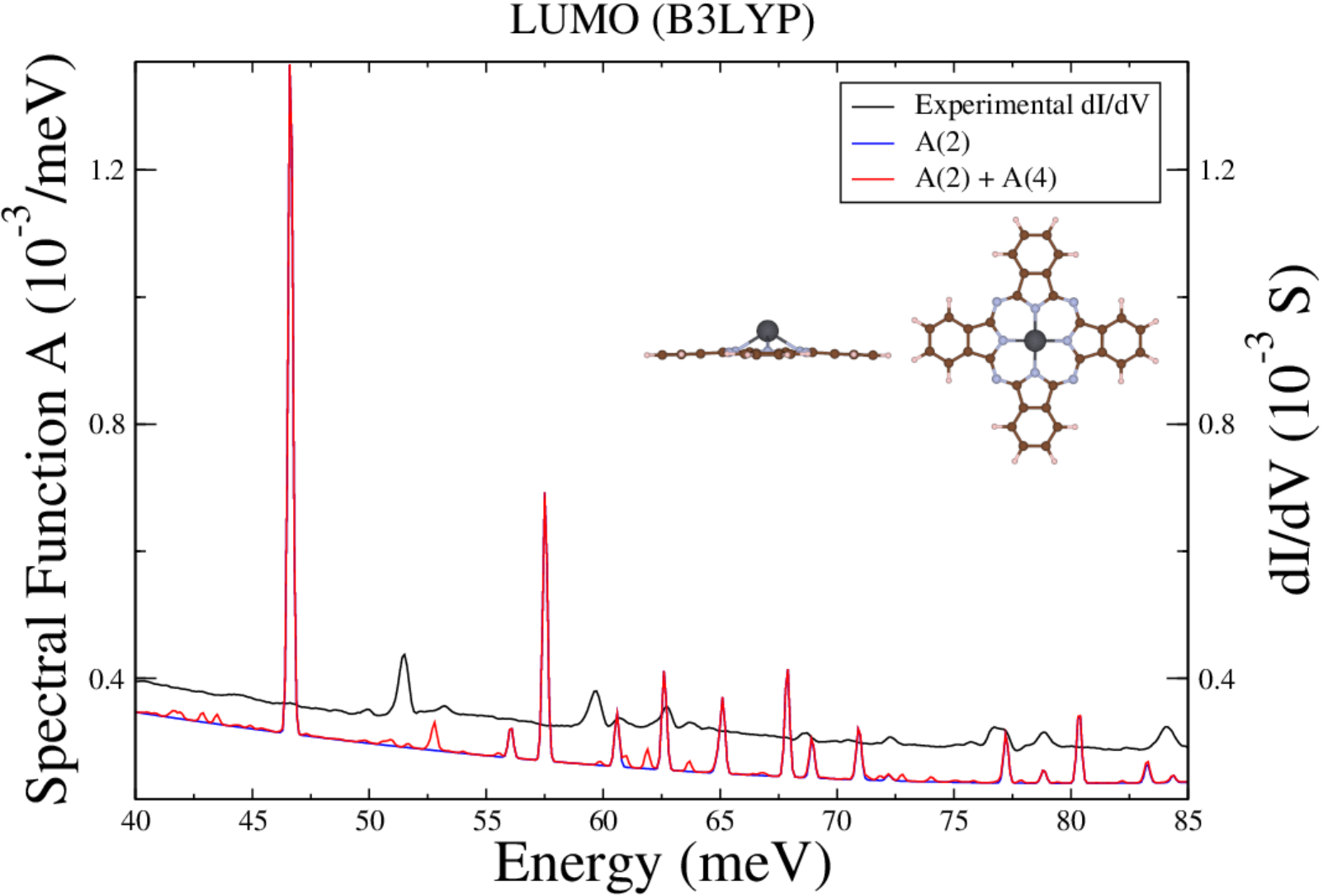}
    \caption{}
    \label{figS8b}
\end{subfigure}
\caption{Spectral function of PbPc describing single \([A^{(2)}_\alpha(\omega)]\) and double \([A^{(4)}_\alpha(\omega)]\) electron-vibrational transitions from the molecular orbitals $\alpha=$HOMO (a) and LUMO (b) using the B3LYP functional, compared with the experimental conductance spectrum. The geometry used was PbPc/surf and it is shown in the inset.}
\label{figS8}
\end{figure}

\begin{figure}[ht]
\centering
\includegraphics[width=0.45\textwidth]{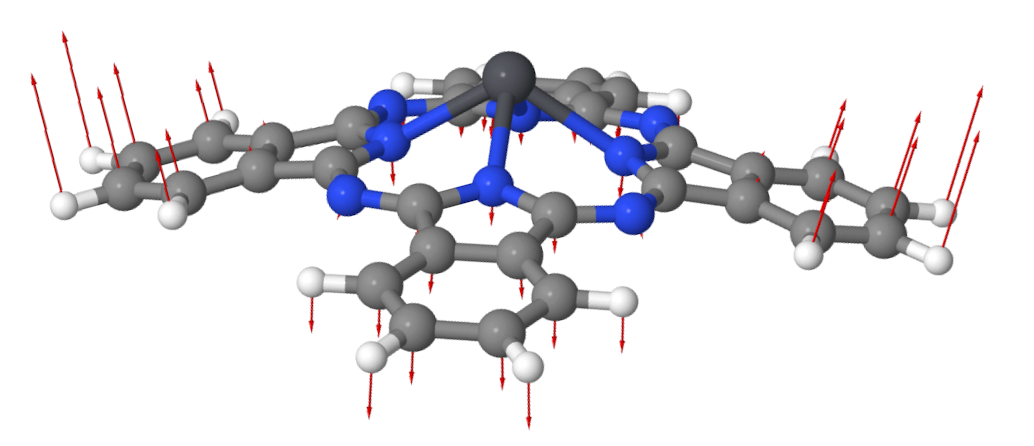}
\caption{Jahn-Teller distortion of PbPc$^{1-}$: Ball-and-stick model of PBE-relaxed geometry of PbPc$^{0}$ with arrows indicating the atomic displacements that take place upon charging with one electron. The magnitude of the arrows is exaggerated for the purpose of illustration. The reduction of symmetry from C$_{4v}$ to C$_{2v}$ is apparent.}
\label{figS9}
\end{figure}

\begin{figure}[ht]
\centering
\includegraphics[width=0.45\textwidth]{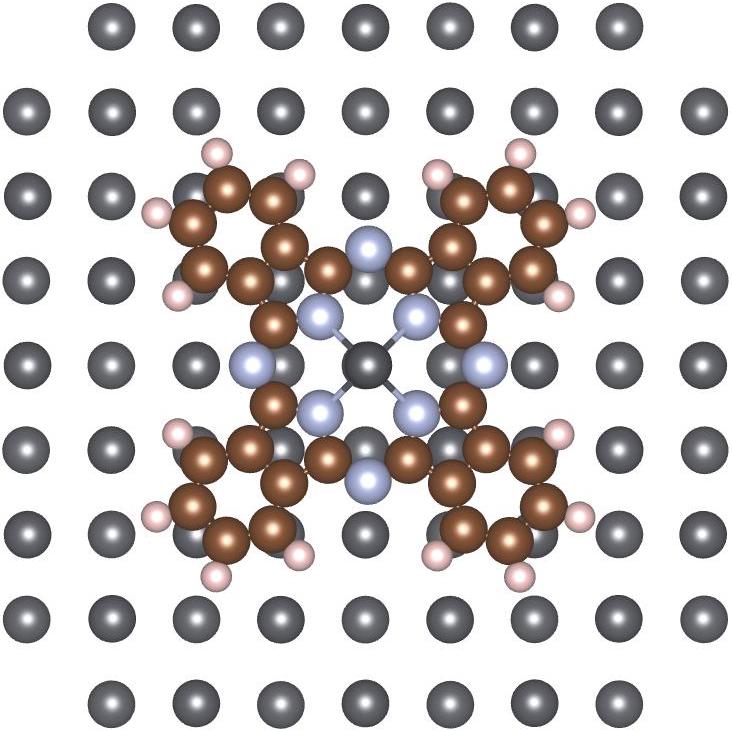}
\caption{Top view of a ball-and-stick model of PbPc relaxed on Pb(100) (Pb=gray, C=brown, N=blue, H=white). The Pb-on-Pb adsorption site is compatible with the C$_{4v}$ symmetry of the neutral PbPc.}
\label{figS10}
\end{figure}

\begin{figure}[ht]
\centering
\includegraphics[width=0.7\textwidth]{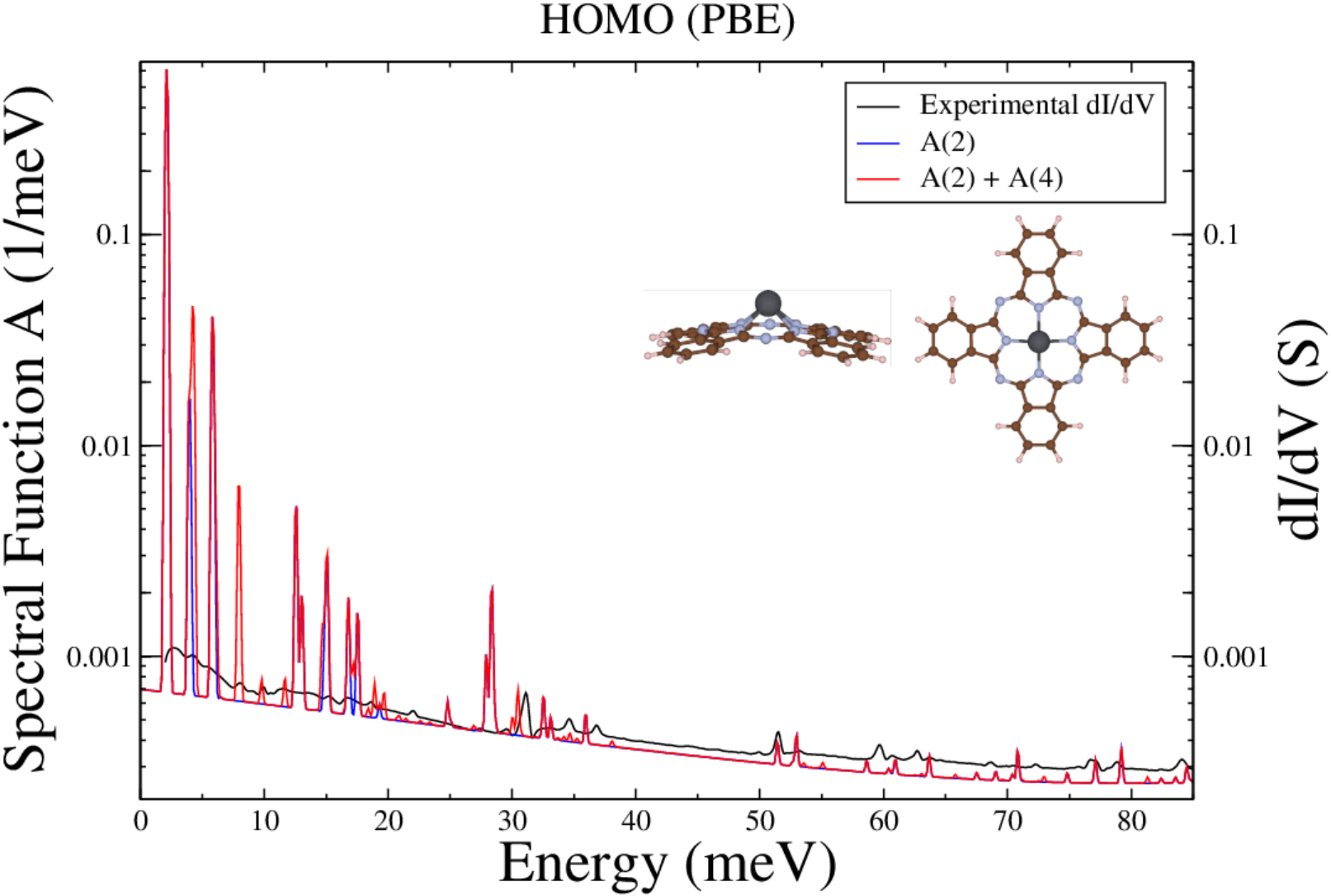}
\caption{Spectral function for the HOMO of the isolated PbPc using the PBE functional, compared with the experimental conductance spectrum. The optimized geometry of the molecule appears in the inset. This graph covers a larger energy range compared to Fig.1 from main text. Due to the differences in order of magnitude of the spectrum in low and high energies, logarithmic scale is used.}
\label{figS11}
\end{figure}

\begin{figure}[ht]
\centering
\begin{subfigure}[b]{0.45\textwidth}
    \centering
    \includegraphics[width=\textwidth]{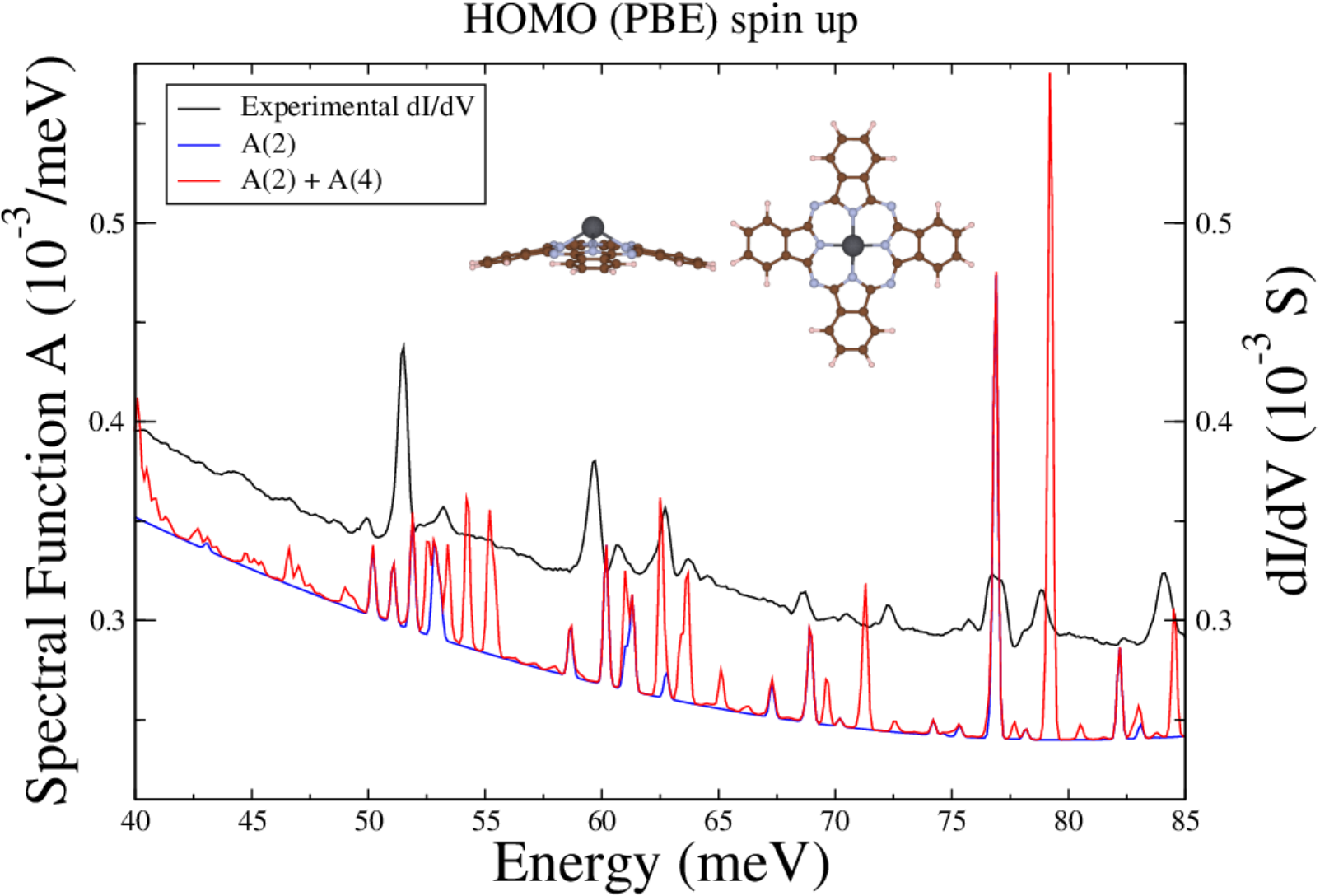}
    \caption{}
    \label{figS12a}
\end{subfigure}
\begin{subfigure}[b]{0.45\textwidth}
    \centering
    \includegraphics[width=\textwidth]{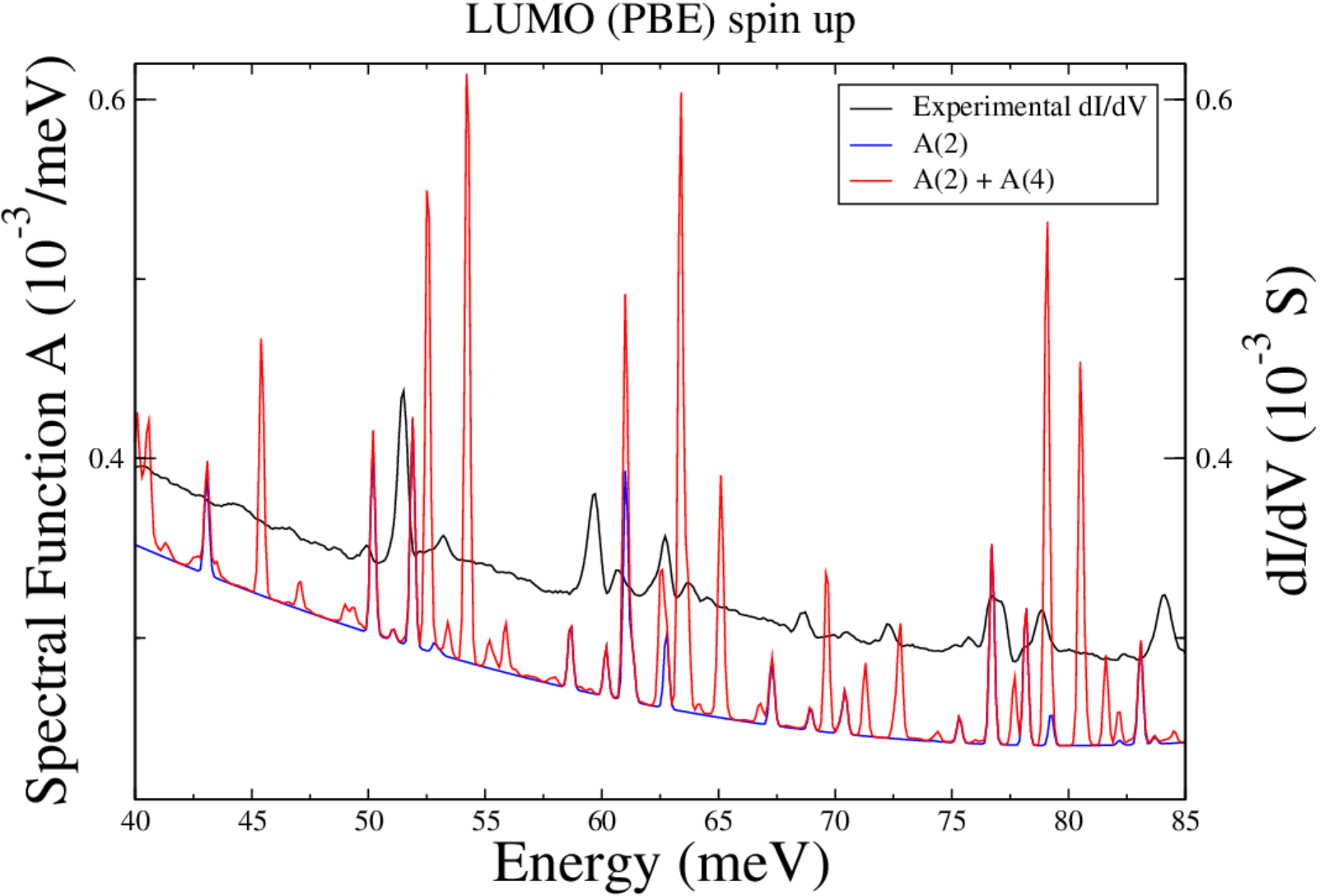}
    \caption{}
    \label{figS12b}
\end{subfigure}
\caption{Spectral function of PbPc$^{1-}$ majority-spin HOMO (a) and LUMO (b) using the PBE functional, compared with the experimental conductance spectrum. The geometry employed was C$_{2v}$ PbPc$^{1-}$; see inset.}
\label{figS12}
\end{figure}

\begin{figure}[ht]
\centering
\begin{subfigure}[b]{0.45\textwidth}
    \centering
    \includegraphics[width=\textwidth]{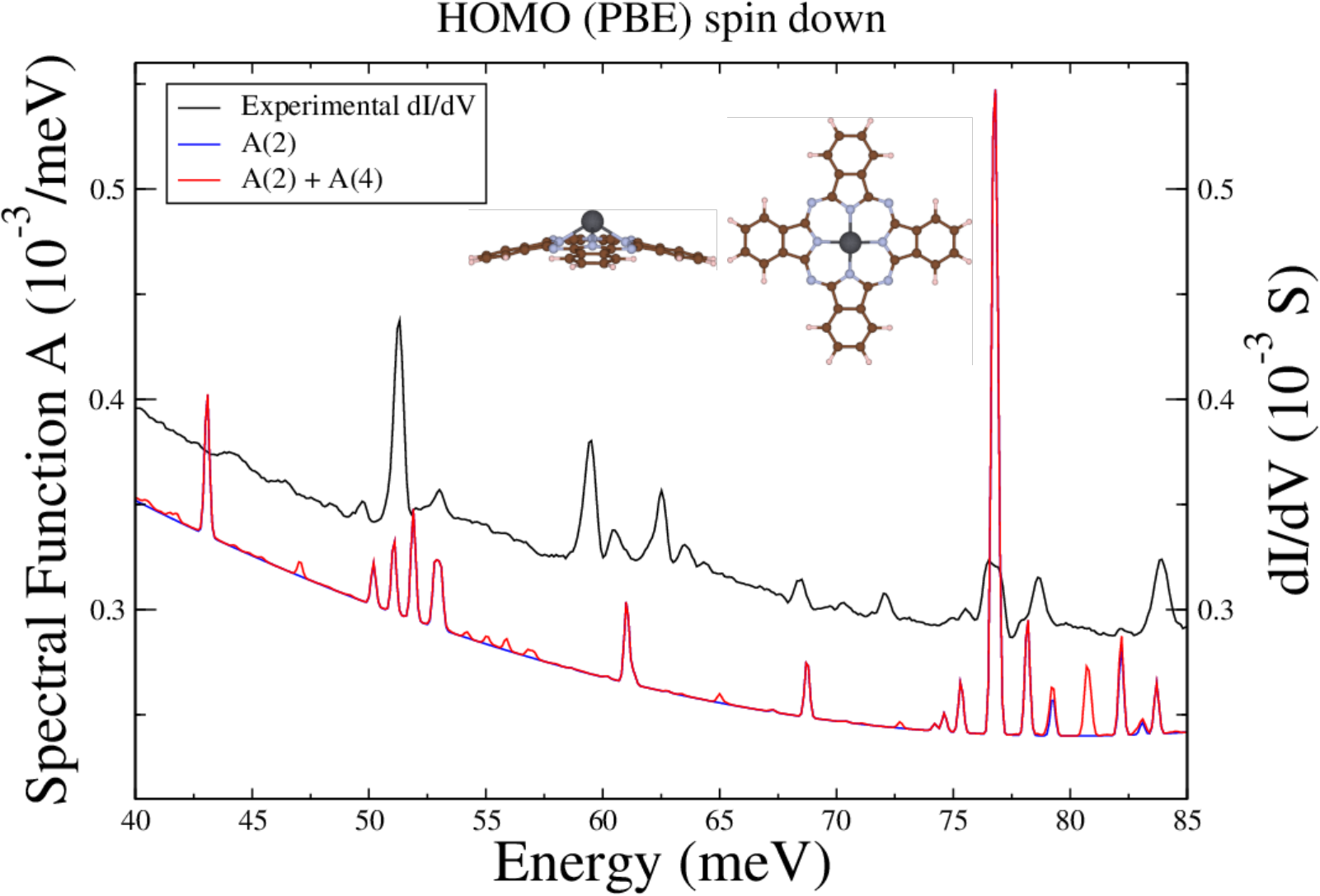}
    \caption{}
    \label{figS13a}
\end{subfigure}
\begin{subfigure}[b]{0.45\textwidth}
    \centering
    \includegraphics[width=\textwidth]{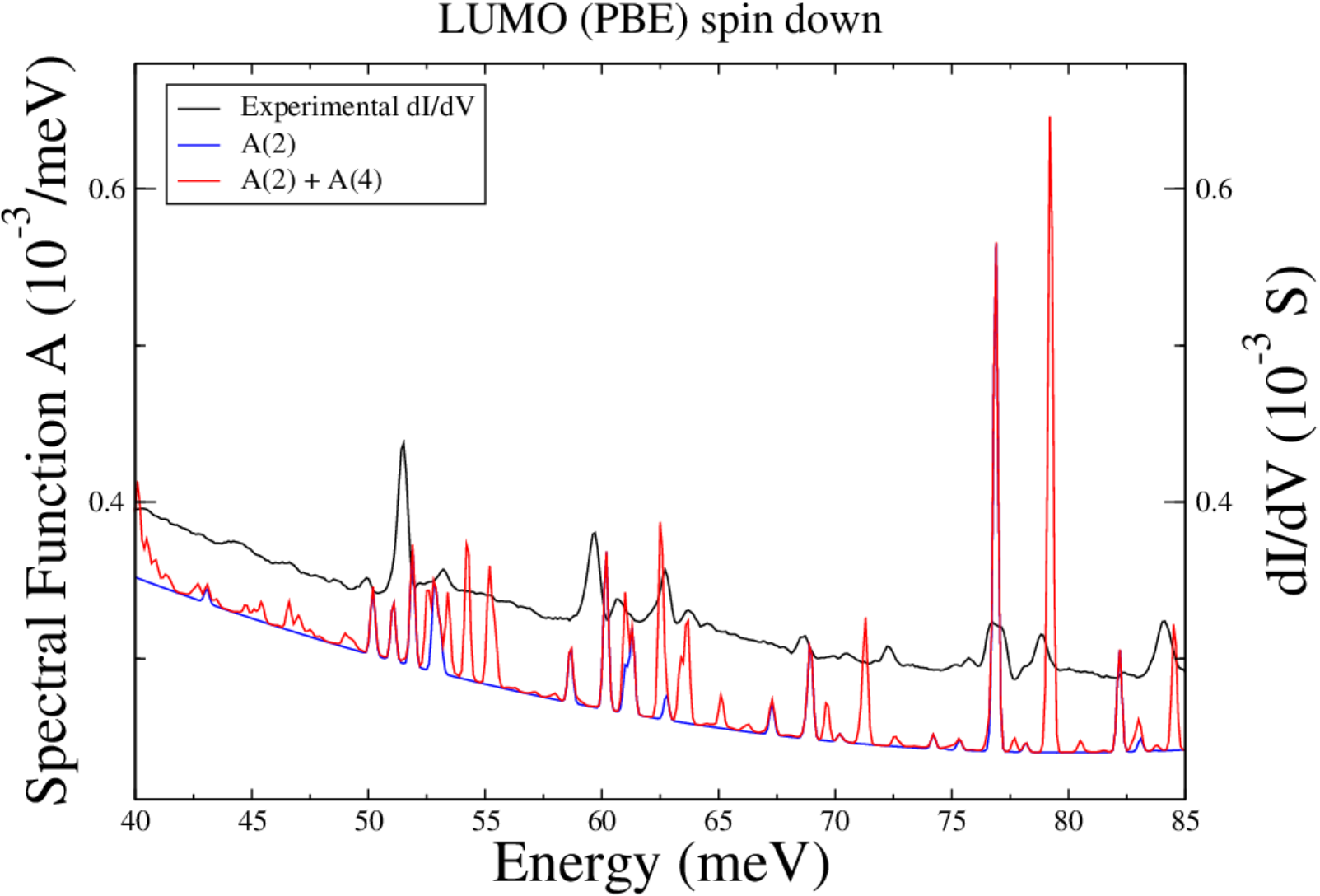}
    \caption{}
    \label{figS13b}
\end{subfigure}
\caption{Spectral function of PbPc$^{1-}$ minority-spin HOMO (a) and LUMO (b) using the PBE functional, compared with the experimental conductance spectrum. The geometry employed was C$_{2v}$ PbPc$^{1-}$; see inset.}
\label{figS13}
\end{figure}

\begin{figure}[ht]
\centering
\includegraphics[width=0.7\textwidth]{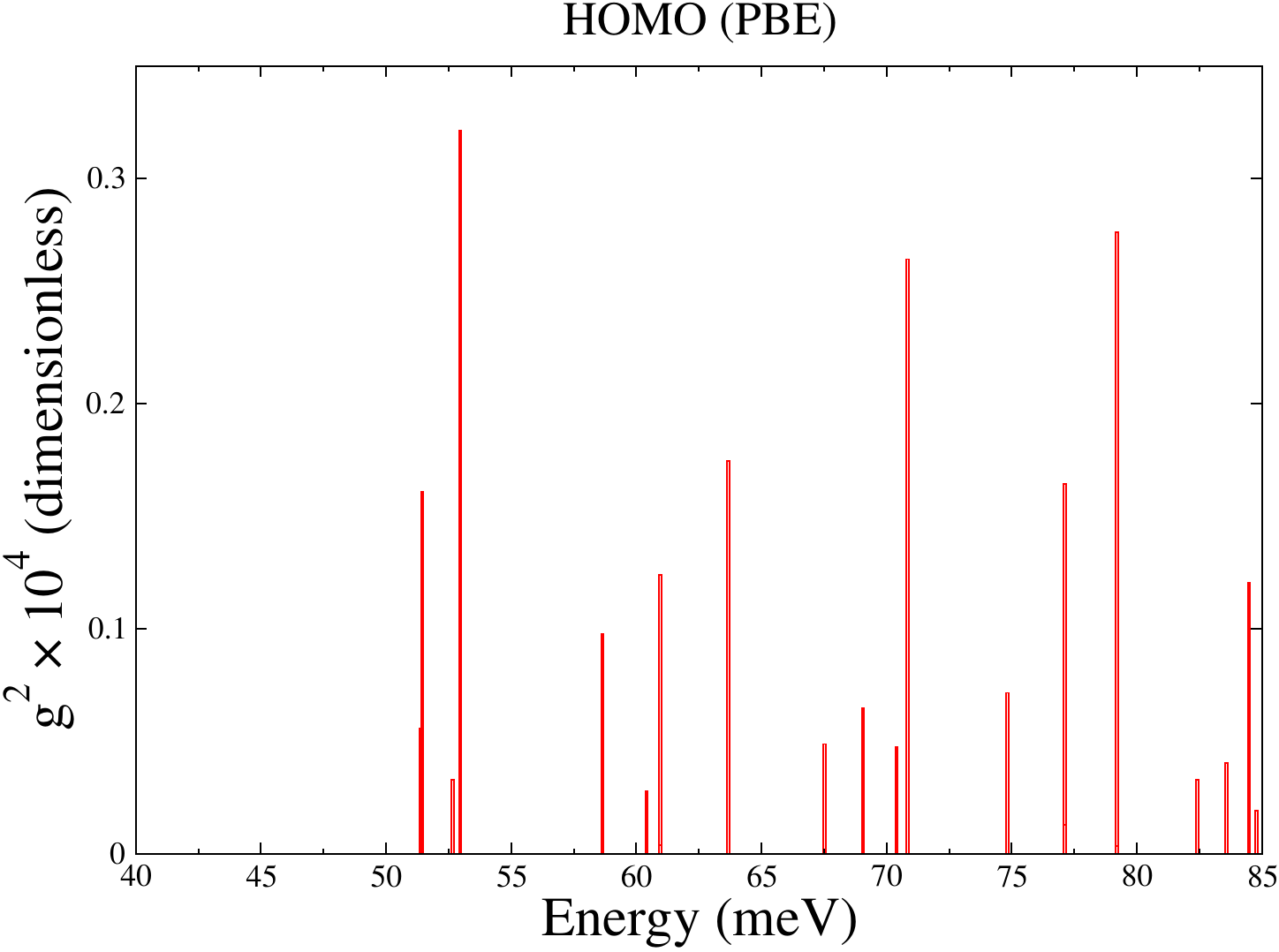}
\caption{Bar plot of electron-vibration coupling constants  $g_v:= {\lambda^{v}_{\alpha\alpha}}/{\hbar\omega_v}$
as a function of vibrational excitation energy for the HOMO of the neutral isolated PbPc.}
\label{figS14}
\end{figure}

\begin{figure}[ht]
\centering
\includegraphics[width=0.7\textwidth]{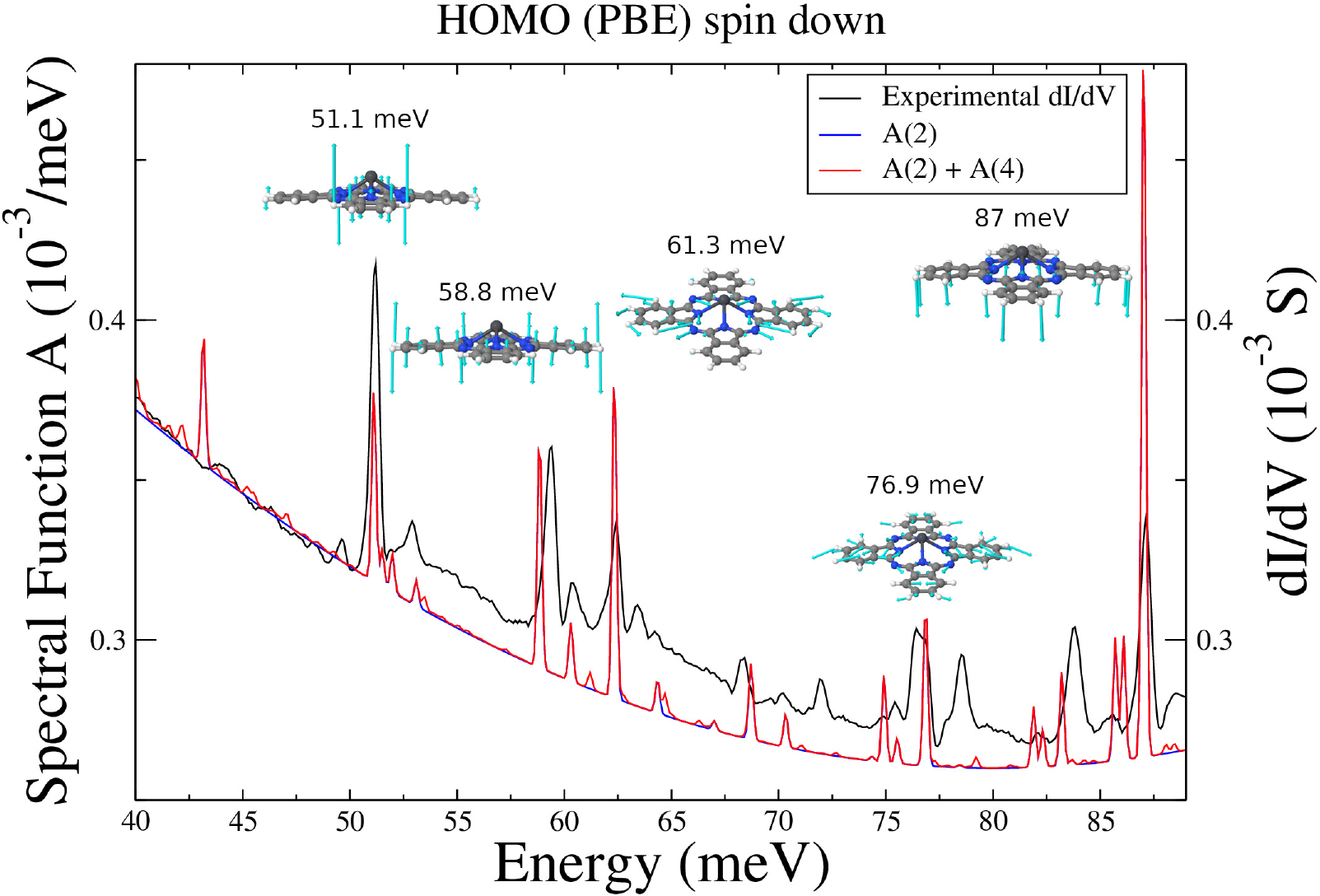}
\caption{Spectral function for the HOMO of PbPc$^{1-}$/surf (minority spin), together with the experimental $dI/dV$ curve. Inside the graph, the vibrational modes of the molecule that correspond to the 5 highest computational EV peaks are shown.}
\label{figS15}
\end{figure}

\begin{figure}[ht]
\centering
\begin{subfigure}[b]{0.45\textwidth}
    \centering
    \includegraphics[width=\textwidth]{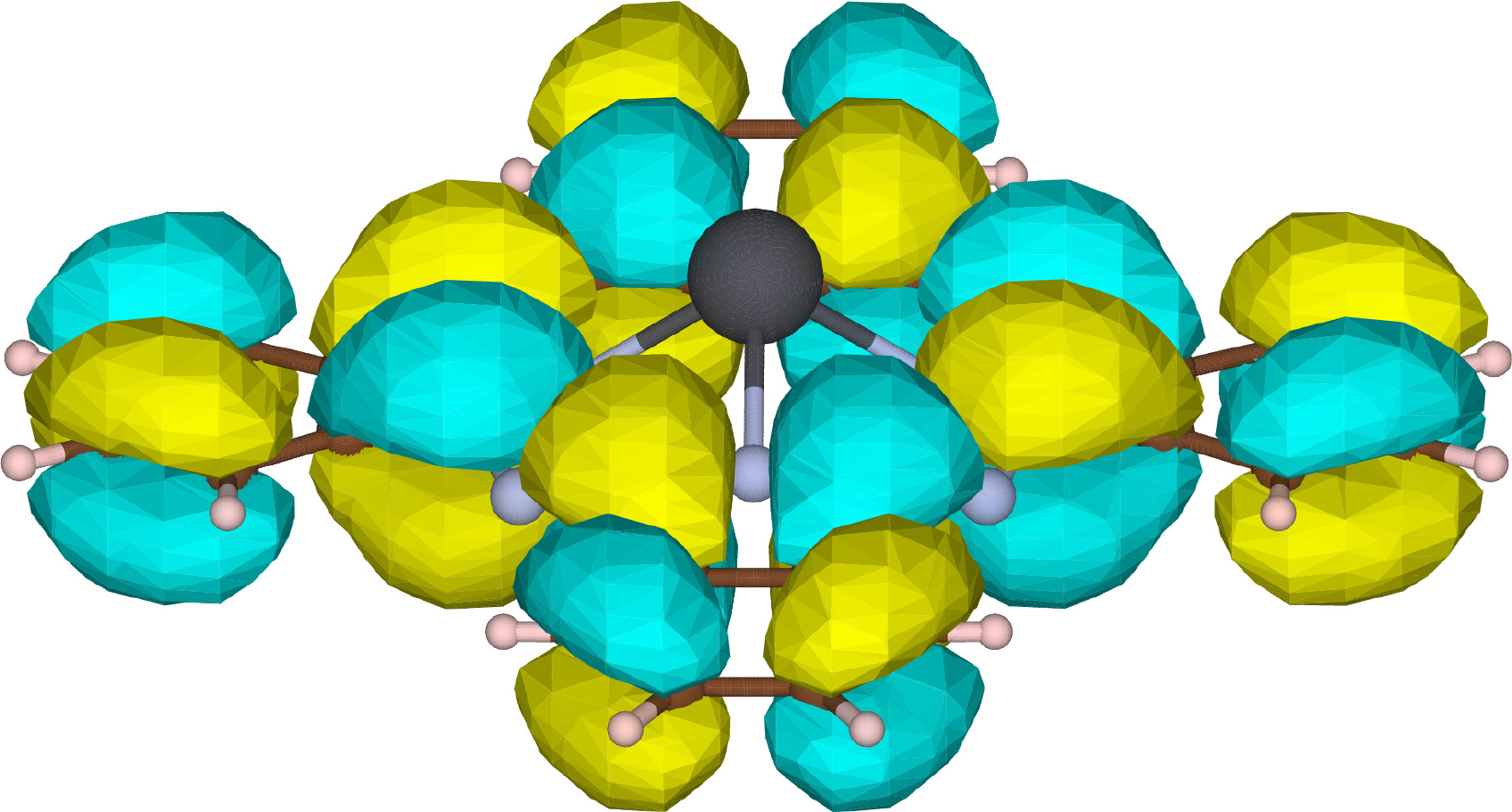}
    \caption{}
    \label{figS16a}
\end{subfigure}
\begin{subfigure}[b]{0.45\textwidth}
    \centering
    \includegraphics[width=\textwidth]{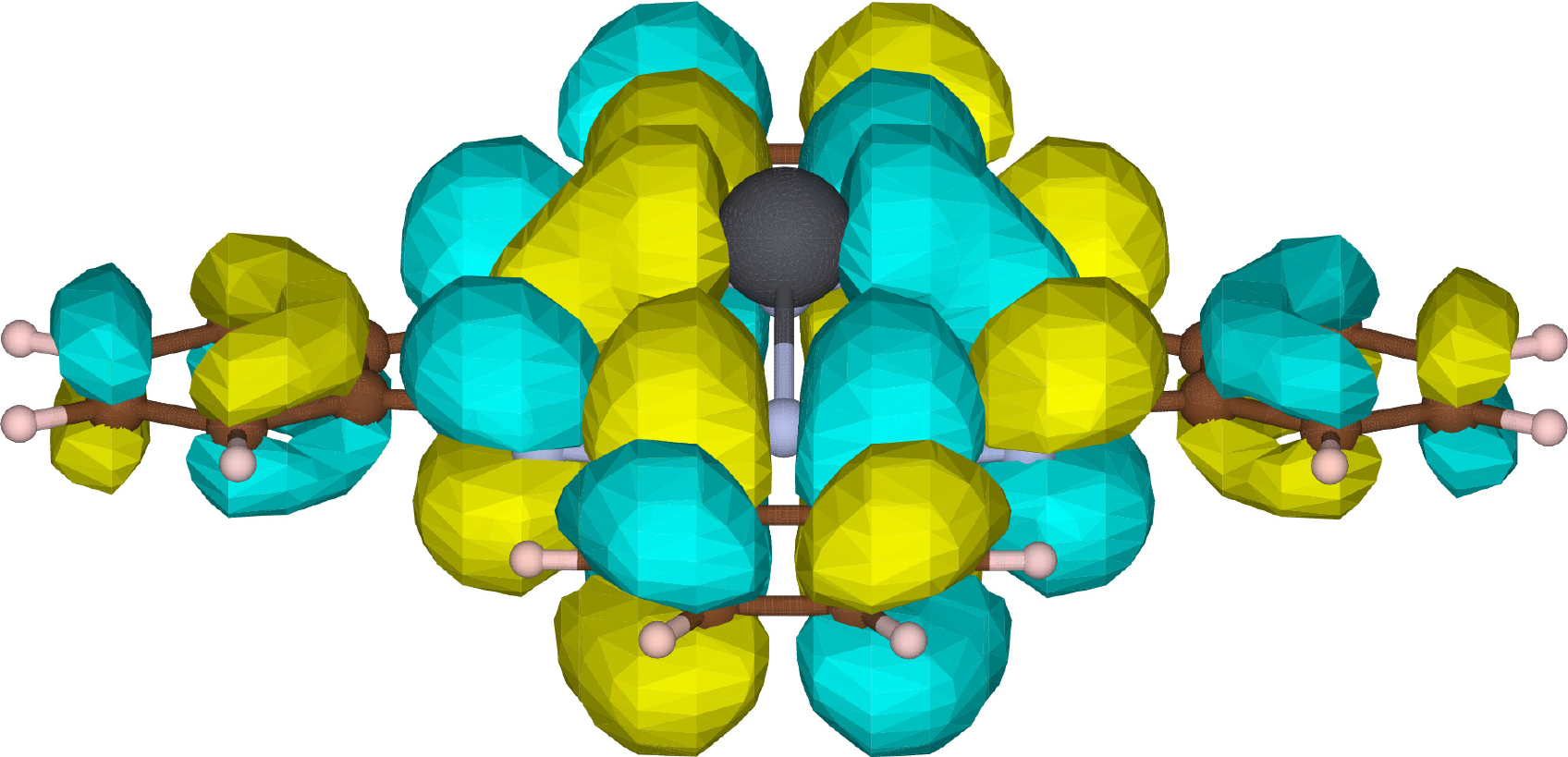}
    \caption{}
    \label{figS16b}
\end{subfigure}
\caption{Isosurface plots of HOMO (a) and LUMO (b) of the minority spin for PbPc$^{1-}$/surf.}
\label{figS16}
\end{figure}

\clearpage

\begin{table}[h] 
    \caption{Data for the 20 peaks indicated in Fig. 2(a). The units of $E$ (energy of the peak) are in meV and $H^{th}_i$ and $H^{exp}_i$ (relative height of the peak from the curve background) are in $10^{-3}$ meV$^{-1}$ and $10^{-3}$ S, respectively. Based on these data and using the equation for the mean deviation ($MD$) shown in the main text, $MD=0.347$.}
    \begin{ruledtabular}
    \begin{tabular}{ccccc}
        Peak ($i$) & $E^{exp}_i$ & $E^{th}_i$ & $H^{exp}_i(\times10^3)$ & $H^{th}_i(\times10^3$)\\
        \hline
        1 & 43.82 & 43.22 & 0.006723 & 0.039599\\
        2 & 51.21 & 51.09 & 0.093277 & 0.059244\\
        3 & 51.99 & 51.87 & 0.001680 & 0.010504\\
        4 & 52.89 & 53.13 & 0.012605 & 0.008404\\
        5 & 59.37 & 58.83 & 0.055042 & 0.065546\\
        6 & 60.39 & 60.27 & 0.012605 & 0.016386\\
        7 & 60.91 & 61.19 & 0.000831 & 0.004818\\
        8 & 62.37 & 62.31 & 0.034454 & 0.096639\\
        9 & 64.18 & 64.30 & 0.003781 & 0.008403\\
        10 & 68.38 & 68.68 & 0.013025 & 0.021428\\
        11 & 70.24 & 70.30 & 0.004201 & 0.009243\\
        12 & 74.86 & 74.92 & 0.002521 & 0.024790\\
        13 & 75.40 & 75.53 & 0.005042 & 0.007563\\
        14 & 76.43 & 76.85 & 0.031932 & 0.045378\\
        15 & 82.07 & 81.83 & 0.03362 & 0.018908\\
        16 & 82.49 & 82.31 & 0.001282 & 0.009341\\
        17 & 83.75 & 83.21 & 0.033194 & 0.028571\\
        18 & 85.55 & 85.67 & 0.005042 & 0.037815\\
        19 & 85.80 & 86.10 & 0.001312 & 0.031109\\
        20 & 87.11 & 86.99 & 0.065546 & 0.212605\\
      
        \end{tabular}
    \end{ruledtabular}
\end{table}